\documentclass[preprint2]{aastex}
\usepackage{amsmath}
\usepackage{natbib}
\usepackage{graphicx}
\usepackage{txfonts}
\usepackage[pagebackref,breaklinks,colorlinks,citecolor=blue]{hyperref}
\usepackage[all]{hypcap}

\usepackage[nameinlink]{cleveref}
\usepackage{epsfig,color}

\newcommand{\eq}[1]{\begin{equation}  #1 \end{equation}}   
\newcommand{\eqa}[1]{\begin{align}  #1 \end{align}} 
\newcommand{\bb}[1]{\left[ #1 \right]}   
\newcommand{\ba}[1]{\left\langle #1 \right\rangle}   
\newcommand{\nn}{\nonumber}

\newcommand{\dd}{{\rm d}}    
\newcommand{\permapj}[1]{\textit{\copyright\ AAS. Reproduced with permission from} \citet{#1}.}  
\newcommand{\permaa}[1]{\textit{Reproduced with permission from} \citet{#1} \textit{\copyright\ ESO.}}  
\newcommand{\permmn}[1]{\textit{Reproduced with permission from} \citet{#1}.}  

\onecolumn

\usepackage{amsmath,amssymb}

\begin{document}
\title{Galaxy alignments: Observations and impact on cosmology}

\author{Donnacha Kirk\altaffilmark{1}, Michael L. Brown\altaffilmark{2}, Henk Hoekstra\altaffilmark{3}, Benjamin Joachimi\altaffilmark{1}, Thomas D. Kitching\altaffilmark{4}, Rachel Mandelbaum\altaffilmark{5}, Crist{\'o}bal Sif{\'o}n\altaffilmark{3}, Marcello Cacciato\altaffilmark{3}, Ami Choi\altaffilmark{6}, Alina Kiessling\altaffilmark{7}, Adrienne Leonard\altaffilmark{1}, Anais Rassat\altaffilmark{8}, Bj{\"o}rn Malte Sch{\"a}fer\altaffilmark{9}}
\email{drgk@star.ucl.ac.uk}

\altaffiltext{1}{Department of Physics \& Astronomy, University College London, Gower Street, London, WC1E 6BT, UK}
\altaffiltext{2}{Jodrell Bank Centre for Astrophysics, School of Physics and Astronomy, University of Manchester, Oxford Road, Manchester M13 9PL, UK}
\altaffiltext{3}{Leiden Observatory, Leiden University, PO Box 9513, NL-2300 RA Leiden, Netherlands}
\altaffiltext{4}{Mullard Space Science Laboratory, University College London, Holmbury St Mary, Dorking, Surrey RH5 6NT, UK}
\altaffiltext{5}{McWilliams Center for Cosmology, Department of Physics, Carnegie Mellon University, Pittsburgh, PA 15213, USA}
\altaffiltext{6}{Scottish Universities Physics Alliance, Institute for Astronomy, University of Edinburgh, Royal Observatory, Blackford Hill, Edinburgh EH9 3HJ, UK}
\altaffiltext{7}{Jet Propulsion Laboratory, California Institute of Technology, 4800 Oak Grove Drive, Pasadena, CA 91109, USA}
\altaffiltext{8}{Laboratoire d'astrophysique (LASTRO), Ecole Polytechnique Fédérale de Lausanne (EPFL), Observatoire de Sauverny, CH-1290 Versoix, Switzerland}
\altaffiltext{9}{Astronomisches Recheninstitut, Zentrum f{\"u}r Astronomie der Universit{\"a}t Heidelberg, Philosophenweg 12, 69120 Heidelberg, Germany}

\begin{abstract}
  Galaxy shapes are not randomly oriented, rather they are
  statistically aligned 
  in a way that can depend on formation environment,
  history and galaxy type. Studying the alignment of galaxies can
  therefore deliver important information about the physics of
  galaxy formation and evolution as well as the growth of structure in the
  Universe. In this review paper we summarise
  key measurements of galaxy alignments, divided by 
  galaxy type, scale and environment.  We also cover the statistics and
  formalism necessary to understand the observations in the literature. With the emergence of weak gravitational lensing as a
  precision probe of cosmology, galaxy alignments have taken on an added
  importance because they can mimic cosmic shear, the effect of
  gravitational lensing by large-scale structure on observed galaxy
  shapes. This makes galaxy alignments, commonly referred to as intrinsic alignments, an important systematic
  effect in weak lensing studies. We quantify the impact of intrinsic alignments on cosmic shear surveys and finish by reviewing practical
  mitigation techniques which attempt to remove contamination by intrinsic alignments. 
\end{abstract}


\vspace{10mm}

\tableofcontents

\section{Introduction} 
\label{sec:introduction}

Galaxies show a wide variation in morphological appearance, due to the complex processes of galaxy formation and evolution. 
Both initial conditions and 
interactions between galaxies can play an important role. For instance, most elliptical galaxies are believed to be the result of major mergers
of galaxies \citep{TT72,S86,BH92}. Consequently the morphology of galaxies is connected to the local environment, as is evidenced by the 
well-established morphology-density relation \citep[e.g.][]{D80}. The connection between morphology and galaxy formation and evolution was made early-on, most notably by \citet{Hubble26} who believed that elliptical galaxies would eventually transform into grand-design spiral galaxies. Although this picture has been reversed in recent years, the importance of morphology has remained and galaxy shape is one of the most important observables that can be used to describe a
galaxy. 

As it was realised that galaxies may influence each other, other questions become relevant for our understanding of galaxy formation and evolution, such as ``Why are galaxies spinning?'' and ``Are the orientations of galaxies correlated?'' These questions have been the main motivator for observational studies of galaxy alignments during the 20th century (a detailed historical overview of the subject can be found in our companion paper \citealt{Paper1}). However,
no consensus was reached on the existence of of alignments between galaxy shapes or spins. For instance, some studies have claimed that cluster galaxies are preferentially 
oriented towards the bright central galaxy, whereas others found no evidence for this. Much of this controversy can be attributed to the quality of the data, but
differences in observational techniques can play a role as well. Weak gravitational lensing as a cosmological tool provided new impetus for the study of galaxy shapes and alignments in the 21st century. Weak lensing measures coherent distortions to the images of background sources that can be mimicked or hidden by galaxy shape alignments.

\begin{figure}[ht!]
    \centering
       \includegraphics[width=3in,height=3in]{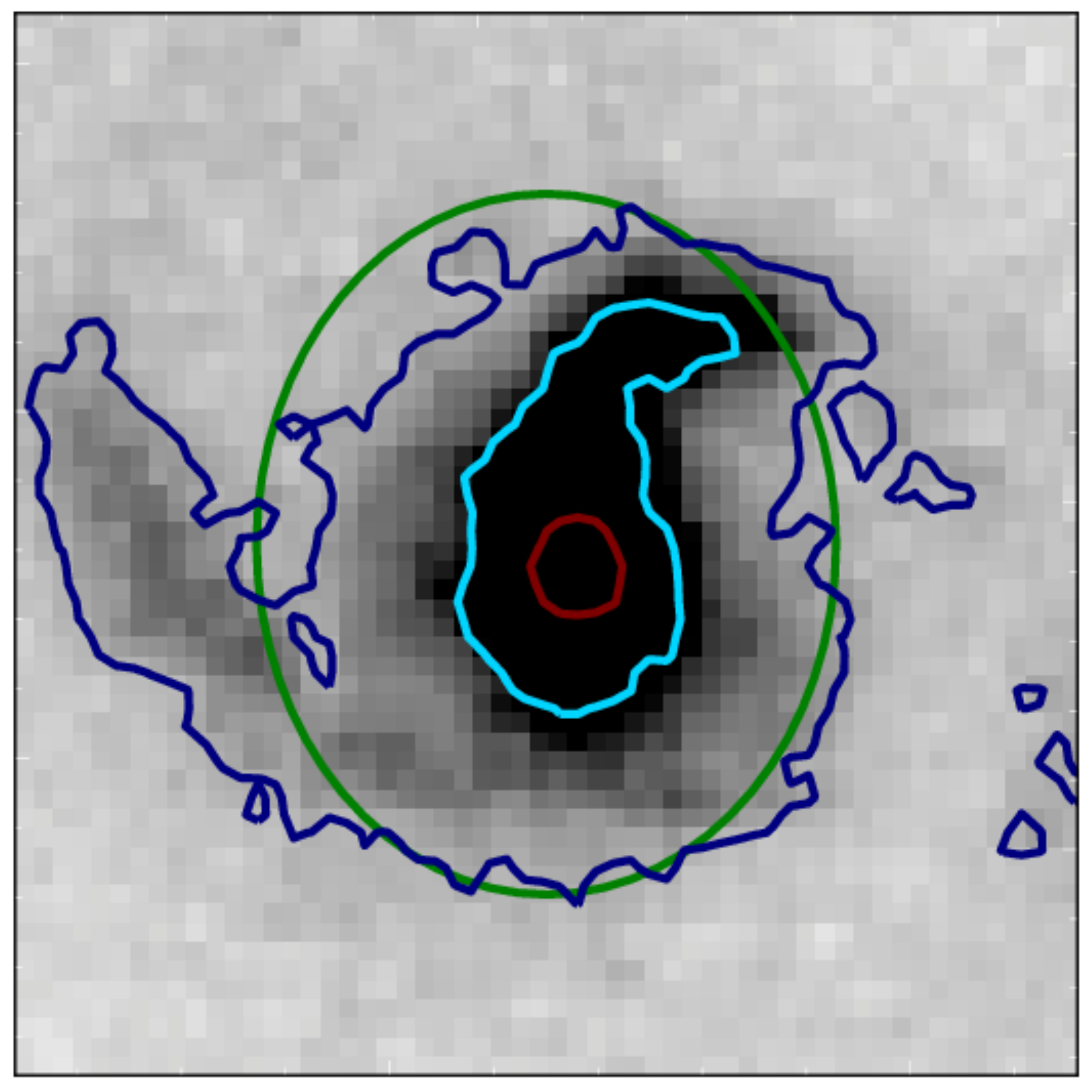}
\vspace{0mm}
\caption{An example of a well-resolved galaxy observed as part of the
  COSMOS survey. The dark blue, light blue and red contours mark 
  isophotes that match 5\%, 20\% and 50\% of the peak flux level,
  respectively. The morphology clearly varies with surface brightness
  and the measured shape will be a strong function of the isophotal limit 
  that is adopted. For comparison the green ellipse indicates the 
  ellipticity of the best fit single S\'{e}rsic model.}
\label{fig:example_cosmos}
\end{figure}

Galaxy shapes and orientations can be measured using different
approaches. For instance, one can consider the region of a galaxy
above a given surface brightness and determine its ellipticity and
position angle. In particular, early studies, based on
photographic plates, tend to fall into this category because the
measurement involved determining the semi-major and semi-minor axis above some
surface brightness limit. Even though the resulting ellipticity might
be biased, due to the particular choice of surface brightness, the estimate of the position angle from these early studies is expected to be
robust, provided the galaxy is much larger than the size of the point
spread function (PSF). The adopted surface brightness limit, which may
itself be determined by the depth of the observations, can affect the
result because low surface brightness features, such as discs or even
tidal tails, only show up if the data are sufficiently
deep. \Cref{fig:example_cosmos} highlights this problem: we show
an example of a well resolved galaxy observed by the Hubble Space Telescope (HST) as part of
the COSMic evOlution Survey \citep[COSMOS,][]{SAB+07}. The different isophotes that are indicated
show how the morphology of real galaxies varies dramatically as a
function of surface brightness level. For reference the green ellipse
in \Cref{fig:example_cosmos} corresponds to the best fit
single S\'{e}rsic model \citep{S68} to the galaxy image.

That this change in morphology with surface brightness can lead to wildly varying conclusions about the level of
galaxy shape alignments can be understood by considering a very simple case:
imagine a scenario where all galaxies are made up of a central bulge component and a broad disc. Now let the bulges of galaxies be strongly aligned
but discs be oriented randomly. If one were to measure the
orientation of some faint isophote, i.e. probing the discs, no
alignment would be measured. On the other hand, shallower data would
probe the brighter bulges, leading to strong alignments. Remember that this is not a particularly physical example as interactions between bulge and disc components could introduce alignments.

Similarly, as we discuss in more detail below, weak lensing studies
usually measure galaxy shapes using algorithms that give more weight to the
brighter, central regions of galaxies, although the precise radial
weighting differs between methods. Biased measurements
may lead to incorrect conclusions: light from bright cluster galaxies
may affect the shape measurements of fainter satellites, resulting in
spurious alignments. Similarly, if the PSF is anisotropic, it will
lead to apparent alignments, especially if the galaxies are poorly resolved. A
comparison of alignment results therefore requires a careful study of
the methodology used to perform the measurement.

Early observational studies focused on understanding the physical origin of the alignments of galaxy positions and shapes, but the first detections of the weak lensing signal by large-scale structure (LSS) \citep{BRE00,KWL00,WME00,WTK00} spurred new
interest in the
topic. Weak gravitational lensing seeks to exploit the alignment in observed galaxy shapes caused by the deflection of light by gravity \emph{en route} from the source galaxy to the observer.
What makes weak gravitational lensing particularly interesting is that the observed galaxy alignments  can in principle be used to reconstruct the 
projected mass distribution or to study the statistical properties of the large-scale structure \citep{TJ04,HJ08,MVW+08}. We will refer to the alignments of galaxy shapes caused by galaxy formation and evolution as \emph{intrinsic alignments} to differentiate them from alignments sourced by gravitational lensing which we will call \emph{cosmic shear}. Weak lensing is a powerful cosmological probe. It is directly sensitive to the dark matter distribution, unlike galaxy redshift surveys which rely on using galaxies as biased tracers of the underlying dark matter field. In addition weak lensing is sensitive to both the growth of structure and the expansion history of the Universe through a geometric term in the lensing kernel. See \citet{BS01,HJ08,MVW+08,Ba10} for general reviews of weak lensing.

For the study of cosmic shear we need accurate estimates for both the ellipticity (also referred to as the third flattening) and orientation on the sky of the galaxies, which can be expressed as
\begin{equation}
\left( \begin{matrix} \epsilon_1 \\ \epsilon_2 \end{matrix} \right) = \frac{1-q}{1+q}\left( 
\begin{matrix} \cos 2\vartheta_{\rm P} \\ \sin 2\vartheta_{\rm P} \end{matrix} \right),
\label{eqn:red_shear}
\end{equation}
\noindent where $\epsilon_\textrm{i}$, with $i=1,2$, are the two ellipticity components, $q=b/a$ is the ratio ($0 \leq q \leq 1$) of the estimated semi-minor and semi-major axes, or axial ratio in short, and $\vartheta_{\rm P}$ is the position angle with respect to some reference axis. The factors of $2\vartheta_{\rm P}$ come from the spin-2 nature of ellipticity due to the symmetry of an ellipse under 180$^\circ$ rotation. Alternatively one can use complex notation, such that $\epsilon=\epsilon_1+\textrm{i}\epsilon_2$ or $\epsilon=|\epsilon|e^{2\mathrm{i}\vartheta_{\rm P}}$. If observed galaxy images are randomly oriented, the ensemble average $\langle\epsilon_i\rangle=0$.

The differential deflection of light rays by the intervening large-scale structure distorts the images of distant galaxies, resulting in  {\it apparent} alignments of the observed shapes. To leading order, i.e. in the weak lensing limit, the effect is to change $\epsilon^{\rm s}$, the intrinsic (or unlensed) ellipticity  of a galaxy, to an observed value \citep{SS95}:
\eq{
\label{eqn:epsgamma}
\epsilon = \frac{\epsilon^{\rm s}+g}{1+\epsilon^{\rm s}g^*} \approx \epsilon^{\rm s}+ \gamma,}
where the asterisk denotes the complex conjugate and $ g $ is the reduced shear, which is related to the weak lensing shear $\gamma$ and convergence $\kappa$ through $g = \gamma/(1-\kappa)$. In the weak lensing regime, we assume that $g \approx \gamma$. A brief introduction to the topic is given in \citet{Paper1}, which also lists references to more thorough discussions of gravitational lensing. The above approximation, \Cref{eqn:epsgamma}, is only true when we take an ensemble average over many galaxies. For individual galaxies there is an additional term of the same order of magnitude. This term is not relevant for the topics covered in the rest of the paper, see \citet{BS01} for more detail. Note that the above formalism highlights the usefulness of expressing the ellipticity and shear in complex notation.  

The measurement of the weak lensing signal involves the correlation, or averaging, of the ellipticity measurements for many galaxies because the
typical lensing-induced  change in ellipticity is $\sim 1\%$ or less, much smaller than the average intrinsic galaxy ellipticity. This demonstrates the relevance of intrinsic alignments for lensing studies: only if such alignments vanish, $\langle \epsilon^{\rm s} \rangle = 0$, is the observed ensemble-averaged ellipticity an unbiased estimator of the lensing shear. Similarly, the galaxy ellipticity correlation function (see \Cref{sec:shape_correlations})
comprises types of contributions (where $i$ and $j$ here indicate a pair of galaxies and the average is over all pairs):

\eqa{
\label{eqn:iadef}
\underbrace{\ba{{\epsilon^i} {\epsilon^j}}} &= \underbrace{\ba{{\gamma^i}
      {\gamma^j}}} + \underbrace{\bigl\langle {\epsilon^{\mathrm{s},i}}
    {\epsilon^{\mathrm{s},j}}} \bigr\rangle + \underbrace{\bigl\langle {\gamma^i}
    {\epsilon^{\mathrm{s},j}} \bigr\rangle+\ba{{\epsilon^{\mathrm{s},i}}
      {\gamma^j}}}{}\;.\\ \nn \mbox{observed} & \hspace*{0.5cm} \mbox{GG} \hspace*{1cm}
  \mbox{II} \hspace*{1.5cm} \mbox{GI}
  }
We adopt the following common shorthand notation: GG for the cosmic shear correlation, which is the quantity of interest to constrain cosmological
parameters, II indicates the correlations between intrinsic ellipticities, and GI denotes correlations between the shear for one galaxy and the
intrinsic ellipticity of the other. In principle one of the GI terms should vanish if foreground and background galaxies can be cleanly separated because the lensing of light from galaxies should not correlate with the intrinsic ellipticity of more distant galaxies.

To understand what these correlations mean physically imagine some region of matter overdensity. This overdensity generates a powerful gravitational potential on scales larger than individual galaxies. This large-scale gravitational potential may cause the orientation and shapes of nearby galaxies to align with the overdensity, due to the effect of the gravitational tidal field. Neighbouring galaxies are affected by the same large-scale tidal field so their mutual alignment with the matter overdensity means that the shapes of those galaxies are now correlated - they point in the same direction. This is what we call the intrinsic-intrinsic (II) correlation. It effects physically close galaxies and the induced correlations have the same orientation as those sourced by weak lensing, meaning the II correlation manifests as a spurious extra signal in cosmic shear studies.

The gravitational-intrinsic (GI) correlation is more subtle. As noted by \citet{HS04}, the II term correlates the shapes of physically close galaxies with some matter overdensity, which in turn is responsible for a lensing effect on background galaxies. The net result is an anti-correlation between the shapes of galaxies in the foreground, whose shape is sourced by the gravitational tidal field from the mass, and those in the background, whose shape is effected by the mass through weak lensing. The GI correlation is a long-range effect and manifests as a negative contribution to measured cosmic shear. The extent to which tidal fields do source alignments, and how this varies between different galaxy populations, is something we explore throughout the rest of this paper. 

If the spurious contributions to cosmic shear caused by intrinsic galaxy alignments are significant compared to the statistical errors of the survey then a naive analysis, which ignores the impact of such alignments, will produce biased estimates of cosmological parameters. Although current work indicates that the intrinsic
alignment signal is too low to have affected the conclusions of early cosmic shear studies, it is also clear that we can no longer ignore this
astrophysical source of bias \citep{HGH+13} and that it will be a significant limitation for future projects, unless we can account for intrinsic alignments
in the analysis.

Although the existence of intrinsic alignments has now been firmly established for luminous red galaxies \citep[LRGs;][]{MHI+06,HMI+07,OJL09,JMA+11,SMM14}, current observational
constraints are not sufficient to correct future cosmic shear surveys \citep{LBK+12,KLB+13,KRH+12}. 
Further progress relies on making observations with sufficient redshift precision and spatial coverage to inform models of intrinsic alignments, or calibration
using the cosmic shear survey data themselves \citep{Bernstein09,JB10}. Importantly, as we will discuss in more detail below, the intrinsic alignment estimates themselves have to be 
closely linked to the shear measurements in terms of shape measurement, galaxy populations and observational strategy. A further complication is that the source and strength of galaxy alignment depends on galaxy type for reasons described in detail below and treated extensively in our companion paper \citet{Paper2} and the papers referenced therein. Although one can attempt to restrict the analysis to a particular type of galaxy, the source sample typically comprises a mix of galaxies thus mixing possible alignment mechanisms, complicating the analysis (see \cite{Paper1} and references therein for more detail). 
We have been fortunate so far that our knowledge of intrinsic alignments, combined with the size of the observed signal, has kept pace with requirements for cosmic shear surveys to remain unbiased \citep{HGH+13}. As powerful next generation surveys become available, their reduced statistical error will require a new level of accuracy in quantifying systematic effects such as intrinsic alignments. The observation of galaxy alignments will remain an important topic as we demand more precise measurements over a wider range of scales and redshifts for all types of galaxies.

In this review we focus on observations of the intrinsic alignments of galaxies (particularly in \Cref{sec:observations_largescale} and \Cref{sec:observations_smallscale}), their impact on cosmic shear measurements and possible mitigation strategies (\Cref{sec:impact}). This current paper is one of three companion papers covering the entire topic of galaxy alignments. Therefore some topics are covered in less detail here or, occasionally, omitted.
For a general introduction and historical review of the subject we refer the reader to \citet{Paper1}, which summarises the basic concepts and highlights the most important developments in theory, modelling and observations. A detailed discussion of the physical theories used to model alignments on a range of scales is presented in our companion paper \citet{Paper2}, which also reviews intrinsic alignment studies conducted through simulations, thus representing the theoretical counterpart of this more observationally-oriented review. \citet{TI15} also provide an independent review of galaxy alignments and \citet{S09} is a detailed review of angular momentum correlations.

This review is structured as follows: In \Cref{sec:observables} we discuss how galaxy alignments are measured. The main statistics
that are used to quantify the galaxy shape alignment signal are reviewed in \Cref{sec:shape_correlations}. 
The most important observations of shape alignment on linear and quasi-linear large scales are discussed in 
\Cref{sec:observations_largescale} and details of environmentally-dependent correlations are reviewed in \Cref{sec:observations_smallscale}. 
In \Cref{sec:impact} we demonstrate the impact that intrinsic alignments can have on attempts to infer cosmological parameters from cosmic shear surveys as well as outlining the most effective ways to mitigate this impact. We summarise in \Cref{sec:summary} and discuss the outlook for future observations of galaxy shape alignments.

\section{Quantifying orientations and shapes}
\label{sec:observables}

According to our current understanding, we can distinguish between two types of alignments. In the case of late-type (disc) galaxies the alignments of the angular
momenta are believed to play the most important role, whereas the orientation of early-type (elliptical) galaxies is thought to be largely determined
by the build-up of the large-scale dark matter distribution that surrounds them. This is a distinction at the level of the theory of alignment origin but it may also have consequences visible in observations. See \cite{Paper1} for a general overview of these mechanisms and \citet{Paper2} for a detailed discussion. 

Although the physical processes at play determine the
strength of the alignment as a function of separation, the method used to quantify the alignment signal plays an important role as well.
Whether or not this is an issue depends on the scientific question that one wishes to address.

In \Cref{sec:obs_orientations} we consider the measurement of galaxy orientations, concentrating on observing galaxy spin alignment in \Cref{sec:obs_spin}. We then cover the measurement of galaxy shapes in \Cref{sec:obs_measuring_shapes}, including discussion of some common methods. Prominent systematic effects in shape measurement studies are described in \Cref{sec:obs_sys} and some specific conclusions on the measurement of galaxy shapes with the aim of mitigating intrinsic alignments in cosmic shear surveys are made in \Cref{sec:obs_IA_cosmic_shear}.

\subsection{Using orientations}
\label{sec:obs_orientations}

Naturally, a critical ingredient in weak lensing studies is the measurement of the alignment of the shapes of galaxies. This measurement is quantified by
a galaxy's ellipticity and position angle with respect to some local coordinate system. In this context it is interesting to note that an observation of an alignment of position
angles (regardless of ellipticity) will be sufficient to imply that weak lensing measurements are contaminated by galaxy intrinsic alignments. However, an accurate estimate of the \emph{level} of contamination still requires
knowledge of the distribution of ellipticities, which itself may vary locally (as the mix of galaxy types depends on environment).  Hence,
while early studies that focused only on the orientations of galaxies have been useful, a successful correction of the intrinsic alignment signal
in weak lensing studies requires more information.

Studies using orientations rather than shapes also suffer a particular ambiguity: what is the
orientation of nearly round galaxies?  For a fixed total signal-to-noise galaxy detection, the
uncertainty on the galaxy orientation is smallest for highly flattened galaxies and largest for
those that are nearly round.  The way this problem is dealt with in practice varies, with some
studies ignoring it entirely \citep[e.g.,][]{PK05,AB06a,FLM+07,FLW+09,OJL09,LJF+13}.  Since ignoring
this issue will tend to dilute any alignments by adding random noise, that simple strategy is in fact a valid
approach when trying to simply {\em detect} alignments.  However, ignoring the dilution of the
orientation correlations due to nearly round galaxies in the sample complicates both the theoretical 
interpretation of the results and also the comparison with results from other
samples (which may have different intrinsic shape distributions and/or levels of noise).

Another approach is to exclude galaxies
with axial ratios 
$b/a \sim 1.0$ on the grounds that their position angles are
meaningless \citep{NSD+10}.  Again, such exclusion does not cause any problem for claiming a detection of intrinsic
alignments but 
this comes at the cost of the
interpretation of the measured alignments in terms of a theoretical model being complicated by
selection biases. 
\cite{SMB+12} give a simple example of a mathematical model for including uncertainties in position
angles in real data in a theoretical model for alignments. 
This
is done in the context of cluster alignments, but the same argument is valid for galaxy alignments.
Unfortunately, in this model, all galaxies are assumed to have, on average, the same position angle
uncertainty; if alignments vary with shape (just as the position angle uncertainties do), then this
prescription would no longer be applicable. An alignment estimate based on shape measurements does not suffer from these problems as there is no ambiguity in assigning a small ellipticity to a nearly circular galaxy, though, of course, signal-to-noise may be lower for galaxies with small ellipticities in real, noisy data.

\subsection{Spin alignments}
\label{sec:obs_spin}

Looking beyond position angles, measurements of the alignments between the angular momenta of galaxies may provide unique insights into the formation of
disc galaxies, especially on the origin of the observed galactic angular momentum. The ellipticities of disc galaxies are the result of the projection of their orientation with 
respect to the observer, combined with any intrinsic ellipticity due to not being a perfectly circular disc. If we assume that 
disc galaxies obtain their angular momentum from tidal torquing, they should spin around their minor axis. Inclination angle, $\xi$, refers to the angle between the observer's line of sight and the symmetry axis of a disc galaxy. Position angle, $\vartheta_{\rm P}$, refers to the angle between the major axis of the ellipse of a projected galaxy image and the north of some coordinate system. Assuming the circular, infinitely thin-disc approximation and
measurements of both of these angles, the components of the unit spin vector are given by
\begin{align}
\hat{J}_{r} &= \pm \cos \xi, \\
\hat{J}_{\theta} &= (1 - \cos^{2} \xi)^{1/2} \sin \vartheta_{\rm P}, \\
\hat{J}_{\phi} &= (1 - \cos^{2} \xi)^{1/2} \cos \vartheta_{\rm P},
\end{align}
where $r$, $\theta$ and $\phi$ are spherical polar coordinates with an origin at the galactic centre and $r$ pointing along the line of sight. The spatial correlation of the spin axes can then be written as
\begin{equation}
\eta (r) \equiv \left< | \hat{J}(x)\cdot \hat{J}(x+r)|^{2}\right> - \frac{1}{3},
\end{equation}
where we are averaging over the position, $x$, of all galaxy pairs separated by a distance $r$ \citep{Lee11}. The value of $1/3$ is subtracted since it is the value of the ensemble average when there is no correlation. 

The direction of the galactic angular momentum (i.e. the spin axis of the galaxy) is expected to be connected to the properties of the
underlying dark matter halo. If the correlation between angular momentum and the distribution of dark matter is retained after galaxy
formation, then the correlation of the density field, which can be predicted using numerical simulations \citep{Paper2}, should be replicated in
correlations of galaxy orientation. Alternatively, a lack of correlation may provide insights into the processes that allow
galaxies to form large rotating discs. For instance, \citet{vdBAH03} showed how pre-heating of the intergalactic medium unbinds baryons
from their dark matter haloes, which may lead to misalignments between the angular momentum of the gas and the dark matter.

For studies that seek to correlate galaxy spins, a fundamental observational problem is the 
``deprojection'' of the observed galaxy shape and the accurate determination of the direction and sign of 
the spin vector. Using a galaxy's observed axial ratio, $q$, and position angle, $\vartheta_P$, it is possible to 
determine the unit spin vector, $\bf \hat{J}$, up to a two-fold ambiguity in the sign of the spin 
i.e. clockwise or anti-clockwise \citep{Lee11}. Early works \citep{LP00} assumed a thin disc geometry 
in this calculation. In reality discs have finite thickness which must be accounted for to make 
an accurate estimate. This involves the assumption of an intrinsic flatness parameter which depends 
on galaxy morphological type \citep{HG84,LE07}. One simplistic way to avoid deprojection 
uncertainties would be to use only galaxies which are edge-on or face-on to the observer 
\citep{TCP06,SW09}, however this greatly reduces the available number density in a given 
sample.

If the right observations are available, it is possible to lift the degeneracy in the sign of the 
spin of a galaxy. The presence of dust lanes \citep{CW00} or the use of kinematic data 
\citep{OMP+15} can both be used to determine clockwise or anti-clockwise spin. In the absence of 
such additional information, authors have adopted a number of strategies. Some assumed the sign of the spin of each 
galaxy was completely independent \citep{KO92}, some assumed all galaxies had the same spin direction \citep{LE07} and some attempted a statistical approach that combined  distributions which assumed all spin signs were 
positive or negative into a single corrected distribution \citep{VBT+11}.  
Each approach resulted in a serious decrease in the available information.
Future mapping of the neutral hydrogen density with 21cm measurements using the Square Kilometre Array (SKA) will be able to deliver unambiguous 
measurements of the spin sign as well as measurements of angular momentum accurate to $3-5\%$ for 
millions of disc galaxies \citep{OG14,OMP+15}.

Reproducing the observed finite thicknesses and sizes of disc galaxies has been a challenge for numerical hydrodynamic simulations, because
the results are sensitive to the implementation of the various processes of baryonic physics. Consequently it is not clear whether the predicted shapes can be compared to observations. On the other hand, the prediction for the orientation of the spin axis should be more robust. Measurements of the alignments of the spin axes of disc galaxies may therefore  be useful to provide insights to the process that aligns the angular momentum. The usefulness of such observations for weak lensing studies, in particular to {\it correct} cosmic shear measurements,  is limited because this would require a prediction for the galaxy shapes as well. We therefore focus for the remainder of this section on the measurement of galaxy shapes.

\subsection{Measuring shapes}
\label{sec:obs_measuring_shapes}

The practical measurement of galaxy shapes is fundamental to both weak gravitational lensing studies and much of the intrinsic alignment literature. Galaxy shapes can be quantified using various approaches and a wide range of tools have been used for intrinsic alignment measurements in the past. In the case of cosmic shear, the galaxies for which shapes are measured are typically faint and have sizes that are comparable to that of the PSF. The effect of the PSF is twofold: (i) because it has a finite size, it leads to observed images that are rounder; (ii) the PSF is typically anisotropic, resulting in alignments in the observed images. Measuring accurate shapes for the source galaxies is challenging, and understanding the limitations and improving shape measurement algorithms has been an area of active research \citep[see e.g. results of the Shear TEsting Programme (STEP) and GRavitational lEnsing Accuracy Testing (GREAT) challenges, ][and references therein]{HVB+06,MHB+07,BBB+10,KBB+12, MRA+14}. As cosmic shear studies are the main driver for current research on this topic, we focus the discussion on the algorithms used to measure the lensing signal. 

One approach, which is gaining popularity thanks to increases in
computing power, is to fit a parametric model to the observed surface
brightness distribution. In the case of weak lensing studies the
initial 
model is sheared and convolved with the PSF. The model
parameters are varied until the resulting image best matches the
observations, which yields an estimate for the ellipticity; examples
are {\tt lensfit} \citep{MillerHK+13} and {\tt im3shape}
\citep{ZKV+13}. However, the choice of suitable models is not
straightforward because galaxies can have complex morphologies. If the
model is too rigid, the resulting shapes will be biased \citep{VB10,MBL+10}, but if the
model is too flexible, the shape will be biased too, because of noise
in the image \citep{RKA+12,KZR+12,MV12}. Forward modelling requires many calculations and is therefore computationally expensive when many parameters are included. For this reason model-fitting algorithms have not yet been extensively used, although {\tt lensfit} was employed to analyse the Canada France Hawaii Telescope Lensing Survey data \citep[CFHTLenS;][]{HvWM+12}.  An additional advantage of these forward modelling
approaches is that various instrumental effects can be incorporated
into a Bayesian framework, with priors imposed on the various model
parameters. However, accurate priors are needed, particularly for faint
galaxies and such information is not always available.

Alternatively, galaxy shapes can be
quantified using the moments of the surface brightness distribution of a galaxy. The quadrupole moments $Q_{ij}$ are defined as
\begin{equation}
Q_{ij}=\frac{1}{F_0}\int {\rm d}{\bf x}\ x_i x_j W(\boldsymbol{x})f(\boldsymbol{x}),
\label{eqn:Qij}
\end{equation}
\noindent where ${\bf x}$ denotes the two-dimensional position on the sky (with $i,j\in \{1,2\}$ denoting each dimension), where $f(\boldsymbol{x})$ is the surface brightness of the galaxy image and $W(\boldsymbol{x})$ is a weight function. Note that the centre is chosen such that the weighted dipole moments vanish and we normalise using the weighted monopole moment, $F_0$ (which corresponds to the flux in the case of unweighted moments). When measuring moments from real data, a weight function is needed to suppress the contribution of noise to the moments. In terms of the signal-to-noise ratio, the optimal choice for the weight function is to match it to the galaxy image. However, other choices can be made to reduce the sensitivity to possible sources of bias, such as the uncertainty in the underlying ellipticity distribution. Similar to the model fitting approach, where models used are often brighter in the center and deviations from the model there affect the fit a lot because of the high signal-to-noise ratio, the effect of the weight function in \Cref{eqn:Qij} is to give more weight to the central (brighter) regions of the galaxy. 

As we discuss below, moments are not only a useful way to quantify shapes; they can also, as was shown in \cite{MHK+13}, give insights into the relative importance of various observational biases, such as those caused by the PSF. The observed surface brightness distribution is the convolution of the true galaxy image and the PSF. For both cosmic shear and intrinsic alignment studies we wish to infer the moments (or shapes) of the former. In the case of unweighted moments the correction for the PSF is straightforward as

\begin{equation}
Q_{ij}^{\rm obs}=Q_{ij}^{\rm true}+Q_{ij}^{\rm PSF}.
\end{equation}
The galaxy ellipticity, or third flattening, can be expressed in terms of the corrected {\it unweighted} quadrupole moments (i.e. adopting $W=1$) through: 
\begin{equation}\label{eqn:flattening}
\left( \begin{matrix} \epsilon_1 \\ \epsilon_2 \end{matrix} \right) = \frac{1}{Q_{11} + 
Q_{22}+2\sqrt{Q_{11}Q_{22}-Q^2_{12}}}\left( \begin{matrix} Q_{11} - Q_{22} \\ 2Q_{12} \end{matrix} 
\right).
\end{equation}
Noise in real data prevents the use of unweighted moments. However, the need to use weighted moments leads to new complications:
the correction for the PSF (and weight function itself) involves knowledge of higher order moments, which themselves are affected by noise.
Limiting the expansion in moments is similar to the model bias mentioned above. Although \Cref{eqn:flattening} can be used to compute
the ellipticity from the best-fit parametric model, moment-based methods tend to make use of the polarisation, also called the third eccentricity or distortion \citep{BJ02}, given by:
\begin{equation}
\left( \begin{matrix} e_1 \\ e_2 \end{matrix} \right) = \frac{1}{Q_{11} + Q_{22}}\left( 
\begin{matrix} Q_{11} - Q_{22} \\ 2Q_{12} \end{matrix} \right),
\end{equation}
which avoids the square root of a combination of moments in the denominator. The two definitions are related through $e=2\epsilon/(1+|\epsilon|^2)$. A discussion of these quantities, and their probability distributions is presented in \citet{MKJ}, see e.g. \citet{HVB+06} for a discussion of the various choices made in actual
implementations.

\cite{KSB95} considered the first-order change in the polarisation, $\delta e_i$, induced by a small constant gravitational shear, for an arbitrary weighting function $W(x)$, and found that this can be expressed as $\delta e_i = P^\gamma_{ij} \gamma_j$, where the indices denote the two components of the ellipticity and shear, respectively, and the Einstein summation convention is assumed. The polarisability tensor $P^\gamma_{ij}$ depends in a rather complicated manner on the morphology and surface brightness distribution of the galaxy. However, it can be directly measured for each individual galaxy, and thereby can be used to calibrate the polarisation measurements: the average of $e_i/P^\gamma_{ij}$ in a particular patch of sky will be directly proportional to the gravitational shear $\gamma_j$ in that region. This can therefore be used to construct an unbiased estimate of the gravitational shear. This is why image simulations are used to not only compare the performance of algorithms \citep[e.g.][]{HVB+06, BBB+10, KBB+12, MRA+14}, but also to calibrate algorithms \citep[e.g.][]{MillerHK+13, HHM+15}. The various definitions for the shapes of galaxies are often used loosely in the literature, which is important to keep in mind when comparing published results.

\subsubsection{Shape measurement systematics}
\label{sec:obs_sys}

A bright, isolated galaxy which subtends a large angle on the sky would be an ideal candidate for 
shape estimation. However, galaxies are clustered on the sky and much of the cosmological signal comes from 
galaxies near dense regions. Hence, most galaxies are not isolated but
``blended'' with other sources and the shape measurements are compromised \citep[see][for the impact 
of very faint galaxies in image simulations]{HHM+15,DST+14}. This is particularly important if we wish to 
study intrinsic alignments as a function of environment. 

For example, \cite{HKF+11} showed that significant detections of satellite galaxy alignments using some shape measurement methods can be 
attributed  to contamination by neighbouring galaxies. Satellite galaxies are particularly prone to 
suffer this effect, being usually relatively dim with many bright neighbours. In addition there is an intra-cluster light contribution from stars bound to the cluster after being stripped from member galaxies \citep{Zwicky51,GZZ07,BCS+12,GAR+12,ADG+13,PGN+14}. 
The amount of bias will depend on the shape measurement method \citep{HKF+11}, with isophotal 
measurements being particularly sensitive, see discussion of \Cref{fig:example_cosmos} above.
Indeed, the papers that report detections of satellite alignments have been made using isophotal measurements 
\citep{PK05,AB06a,YBM+06,FLM+07} which weight the outskirts of  galaxies more strongly than weak 
lensing-optimised measurements. The later results, which report no significant alignment \citep{HKF+11,HE12,SCF+13,CMS+14,SHC+15}, are therefore less prone to these effects. A conclusion to the same effect was reached by \cite{SCF+13}.

The correction for smearing by the PSF is critical in any lensing analysis: the finite size of the PSF leads to rounder images and
 observed ellipticities lower than the true values. This bias is commonly referred to as multiplicative bias as it merely scales the
amplitude of the signal. However, due to (inevitable) misalignments of optical elements and atmospheric turbulence, the PSF is never perfectly round but tends to have a preferred direction, which may vary spatially and with time. This leads to an additional signal (therefore referred to as additive bias), which can mimic the cosmological
or intrinsic alignment signal. For this reason cosmic shear studies take great care in characterising the PSF and quantifying any residual systematics \citep[see e.g.][]{HvWM+12}. Similar rigour is required for accurate measurements of intrinsic alignments. Most recent results (see \Cref{sec:observations_largescale}) are based on weak lensing pipelines, but we note that this is not the case for older studies of intrinsic alignments.

The change in observed shape depends on the size of the galaxy relative to the PSF. 
How various biases propagate was studied in detail in \cite{MHK+13}. 
If the PSF is sufficiently well understood then it can be used 
to correct the observed shapes 
either using the observed 
moments of the surface brightness distribution or by convolving galaxy models with the PSF model. A 
complication is that the PSF varies in time, due to turbulence in the atmosphere 
\citep[e.g.][]{HRH+12} or variations in the thermal and gravitational load on the optical elements. 
However, if a sufficiently large number of stars are visible in the field-of-view, these can be used 
to quantify the spatial variation of the PSF \citep{Hoekstra04,JJB06}.
The PSF is not the only instrumental source of bias. Imperfections in the detector 
can also affect the observed shapes.
In the case of space-based observations, radiation damage causes charge traps leading to 
charge-transfer inefficiency during read-out
\citep[e.g.][]{RLS+10,MSC+14}. This produces charge trails, which result in 
alignments of the observed images. This is less relevant for ground-based 
observations where the brighter sky fills the charge traps but other effects persist, including charge-induced pixel shifts \citep{GBJ+15}, the 
brighter/fatter effect for individual charge wells \citep{AAD+14} and others. 

\subsubsection{Intrinsic alignment measurements and cosmic shear}
\label{sec:obs_IA_cosmic_shear}

We discuss the issue of mitigating the impact of intrinsic alignments on cosmic shear measurements in more detail in \Cref{sec:impact}. Note for now that in the typical situation all terms in \Cref{eqn:iadef} are relevant, hence the observed signal is  a combination of the lensing signal itself and the 
II and GI contributions. We have discussed how shape measurement can depend on details of the algorithm deployed; accurately accounting or correcting for the contribution of the intrinsic alignment signal in a cosmic shear measurement therefore requires that the same shape measurement
algorithm is used for both the weak lensing measurement and the estimate of the alignment signal. This condition may be trivially satisfied if the 
intrinsic alignment signal is determined from the cosmic shear survey itself, but one can also imagine scenarios where the intrinsic alignment signal is modelled using external data. Attempts to employ intrinsic alignment measurements acquired using different data or a different shape measurement algorithm should only be undertaken with great care.

For instance, for low redshift galaxies the shear is low and the correlations between galaxy shapes are dominated by the II term. As such
galaxies are bright and large compared to the PSF, their shapes can be measured reliably using deep modern observations. A limitation
is that the number of sufficiently large and bright galaxies is small, giving rise to a large shot noise. However, a more serious concern is that it is not clear how to
relate such measurements to predictions for the intrinsic alignment signal for galaxies at higher redshifts. As shown in \Cref{sec:introduction}, the estimated galaxy shape depends on the weight function applied to the galaxy light profile, which might differ for a well-resolved low-redshift observation and a
poorly-resolved high-redshift observation of two very similar galaxies. 
A further complication is that, even if a robust shape measurement method can be found, which can measure shapes well regardless of redshift, the mix of galaxy types and properties may evolve and the intrinsic alignments themselves may vary with time. 

Therefore direct measurements of the alignments of distant galaxies are needed. In fact, such studies will have to use the actual cosmic shear survey data.
This naturally implies that the requirements on the accuracy of the shape measurements are similar, including a careful correction for the PSF.
Moreover, to be able to extract the intrinsic alignment contributions with sufficient precision, good photometric redshift information is required. As the lensing kernel is broad in redshift, the required precision of photometric redshifts for the next generation of cosmic shear surveys is actually driven by our desire to model intrinsic alignments \cite[][see \Cref{sec:projected_corr_fns} for more details on the importance of redshift precision]{amara08,JB10,LBK+12}.

\section{Shape correlations}
\label{sec:shape_correlations}

The average shear and intrinsic alignment signal should vanish on very
large scales because of statistical isotropy of the Universe. However, both effects cause local coherent variations in observed ellipticity that can be used to measure cosmic shear and intrinsic alignments over a range of scales. In this section we will introduce some of the statistics used to describe these correlations and measure them in data.

Both weak gravitational lensing and intrinsic alignments produce a correlation between galaxy shape and matter density. 
In addition to observed galaxy ellipticity, it is useful to consider another quantity, the projection of the ellipticity of a galaxy 
perpendicular to the line connecting the position of that galaxy to some point. This is called the tangential shear, $\gamma_+$, and is
related to the observed ellipticity parameters through
\begin{equation}
\gamma_+=-(\epsilon_1\cos 2\vartheta_{\rm P}+\epsilon_2\sin 2\vartheta_{\rm P})=-{\rm Re}
\{\epsilon\exp(2\mathrm{i}\vartheta_{\rm P})\},\label{eqn:gammat}
\end{equation}
where $\vartheta_{\rm P}$ is the position angle with respect to the centre of the lens. The sign convention in \Cref{eqn:gammat} is such that $\gamma_+<0$ implies tangential alignments while $\gamma_+>0$ implies radial alignments. As an example, tangential shear is important 
in determining the masses of galaxy clusters \citep{OTU+10,AvdLK+14,HHM+15}, where ensemble-averaged tangential shear as a function of cluster-centric
radius can be related to the projected mass, or fit by a parametric
model for the mass distribution \citep[see][for a recent
review]{HBD+13}.

In this section we will concentrate on the most frequently used statistics in the intrinsic alignment literature: two-point correlations over large scales. These are particularly relevant for the large-scale measurements presented in \Cref{sec:observations_largescale} and for understanding the impact of intrinsic alignments on cosmic shear surveys detailed in \Cref{sec:impact}. For more detail on environment-dependent measurements see \Cref{sec:observations_smallscale} and some discussion of higher-order statistics can be found in \Cref{sec:three_point}. 

In \Cref{sec:2_two-point} we introduce the relevant two-point correlation functions before describing practical estimators for the same correlation functions in \Cref{sec:2point_estimators}. In \Cref{sec:projected_corr_fns} we describe the projection of three-dimensional statistics along the line of sight before relating these observables to intrinsic alignment models in \Cref{sec:corr_fn_to_IAs}. In \Cref{sec:null_tests} we describe some common systematic and null tests used when making measurements of intrinsic alignments.

\subsection{Two-point correlation functions}
\label{sec:2_two-point}

If the density or ellipticity (or shear) field is Gaussian, then all
the cosmological information is contained in the correlations between
galaxy positions, galaxy shapes, and the (cross-)correlations between
positions and shapes, averaged over pairs of galaxies as a function of separation. These are known as two-point correlation statistics. If redshift 
information is available, the correlations can be computed for galaxies binned in redshift. This allows the calculation of auto- and cross-correlations of redshift bins. This is referred to as a \textit{tomographic} analysis. In the case of non-Gaussian fields, due, for instance, to non-linear structure
formation on small scales, higher-order statistics, such as the
bispectrum, can be used to extract further information.

The statistical properties of the projected mass distribution are most
easily quantified using the correlations between galaxy shapes as a
function of their separation, i.e. in configuration space. This approachs allows for the treatment of complicated masks and survey boundaries. The
corresponding ellipticity autocorrelation function is defined as the
excess probability that any two galaxies are aligned (with respect to
some arbitrarily defined coordinate system): 

\begin{equation}
  \langle \epsilon^{i}\epsilon^{j}\rangle(\theta) = 
  \langle \epsilon^{i}(\boldsymbol{\theta}')\epsilon^{j}(\boldsymbol{\theta}'+\boldsymbol{\theta})\rangle,
  \label{eqn:eecor}
\end{equation}
where $i,j\in\{1,2\}$ denote pairs of galaxies and the angle brackets represent averaging over all pairs separated by angle $\theta =
|\boldsymbol{\theta}|$. Because of parity the correlation vanishes if $i\neq j$ and isotropy ensures that 
the correlation function is a function
only of the separation $|\boldsymbol{\theta}|$ for $i=j$.

\Cref{eqn:eecor} is defined with reference to (local) coordinate axes which are somewhat arbitrary. 
Instead it is more
convenient to consider the ellipticities with respect to axes oriented tangentially ($+$; see 
\Cref{eqn:gammat}) or at 45 degrees
($\times$) 
to a line joining each pair of 
galaxies. For convenience, it is common to
define the ellipticity correlation functions \begin{equation}
\xi_{\pm}(\theta) = \langle\epsilon_+\epsilon_+\rangle(\theta) \pm
\langle\epsilon_\times\epsilon_\times\rangle(\theta).
\end{equation}
We note that this notation is used in cosmic shear studies, but that it is different from the 
conventions commonly used
in clustering studies, where the symbol $\xi$ indicates the correlation function in 3D, and $w$ is 
used for projected
quantities. 
We will try to clarify these differences where needed.

When ellipticity correlation functions 
are estimated from real, noisy data, a weighted combination of observed 
ellipticities is employed:
\begin{equation}
\hat{\xi}_{\pm}(\theta) = \frac{\sum_{ij} W^{i}W^{j}\left[\epsilon_{+}(i|j)\epsilon_{+}(j|i) \pm 
\epsilon_{\times}(i|j)\epsilon_{\times}(j|i)\right]}{\sum_{ij}W^{i}W^{j}},
\end{equation}
where 
the weight $W_i$ typically accounts for the measurement uncertainties and we use $\epsilon_{+}(i|j)$ to mean the $+$ component of the ellipticity of a galaxy $i$, measured relative to the vector linking it to galaxy $j$, $\epsilon_{\times}(i|j)$ is the same for the $\times$ component. In the absence of intrinsic 
correlations these estimators are unbiased tracers of the weak lensing shear correlation functions, $\langle 
\epsilon_{+}(i|j)\epsilon_{+}(j|i) \pm \epsilon_{\times}(i|j)\epsilon_{\times}(j|i)\rangle = 
\sigma_{\epsilon}^{2}\delta_{ij} + \xi_{\pm}(|\boldsymbol{\theta}^{i}-\boldsymbol{\theta}^{j}|)$, where $\delta_{ij}$ is the Kronecker delta, $\sigma_{\epsilon}$ denotes the total intrinsic ellipticity dispersion, and the angle brackets indicate an ensemble average over the intrinsic ellipticity distribution and over the cosmic shear field, assuming randomly-oriented intrinsic ellipticities \citep{SWK+02}. Though of 
course, due to intrinsic alignments, this is not the case in reality. 

The ellipticity correlations between galaxies with similar redshifts can be used to determine the II 
signal because it is something that affects physically close galaxies. Correlations in the shapes of physically close galaxies due to a shared local tidal field boosts the intrinsic alignment signal, 
compared to the gravitational lensing contribution. This requires very good redshift information for the 
sources, and even in this case the observed signal contains a contribution from gravitational lensing itself, 
unless one restricts the analysis to $z \lesssim 0.1$, where the lensing signal is very small \citep{HH03}. 
Alternatively one can remove such  close pairs from the lensing analysis, thus efficiently 
suppressing the II contribution. 

The GI contribution, on the other hand, cannot be easily removed as 
it results from correlations between the shapes of galaxies that are separated in redshift, as described in \Cref{sec:introduction}. To 
estimate the IA signal directly from data, we need to determine the cross-correlation of tangential galaxy ellipiticity with the 
matter overdensity,  $\xi_{\delta +}(r_{\rm p},\Pi,z)$, or its projection, $w_{\delta+}(r_{\rm p})$. The subscript $\delta+$ indicates that we are correlating 
the density $\delta$ with  the tangential ellipticity $\epsilon_+$ for pairs separated by a transverse separation $r_{\rm p}$ 
and a radial distance $\Pi$ along the line of sight.  In general it is not possible to directly estimate the matter overdensity 
field, $\delta$, because the bulk of the matter in the Universe consists of dark matter. Instead 
galaxies are used as (biased) tracers of the density field. The cross-correlation of galaxy position 
with tangential ellipticity is indicated by $\xi_{\rm g+}(r_{\rm p},\Pi,z)$. 
A positive 
$\xi_{\rm g+}(r_{\rm p},\Pi,z)$ is a signal of coherent radial alignments of galaxy ellipticity with galaxy density. Assuming galaxy density traces the matter density, this is the correlation which sources intrinsic alignments of galaxy ellipticity, in both the II and GI flavours. Negative $\xi_{\rm g+}(r_{\rm p},\Pi,z)$ indicates tangential alignments induced by gravitational lensing. See \Cref{eqn:modified_LS} for a practical estimator of this correlation. The matter-ellipticity correlation is related to the galaxy-ellipticity correlation by the galaxy bias which can, in general, be a function of position and redshift, $\xi_{\rm \delta +}(r_{\rm p},\Pi,z)=b_{\rm g}(r_{\rm p},\Pi,z)\xi_{\rm g+}(r_{\rm p},\Pi,z)$.

The shear field can be decomposed into a gradient and a curl
component. The curl-free component is commonly referred to as the
``E''-mode, whereas the pure curl component is called the ``B''-mode,
analogous to the polarisations of the electric and magnetic field. If weak gravitational lensing was the only source of correlations in the shapes of galaxies, then 
one would expect to observe
$\xi_{\rm BB}(\theta)=0$. Although this is a good assumption for current surveys, a number of effects can 
introduce B-modes. For instance,
\cite{SvWM02} showed that B-modes are introduced if the source galaxies are clustered. However, most 
of these effects are expected to
be small, and the measurement of the B-mode has been used as a measure of residual systematics (as 
instrumental effects tend to include
B-modes).  However, intrinsic alignments can also introduce B-modes. Both the linear and quadratic alignment models are believed to source B-modes \citep{HS04}, while \cite{CNP+01} and 
\cite{CNP+2002} showed that spin alignments are not curl-free. Although the level of actually existing B-modes sourced by IAs (rather than instrumental systematics etc.) remains 
uncertain, it is too small to be detected
in current surveys.

\subsection{Estimators of the two-point correlation functions}
\label{sec:2point_estimators}

The ellipticity correlation functions are straightforward to compute from observations and are 
insensitive to the survey
geometry. This geometry is usually rather complex because of areas that need to be masked due to the 
presence of bright stars or other artefacts in
the data and must be accounted for when estimating the galaxy correlation function $\xi_{\rm gg}$. Nonetheless, it is important that the estimator that is employed accounts for the fact 
that some measurements are noisier than others. Alternatively, one may want to define estimators 
that minimise certain biases. In practice the correlation functions are computed from entries in a 
galaxy catalogue, which lists their positions, shapes, etc. 

Assume that $D$ is a catalogue of $N_{D}$ galaxies with positions from which we can compute 
$P_{DD}(r_{\rm p},\Pi)$, the number of pairs as a function of 
separation. It is convenient to normalise the result by the total number of pairs, given by $N_{D}(N_{D}-1)/2$, and the volume fraction, to define:
\begin{equation}
DD(r_{\rm p},\Pi) = \frac{P_{DD}(r_{\rm p},\Pi)}{N_{D}(N_{D}-1)V_{\rm bin}/(2V_{\rm survey})},
\end{equation}
where $V_{\rm survey}$ is the total volume of the survey and $V_{\rm bin}$ is the volume of some three-dimensional bin in $r_{\rm p}$ and $\Pi$. The volume of the bin is given by $V_{\rm bin} = 2 \pi r_{\rm p} \Delta r_{\rm p} \Delta \Pi$, in the limit that $\Delta r_{\rm p}$ is small. If the galaxies are clustered then $DD$ will be larger than unity on small scales. If we were considering a cross-correlation, rather than an auto-correlation, the equivalent normalisation would be $N_{D1} N_{D2} V_{\rm bin}/V_{\rm survey}$. 

However, we have not so far taken into account the
survey geometry - the fact that magnitude limits, source density etc vary across the survey field of view due to observational effects. The simplest way to do this is to consider a catalogue of objects with random 
positions, but to which the mask (which excludes those regions not observed by the survey) has been
applied. This random catalogue is indicated by $R$ and should contain many more 
entries than the data to avoid introducing
unnecessary noise. $RR$ denotes a pair of galaxies where both are drawn from the random catalogue.  
The 
most commonly used estimator for modern studies is the Landy-Szalay estimator \citep{LS93}. For the galaxy position auto-correlation this takes the form:
\begin{equation}
\hat{\xi}_{\rm gg}(r_{\rm p},\Pi) = \frac{DD-2DR+RR}{RR},
\end{equation}
where $DR$ means a pair of galaxies with one drawn from the data and one from the random catalogue.

A version of this estimator can be adopted for the GI cross-correlation function. In this form it is 
referred to as the modified Landy-Szalay estimator \citep{MBB+11}:
\begin{equation}
\hat{\xi}_{g+}(r_{\rm p},\Pi) = \frac{S_{+}(D-R)}{R_{S}R} = \frac{S_{+}D-S_{+}R}{R_{S}R}.
\label{eqn:modified_LS}
\end{equation}
Here we have assumed that, as well as the galaxy population, $D$, we have some other set of galaxies, $S$, with good shape 
measurements (note these could be the same population, or $S$ could be some sub-set of $D$). $R$ and $R_S$ are now sets of random positions corresponding to the position sample and shape sample 
respectively. $S_{+}D$ is the sum of the $+$ component of the ellipticity over all pairs of 
galaxies with separations $r_{\rm p}$ and $\Pi$, where one galaxy is in the good shape sample and 
one is 
in the position sample,
\begin{equation}
S_{+}D = \sum_{i \neq j | r_{\rm p},\Pi} \epsilon_{+}(j|i),
\end{equation}
for a pair of galaxies $i$ and $j$. 
Similarly we can define the estimators for the tangential and cross ellipticity autocorrelations,
\begin{align}
\hat{\xi}_{++}(r_{\rm p},\Pi) &= \frac{S_{+}S_{+}}{R_{S}R_{S}}, \\
\hat{\xi}_{\times\times}(r_{\rm p},\Pi) &= \frac{S_{\times}S_{\times}}{R_{S}R_{S}},
\end{align}
where
\begin{align}
S_{+}S_{+} &= \sum_{i \neq j | r_{\rm p},\Pi} \epsilon_{+}(j|i)\epsilon_{+}(i|j),\\
S_{\times}S_{\times} &= \sum_{i \neq j | r_{\rm p},\Pi} 
\epsilon_{\times}(j|i)\epsilon_{\times}(i|j).
\end{align}

Alternatively one can define correlation functions of spin or position angles but, as mentioned 
earlier, a direct relation to the ellipticity 
correlation function then requires knowledge of the underlying ellipticity distribution. Hence, to 
be useful to mitigate the impact
of intrinsic alignments in weak lensing studies, shapes need to be measured regardless.

\subsection{Projected correlation functions}
\label{sec:projected_corr_fns}

Although three-dimensional correlation functions are conveniently computed from theory, the 
observations are most commonly presented as projected quantities. Here we describe the projection of a general correlation function but it applies specifically to each of the correlations described above. 

Consider a three-dimensional correlation function, $\xi_{\rm ab}(r_{\rm p},\Pi,z)$, where the separation of our pair, $ab$, under consideration has been split into components parallel, $\Pi$, and perpendicular, $r_{\rm p}$ to the line of sight. The presence of $z$ is because the correlation function itself may depend on redshift. 
The corresponding projected
correlation function, $w_{ab}(r_{\rm p})$, for objects in a particular redshift bin, separated by a 
distance $r_{\rm p}$, transverse to the line of sight, is obtained by integrating the
equivalent 3D correlation function $\xi_{\rm ab}(r_{\rm p},\Pi,z)$ along the line-of-sight:

\begin{equation}
w_{\rm ab}(r_{\rm p}) = \int dz \ W(z) \int d\Pi \ \xi_{\rm ab}(r_{\rm p},\Pi,z),
\end{equation}
where $\Pi$ is the distance along the line of sight coordinate, and  $W(z)$ is the redshift weighting \citep{MBB+11},
\begin{equation}
W(z) = \frac{p^{2}(z)}{\chi^{2}(z)\chi'(z)}\left[ \int \mathrm{d}z \frac{p^{2}(z)}{\chi^{2}(z)\chi'(z)} \right]^{-1},
\label{eqn:z_weighting}
\end{equation} 
where $p(z)$ is the unconditional probability distribution of galaxy redshifts. $\mathrm{a,b}$ represent any combination of observables, $\mathrm{a},\mathrm{b} \in \{\delta,\mathrm{g},\epsilon,+,\times 
\}$, where $\delta$ is the matter overdensity, g is galaxy position, $\epsilon$ is galaxy ellipticity and $+, \times$ are the components of ellipticity parallel/perpendicular or at 45$^\circ$ to the vector connecting the pair of positions. If redshift information is available, it is convenient and straightforward to express the 
measurements as a function of $r_{\rm p}$ in physical coordinates. Alternatively one can show results as a function of 
angular separation, although this only applied to early results in practice.

The most precise results are 
obtained if spectroscopic redshifts \citep{A13} are available. However, this requires relatively large 
investments of observing time on large telescopes,
especially for the faint galaxies typically used in weak lensing studies. Alternatively we can use 
photometric observations in multiple filters which probe features in the spectral energy distribution, which in turn can be 
used to estimate a photometric redshift \citep[see e.g.][for their application to the CFHTLenS dataset]{HEK+12}. Compared to spectroscopy, photometry is less precise but 
faster and therefore cheaper. 
Most of the observations discussed here use spectroscopic redshifts but the larger number density 
available from photometric surveys makes their use desirable, even at the cost of lower redshift accuracy. 

\begin{figure}[ht!]
  \begin{flushleft}
    \centering
       \includegraphics[width=3.5in,height=4.5in]{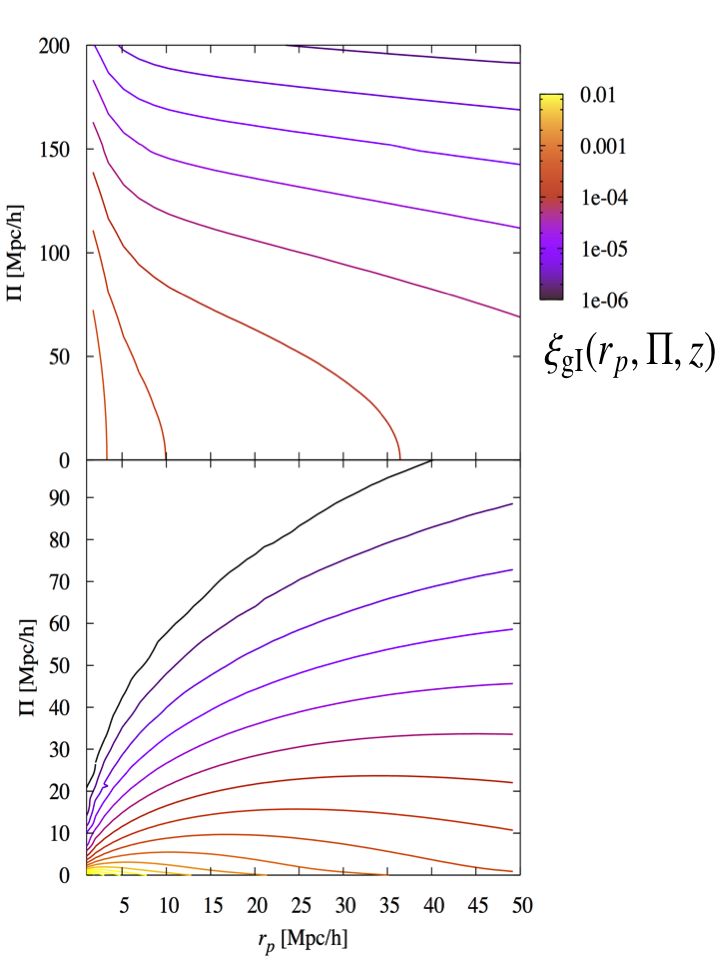}
\vspace{0mm}
\caption{Three-dimensional galaxy position-ellipticity correlation function, $\xi_{\rm g+}(r_{\rm p},\Pi)$, as a function of comoving line-of-sight separation $\Pi$ and comoving transverse separation $r_{\rm p}$ at $z \sim 0.5$. Contours are logarithmically spaced between $10^{-2}$ (yellow) and $10^{-6}$ (black) with three lines per decade. \textbf{Top panel:} Applying a Gaussian photometric redshift scatter of width 0.02. \textbf{Bottom panel:} Assuming exact redshifts. Note the largely different scaling of the ordinate axes. The galaxy bias has been set to unity, and the linear alignment model with SuperCOSMOS normalisation (see \Cref{sec:observations_largescale}) has been used to model $P_{\rm \delta I}$ in both cases. Redshift-space distortions have not been taken into account. \permaa{JMA+11}}
\label{fig:photoz_scatter}
  \end{flushleft}
\end{figure}

Unsurprisingly the photometric redshift scatter tends to smear the intrinsic alignment signal along the line-of-sight 
direction, $\Pi$ \citep{JMA+11}. When calculating projected two dimensional correlation functions, the full intrinsic alignment signal can be 
retained by extending the range of $\Pi$ considered in a measurement. This does however reduce the 
measured signal-to-noise ratio, because the signal has become more spread out, and increases the contamination by gravitational shear. In practice the 
line-of-sight integral gets truncated and some portion of the intrinsic alignment signal is lost. This effect can be seen in \Cref{fig:photoz_scatter}. In the lower panel, where exact redshift information is assumed, the power of the galaxy position-ellipticity correlation falls off very quickly with line-of-sight separation. In the upper panel, where a Gaussian photometric redshift scatter of width $\sigma_{\rm z} = 0.02$ is assumed, there is still significant correlation, even at line-of-sight separations of $>100 \mathrm{Mpc}/h$.

Careful modelling of the expected signal is even more important when using photometric redshifts. 
The large line-of-sight spread means the effect of contributions from the galaxy position-gravitational lensing cross-correlation and 
lensing magnification cross-correlations is more pronounced, see \Cref{sec:self_calib} for further discussion.

\subsection{Using correlation functions to test intrinsic alignment models}
\label{sec:corr_fn_to_IAs}

A detailed physical understanding of our measurements requires comparison with theoretical alignment 
models, such as the ones that are detailed in \citet{Paper2}. From a theory 
perspective, it is often more convenient to calculate correlations in Fourier space or in spherical 
harmonic space. The resulting power spectrum can be directly related to the real space statistics, 
however the choice of space for the measurement can depend on the survey geometry (e.g. for a wide-field survey a 
spherical harmonic expansion is natural on the sphere), or related to the strength of the signal 
(e.g. the presence of bad pixels mean configuration space can be preferred), or on the numerical tools available. In this section we will concentrate on the galaxy-ellipticity correlation but the relations generalise to other observables, see \citet{Paper2} for the full range of expressions.

We can relate the projected galaxy position-ellipticity correlation function, $w_{\rm g+}(r_{\rm p})$, directly to the three-dimensional density-intrinsic ellipticity function in Fourier space, $P_{\rm \delta I}(k_{\perp},z)$, via
\begin{equation}
w_{\rm g+}(r_{\rm p}) = -b_{\rm g} \int \mathrm{d}z W(z) \int_{0}^{\infty} \frac{\mathrm{d}k_{\perp}k_{\perp}}{2\pi} J_{2}(k_{\perp}r_{\rm p})P_{\rm \delta I}(k_{\perp},z),
\end{equation}
where $b_{\rm g}$ is the galaxy bias, $J_{2}(k_{\perp}r_{\rm p})$ is the second-order Bessel function of the first kind, $k_{\perp}$ is the wavevector perpendicular to the line-of-sight and $W(z)$ is the weighting over redshifts as derived by \citet{MBB+11}, see \Cref{eqn:z_weighting}.

The contribution from intrinsic alignments is encoded in the three-dimensional power spectrum $P_{\rm \delta I}(k_{\perp},z)$. One model we will refer to throughout this paper is the linear alignment model \citep{HS04},
\begin{equation}
P_{\rm \delta I}\left(k_{\perp},\chi\right) = -\frac{C_{1}\bar{\rho}(\chi)}{\bar{D}(\chi)}a^{2}(\chi)P_{\rm \delta \delta}\left(k_{\perp},\chi\right) ,
\label{eqn:LA} 
\end{equation}
where $C_{1}$ is a constant setting the amplitude of correlation, $\bar{\rho}(\chi)$ is the mean matter density of the Universe, $\bar{D}(\chi) = D(\chi)(1+z)$, $D(\chi)$ is the linear growth function, normalised to $D(\chi=0)=1$, $a(\chi)$ is the scale factor and $P_{\rm \delta \delta}\left(k_{\perp},\chi\right)$ is the linear matter power spectrum.
We refer to \citet{Paper2} for more information on this and other
specific models. We can also construct projected angular correlation functions, $C(\ell)$, directly from the three-dimensional power spectra, see \Cref{sec:impact} for more details.

While several terms in \Cref{eqn:LA} depend on redshift, these dependencies cancel out meaning that, overall, there are no explicit dependencies on redshift or galaxy luminosity in the linear alignment 
model but it is often thought useful to check for these when fitting to data. This is because the strength of coupling between dark matter and galaxies is unknown. These extra terms help describe the dependence of the coupling. A common approach in 
the literature is to insert power-law dependences on redshift, $z$, and the 
luminosity, $L$, where the index of the power-law is a free parameter that can be fit to data. 
Putting these together produces a model of the form

\begin{equation}
P^{\textrm{model}}_{\rm gI}(k,z,L) = A_{\rm I} \ b_{\rm g} \ P_{\delta I}(k,z)\left(\frac{1+z}{1+z_0} 
\right)^{\eta_{\textrm{other}}} \left( \frac{L}{L_{0}} \right)^\beta,
\label{eqn:power_laws}
\end{equation}
where $P_{\delta I}$ is the power spectrum for the mass density - intrinsic ellipticity correlation, 
provided by some intrinsic alignment model. $A_{\rm I}$ is a free amplitude term, $b_{\rm g}$ is the (linear, 
deterministic) galaxy bias, $z_0$ is a reference pivot redshift, $L_0$ is a reference pivot 
luminosity and $\eta_{\rm other}, \beta$ are the free power-law indices for the redshift and 
luminosity dependence respectively. The power-law index for redshift has been called $\eta_{\rm 
other}$ because it attempts to capture redshift evolution due to any ``other'' physical processes 
beyond the linear alignment model.

\subsection{Tests for systematics}
\label{sec:null_tests}

Some of the correlations are expected to be consistent with no signal in the absence of systematics 
errors. Such {\it null tests} can be used to test for the presence of systematics in the data, and a 
significant detection of a signal is a warning that the measurements of real interest may be biased. 
For instance, we already saw that the ellipticity auto-correlations can be written in terms of 
curl-free ``E'' and divergence-free ``B'' modes. Although the signal caused by spin alignments is not 
curl-free, the much stronger signal from the linear alignment model as well as weak lensing itself 
comprise only E-modes to first order. Therefore such a decomposition can be a useful diagnostic in studying the effects of 
systematics,
even though the B-mode signal is not expected to vanish completely.

A common null test in the literature is the measurement of $w_{\rm g\times}$, the 
correlation of the  density sample with the cross-component of the shear from the shape sample, i.e. 
the ellipticity measured at 45$^{\circ}$ to the line connecting the pair of galaxies under 
consideration, one from the shape sample, the other from the density-tracer sample. This statistic 
is of course very closely related to the measurement of $w_{\rm g+}$, the correlation of the density 
sample with the ellipticity measured along the connecting vector, and requires no additional data 
products, random catalogues or statistical tools. Parity symmetry means $w_{\rm g\times}$ is expected to be zero. A non-zero measured value might indicate the presence of a range of systematic 
effects including residual PSF distortions. The cross-component of the shear is useful generally as a systematic. 
Another related statistic is $w_{+\times}$, which is 
expected to be consistent with zero in the absence of systematics because intrinsic alignments only 
induce alignments in the radial/tangential direction. 

A different null test is the calculation of $w_{\rm g+}$, the same statistic used to measure shape 
alignment, but where only certain pair separations are considered. The chosen scales should be such that this correlation function will 
vanish because the line-of-sight separation is sufficiently large that intrinsic alignments are negligible, being a local effect, but small enough that gravitational lensing shear is still negligible. Spurious galaxy alignment, whether from optical distortion 
in the telescope, deblending, mistaken sky subtraction or 
photometric redshift errors could generate a $w_{\rm g+}$ signal at large (apparent) separation. 

These various tests were applied to the observational results that are reviewed in more detail in 
the next few sections. Many of these studies found $w_{\rm g+}$ consistent with zero at large 
line-of-sight separation, and $w_{\rm g\times}$ consistent with noise at all scales 
\citep[e.g.][]{HMI+07,MBB+11,JMA+11, SMM14}. While as yet undiscovered systematic effects cannot entirely be ruled out, these results give us hope that the intrinsic alignment signal can be reliably determined from current and near-future data.

Multiple shape measurement codes can be applied to the same data. Of course common systematic effects will manifest in both the resulting shape catalogues, but any method-specific systematics can be detected by looking at differences in correlations using the different shape estimates. For example, \citet{SHC+15} shows results using two different shape measurement pipelines for this reason.

\section{Observations of alignment in large galaxy samples}
\label{sec:observations_largescale}

On small scales, intrinsic alignments are expected to be intimately connected to the environment of the galaxy, in terms of the morphology of the local cosmic web. Details of the galaxy's evolution, including feedback processes, mergers etc. may also be expected to play a role. In contrast it is believed that the mechanisms that give rise to alignments which persist in pairwise correlations of galaxies on large scales can be related to the large-scale gravitational tidal field.
Examples of such mechanisms are introduced in \citet{Paper1} and discussed in more technical detail 
in \citet{Paper2}, but many open questions about galaxy alignments including their amplitude, 
dependence on galaxy type and luminosity and their evolution with redshift, have no obvious analytic 
answer. Precise and accurate observations are therefore critical. 

To understand alignments on small scales we need to consider the dependence on the 
environment of the galaxies under consideration, which we do in the next section.  In this section we 
start by reviewing the observational status of alignments from the linear regime ($>10 {\rm Mpc}/h$)
into the quasi-linear regime ($\sim 5-10 {\rm Mpc}/h$).  On these scales the matter power spectrum 
of density fluctuations is fairly well understood from linear theory and largely unaffected by baryonic physics 
\citep[e.g.][]{semboloni11}. Furthermore, the measurements are based on 
datasets and methods that are similar to those used for cosmic shear studies.

The first large scale study of intrinsic alignments in the cosmic shear era was \citet{BTH+02}. This paper used $2\times 10^{6}$ galaxies from the SuperCOSMOS sky survey \citep{HMR+01} with a median redshift of $z \sim 0.1$. An observed excess correlation above that expected from cosmic shear was seen as evidence of intrinsic alignments. The observed amplitude was subsequently used widely to normalise intrinsic alignment models at low redshift to the value $C_{1}=5 \times 10^{-14}(h^{2}\mathrm{M}_{\odot} \mathrm{Mpc}^{-3})^{-1}$ \citep{BK07}. $C_{1}$ is the normalisation factor setting the amplitude of \Cref{eqn:LA} and \Cref{eqn:power_laws}. The free parameter $A$ allows the amplitude to vary around this constant. The \citet{BTH+02} observations immediately required several popular IA models \citep{HRH2000, CM00, CKB01} to be revised downwards in amplitude as they had over-predicted the SuperCOSMOS signal. \citet{BTH+02} also offered the first observation-based assessment of the likely impact for cosmic shear measurements, see \Cref{sec:impact} for more discussion.  

In the following we split the results by galaxy type because the leading theories predict that 
different processes dominate for late- and early-types
\citep[see][for more details]{Paper2}. For example, the linear alignment model for dispersion-dominated 
galaxies \citep{HS04} and tidal torque theory for angular momentum-dominated galaxies 
\citep{Peebles69,Doroshkevich70,White84} motivate a split into early- and late-type galaxies. 
We 
therefore review results for late-type galaxies in \Cref{sec:observations_largescale_latetype} and 
for early-type galaxies in  \Cref{sec:observations_largescale_earlytype}. Alternatively, the samples 
are split by rest-frame colour or spectral energy distribution.
Although we note that a split into blue and red galaxies is not exactly 
the same as a morphological selection, we consider this implicitly to be the case when reviewing the 
different galaxy samples. Finally, in \Cref{sec:observations_largescale_indirect}, we review 
indirect methods of measuring intrinsic alignments for both early- and late-type galaxies.

\subsection{Late-type galaxies}
\label{sec:observations_largescale_latetype}

The most commonly accepted scenario for the alignments of disc galaxies is the quadratic alignment model, 
which describes how the angular momentum of dark matter haloes is spun up to produce correlations 
between the orientations of galaxies \citep{HS04}. 
Disc-galaxies are believed to be nearly circular when viewed face-on, they appear elliptical as a result of projection due to their orientation with respect to the observer. Total observed ellipticity is the sum of this projection effect and any small intrinsic ellipticity the galaxy may have. If the orientation of 
the disc is determined by the spin vector of the galaxy, and these are correlated between galaxies, 
then there will be a correlation between the observed intrinsic ellipticities i.e. intrinsic \emph{alignment}. These can be observed 
by simply measuring the correlations in measured shapes \citep{HMI+07,MBB+11}. Before reviewing 
these results, we note that this paradigm gives us another avenue to study intrinsic alignments of 
disc galaxies: we can measure the correlation of galaxy spins. Note that the quadratic alignment model is so-called because the II three-dimensional power spectrum depends on the square of the linear matter power spectrum and therefore the alignment signal is expected to be suppressed compared to the linear alignment model which is believed to apply to early-type galaxies and has a linear dependence on the matter power spectrum.

\citet{SLB+09} presented an early example of such a measurement using data from the Galaxy Zoo citizen 
science project\footnote{\texttt{http://www.galaxyzoo.org/}}. In Galaxy Zoo, spiral galaxies are classified as 
clockwise, anti-clockwise or edge-on. For each face-on galaxy there is therefore one bit of 
information corresponding to the sign of the galaxy spin vector projected along the line-of-sight. 
This information enabled the measurement of the correlation function of spin chirality of face-on 
spirals. The authors tentatively reported that galaxy spin directions are correlated at very small 
scales ($<0.5 $Mpc), albeit with low significance ($2-3\sigma$). There is no obvious reason, under the tidal torquing model, why this chiral correlation should exist. If confirmed to high significance in future studies, it might provide useful insight into the sourcing of galaxy angular momentum.

Going beyond correlations in spin chirality (clockwise vs. anticlockwise), \citet{CHP10} searched for 
correlations between the orientations of the spin vectors themselves in pairs of spiral galaxies from the SDSS 
survey. They computed correlations in the spin parameter, $\lambda$ \citep{Peebles69}:
\begin{equation}
\lambda = \frac{J|E|^{1/2}}{GM^{5/2}},
\end{equation}
where $G$, $E$, $M$ and $J$ are Newton's constant, the total energy, mass and angular momentum of the configuration, 
respectively. $\lambda$ accounts for the magnitude of the spin, while the position angle was used as an estimate of direction. 

\citet{CHP10} reported a weak ($1.5\sigma$) correlation between the spin magnitude of neighbouring 
galaxies, but, contrary to \citet{SLB+09}, no
clear alignment between their orientation. The authors suggested that this is due to some late-time 
dilution of a primordial correlation laid down at the time of galaxy formation. They suggest that interactions 
with close neighbours can significantly redistribute angular momentum through clumpy and irregular 
mass accretion, reducing the value of $\lambda$.

\citet{Lee11} presented another measurement of intrinsic alignments using spin statistics. This paper 
used large (angular size $\geq 7.92$ arcsec) late-type spiral galaxies from the SDSS DR7 over $0 \leq z \leq 0.02$. The SDSS observations provide 
information on the galaxy's axial ratio, $q$,  and position angle, $\vartheta_{\rm P}$, from which the 
unit spin vector for each galaxy is reconstructed. These spin vectors are 
combined to form the two-point spatial correlation function for galaxy spin axes. For this sample a 
positive spatial correlation is detected at $3.4\sigma$ (separation $r \lesssim 1 $Mpc/$h$)  and 
$2.4\sigma$ (separation $r \lesssim 2 $Mpc/$h$). The correlations are stronger for galaxies located in 
dense regions, which have more than 10 neighbours within 2 Mpc/$h$. The measured correlations are 
consistent with the predictions of the quadratic alignment model that the spin two-point correlation 
should follow a quadratic scaling with the linear density correlations. We note that the estimation 
of the spin vector relies on the assumption that galaxies form thin discs. If this is not the case 
across the galaxy sample, this assumption can introduce a systematic error of order 10\% in the 
measured spin correlation \citep{LE07,Lee11}. 

 \citet{AJ11} also used SDSS data to analyse angular momentum correlations of disc galaxies. They 
found that positive correlations of spiral-arm handedness and angular momentum orientations on 
distance scales of $1$ Mpc/$h$ are plausible but not statistically significant. Furthermore, they 
suggested that previous studies such as the ones presented by \citet{SLB+09} and \citet{Lee11} 
overestimated the correlation of spins for spiral galaxies because of bias in ellipticity estimates 
based on second moments due to galactic bulges. This highlights the importance of how orientations 
are determined.

Thanks to large imaging surveys, pre-eminently the SDSS, much progress has been made over the last decade in the correlation of galaxy ellipticities. The first to take advantage of SDSS, in the context of late-type galaxies, is the study by \citet{HMI+07} who attempted to measure intrinsic alignments for a 
low-redshift sample of blue galaxies, selected from
the SDSS main spectroscopic redshift sample. 
This is a flux-limited sample primarily covering the 
range $0.05 < z < 0.2$.  \citet{HMI+07} split a colour-selected blue subset of this flux-limited sample into four luminosity bins for analysis, but detected no significant signal in any of those 
bins (\citet{HMI+07} did make a positive detection for intrinsic alignments in SDSS red galaxies). While this null result for blue galaxies can be interpreted as beneficial
for cosmic shear studies, the statistical uncertainties are relatively large due to the small number of blue galaxies in the high luminosity bins. Furthermore, given the 
low redshift of the sample, applying the results to
cosmic shear surveys, which target galaxies at much higher redshifts, requires a large 
extrapolation. This motivated attempts to repeat a similar
measurement at higher redshifts.

\begin{figure}[ht!]
  \begin{flushleft}
    \centering
       \includegraphics[width=3in,height=3in]{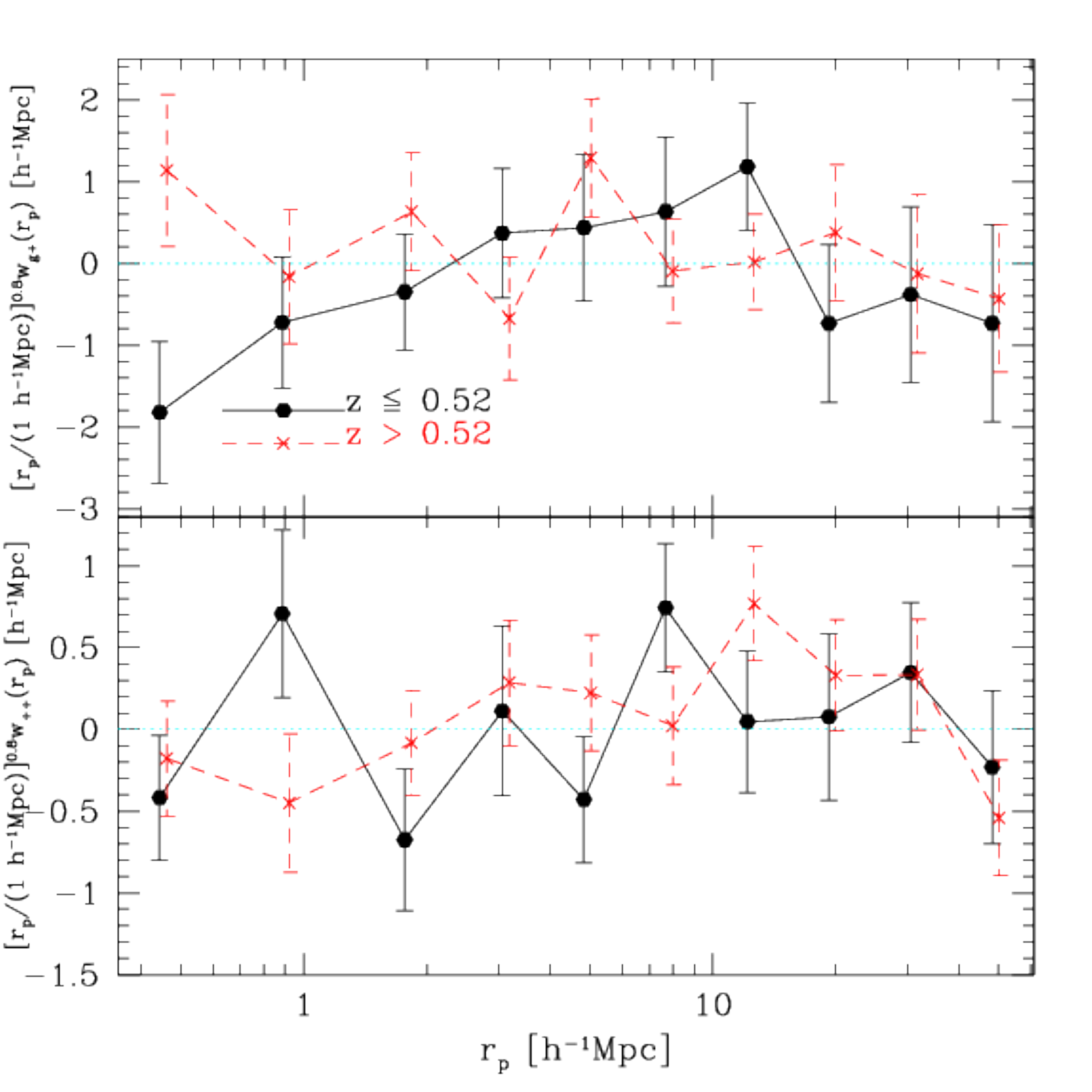}
\vspace{0mm}
\caption{\textbf{Top panel:} Projected galaxy-tangential ellipticity cross-correlation signal $w_{\rm g+}(r_{\rm p})$, multiplied by 
$r_{\rm p}^{0.8}$, from the WiggleZ data of \citet{MBB+11}. Results are shown averaged over all 
WiggleZ regions, for the two redshift subsamples. Points at a given value of $r_{\rm p}$ are 
slightly horizontally offset for clarity. \textbf{Bottom panel:} Same as the top, but for the ellipticity-ellipticity 
cross-correlation signal $w_{++}(r_{\rm p})$. \permmn{MBB+11}}
\label{fig:wigglez_ia}
  \end{flushleft}
\end{figure}

Such a sample was provided by the overlap of the SDSS imaging and the WiggleZ redshift survey 
\citep{DJB+10}, which targeted a population of blue galaxies, whose mean redshift was $z_{\rm mean}\sim 0.6$, with a primary science goal of measuring Baryon Acoustic 
Oscillations (BAO). \cite{MBB+11} measured the intrinsic alignments of the subset of galaxies for 
which ellipticities were determined from the SDSS. The full WiggleZ sample was used to trace the 
density field, assuming a linear galaxy bias. While previous papers, notably 
\citet{HMI+07}, had measured intrinsic alignments for blue galaxies at low redshifts, this was the first paper to 
push the measurements to intermediate redshifts. 

\Cref{fig:wigglez_ia} shows the resulting measurements from \citet{MBB+11} for  $w_{\rm g+}$ and 
$w_{++}$ as a function of transverse separation for the WiggleZ dataset, split into two redshift 
bins. The correlation functions are consistent with zero within the statistical uncertainties, which 
are relatively large.
By combining with the results for some lower-redshift blue galaxy samples from 
SDSS that were previously presented in \cite{MHI+06}, the null detection was used to place upper 
limits on how blue galaxy intrinsic alignments could contaminate weak lensing measurements from a
CFHTLenS-like survey, with the result being a bias in the amplitude of the (linear) power spectrum 
on the scale of 8 Mpc$/h$, $\sigma_8$,
of $^{+0.02}_{-0.03}$ at the 95\% confidence level. See \Cref{sec:impact} for more details on how 
intrinsic alignments impact estimates of cosmological parameters from cosmic shear.

\begin{figure}[ht!]
  \begin{flushleft}
    \centering
       \includegraphics[width=3in,height=3in]{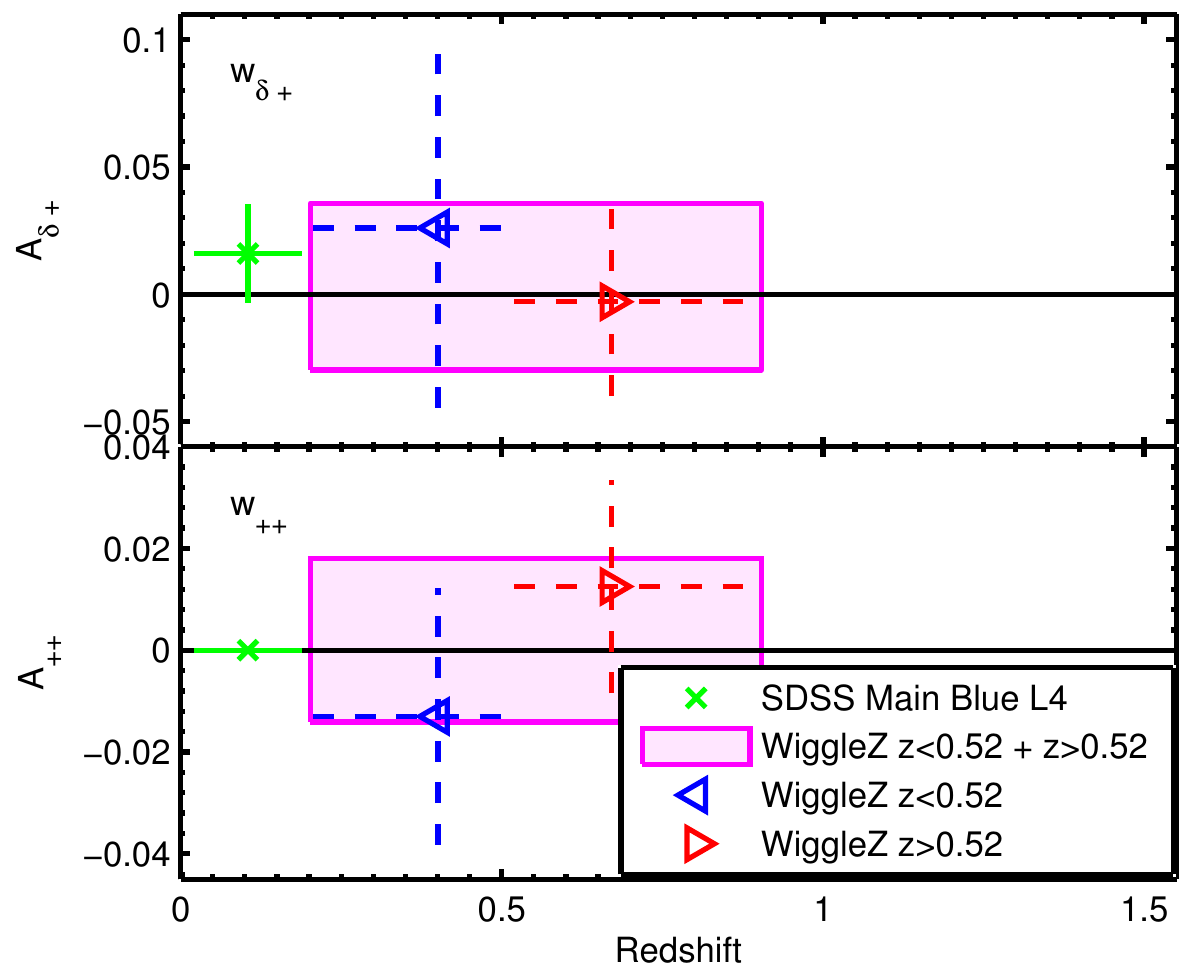}
\vspace{0mm}
\caption{Constraints on the power-law amplitude $A$ as a function of redshift from the WiggleZ data 
of \citet{MBB+11}. This analysis used line-of-sight range, $\Pi_{\textrm{max}} = 60 $Mpc/$h$, 
and fixed the power-law slope at $\alpha = −0.88$. \textbf{Upper panel:} Constraints from $w_{\rm 
g+}$. \textbf{Lower panel:} Constraints from $w_{++}$. From left to right, the points show 
constraints from SDSS Main Blue L4 (green cross); WiggleZ $z < 0.52$ (blue triangle); and WiggleZ $z > 0.52$ 
(red triangle). The horizontal lines indicate the redshift range of the observations. The shaded pink 
rectangle indicates the constraint from the full WiggleZ sample (both redshift ranges). 
\permmn{MBB+11}}
\label{fig:MBB11_fig9}
  \end{flushleft}
\end{figure}

\citet{MBB+11} also placed constraints on the redshift evolution of blue galaxy intrinsic 
alignments, taking advantage of the broad redshift range of
the WiggleZ sample. The full sample was split into two redshift slices containing galaxies below and 
above $z = 0.52$. \citet{MBB+11} fit two simple models to their intrinsic alignment measurements: a power-law in 
transverse separation and the non-linear alignment model with an additional free power-law dependence on redshift \citep{BK07}. The non-linear alignment model is simply the linear alignment model of \citet{HS04} with the non-linear three-dimensional matter power spectrum substituted in place of the linear matter power spectrum. It is not a truly non-linear formulation of IAs, indeed there is no theoretical justification for this model but, as we shall see, it fits the data better than the linear alignment model
(see Section 3.3 of \citealp{Paper2} for more details on the non-linear alignment model).  The 
power-law in redshift takes the form given by \Cref{eqn:power_laws} with the free parameter 
$\eta_{\rm other}$ and a pivot redshift of $z_{p}=0.3$. Constraints on the amplitude of the power 
law are shown in \Cref{fig:MBB11_fig9}. They are consistent with zero for all samples from both 
$w_{\rm g+}$ and $w_{++}$. This tells us that there is no evidence for evolution of the intrinsic alignment signal with redshift. 

Note that the most likely sign of contributions of both PSF systematics and intrinsic alignments to $w_{++}$ is positive, so the null detection itself indicates that there are no substantial observational systematics contaminating the 
measurement. The primary uncertainty in the application of these upper limits to future surveys is 
that the sample does not correspond to the entire blue cloud \citep{KHW+03,WMS+07}, but rather to a specific subset of it 
that appears to consist of morphologically disturbed starburst galaxies. The signal might therefore 
be compromised by recent mergers and other environmental effects that apply to the galaxies selected 
by WiggleZ, but which do not apply to the general population of late-type galaxies. 

Much more observational progress is needed before models of intrinsic alignments for late-type 
galaxies can be confronted with precise observational constraints. Current constraints come from 
spectroscopic studies, which mostly target the bright early-type galaxies we will discuss next. 
Unfortunately photometric redshifts are typically the least reliable for late-type galaxies as well, as 
their 4000$\buildrel _\circ \over {\mathrm{A}}$ 
break is less pronounced. However, the situation may improve thanks to new  surveys 
that aim to cover a significant wavelength range with a large number of narrow-band filters 
\citep{MMC+14}.

\subsection{Early-type Galaxies}
\label{sec:observations_largescale_earlytype}

Compared to the null-detections for late-type galaxies, the situation is markedly different for 
early-type galaxies. Luminous Red Galaxies (LRGs) make easier targets for BAO surveys. Thanks to their red colours, indicative of old stellar populations, they are readily 
identified out to high redshifts. Secondly, they are pressure-supported systems (rather than 
rotationally-supported), and this lack of angular momentum has led to the view that their alignments 
are described by the linear alignment model \citep[see Section~3 of][for more detail on this 
model]{Paper2}. Therefore the signal is expected to be larger than that of disc galaxies, which is 
indeed the case, as shown already by \citet{HMI+07}.

\begin{figure}[ht!]
  \begin{flushleft}
    \centering
       \includegraphics[width=6in,height=6in]{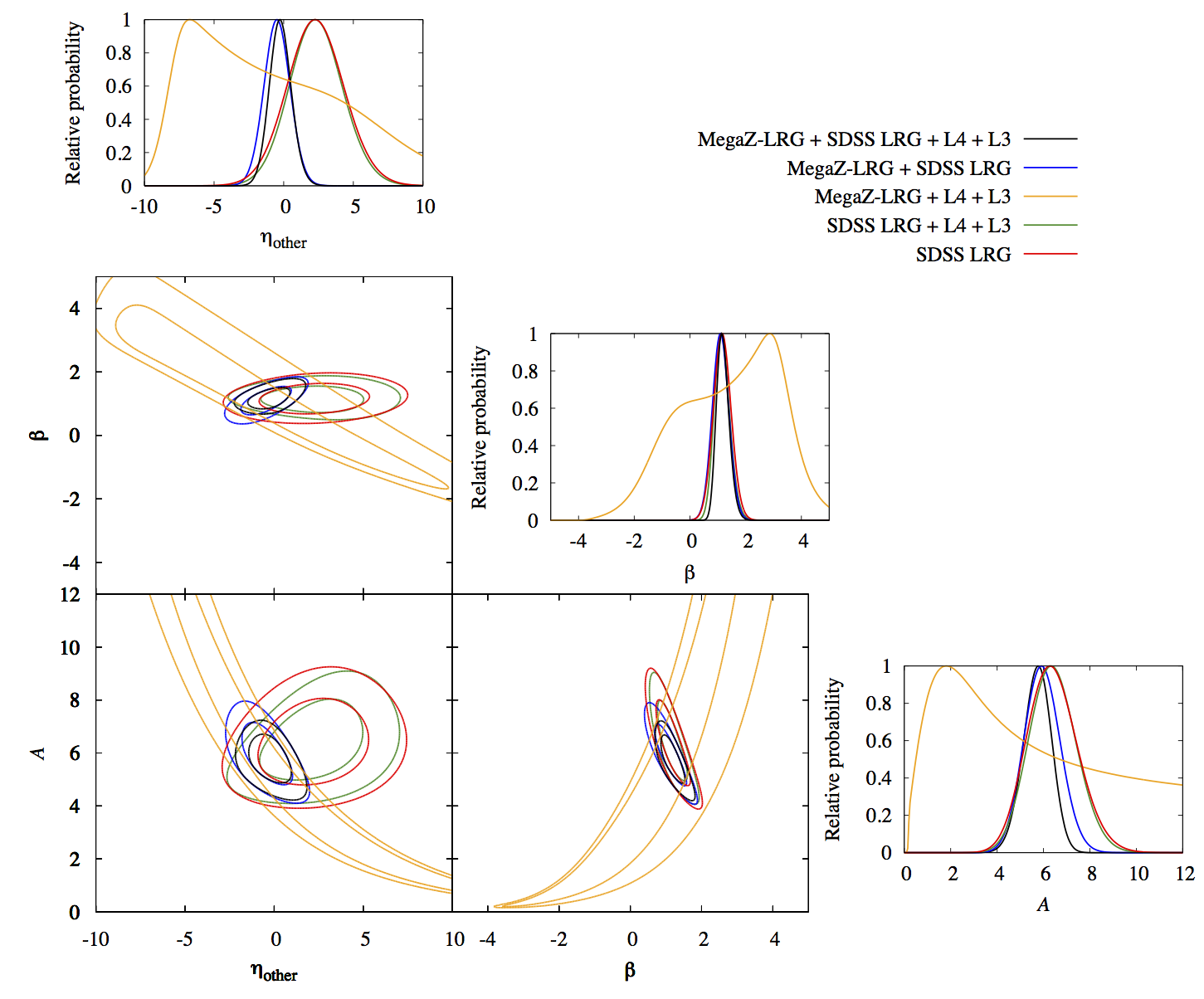}
\vspace{0mm}
\caption{Posterior probabilities from the joint fit to the MegaZ-LRG and SDSS spectroscopic samples from 
\citet{JMA+11}. Fits are shown for the amplitude $A$ of the intrinsic alignment model (which is the same as the $A_{\rm I}$ we use throughout this paper), the 
extra redshift dependence with power-law index $\eta_{\textrm{other}}$ and the index $\beta$ of the 
luminosity dependence, as in \Cref{eqn:power_laws}. \textbf{Lower left triangle of panels:} The two-dimensional 1$\sigma$ and 
2$\sigma$ confidence contours, marginalised in each case over the parameter not shown with flat 
priors in the range $A \in [0; 20]$, $\eta_{\textrm{other}} \in [-10 : 10]$, and $\beta \in [-5;5]$. 
\textbf{Upper right diagonal panels:} Posterior probabilities for $A$, $\eta_{\textrm{other}}$, and $\beta$, each 
marginalised over the two remaining parameters. The different coloured lines result from using different samples in the fitting process (see legend for details).
\permaa{JMA+11}
}
\label{fig:JMA11_fig14}
  \end{flushleft}
\end{figure}

\citet{JMA+11} studied intrinsic alignments in LRGs from the SDSS MegaZ-LRG dataset. Compared to the 
other studies discussed in this
section, \citet{JMA+11} studied galaxies at higher redshifts (closer to the typical redshifts of 
cosmic shear surveys) and were the first to
measure large-scale intrinsic alignments in a dataset with photometric redshifts (see \citealp{CBC+03} 
for a seminal example of the use of photometry for redshifts). The MegaZ-LRG sample used in 
\citet{JMA+11} contained more than 800,000 LRGs up to $z \sim 0.7$. Compared to spectroscopic 
redshifts, the advantage of photometric redshifts is that these are cheaper to obtain for large 
samples, albeit at the expense of redshift precision. Consequently,
it is necessary to explicitly model out the gravitational lensing ``contamination'' due to the 
intrinsic alignment signal. For more discussion of the use of photometric redshifts, see \Cref{sec:projected_corr_fns}.

\citet{JMA+11} measured $w_{\rm g+}$  and used the results for $r_{\rm p} > 6 $Mpc/$h$ to place 
constraints on non-linear alignment model parameters (see
\Cref{fig:JMA11_fig14}) by combining this sample with those used by \citet{HMI+07}. Power-laws with 
an extra free parameter each were introduced to allow for  redshift and luminosity dependence, just as in \Cref{eqn:power_laws}.   The 
normalisation of the intrinsic alignment model was found to be $A_{\rm I} = (0.077 \pm  0.008)\rho_{\rm crit}^{-1}$ 
(where $\rho_{\rm crit}$ is the critical density)  for galaxies at $z=0.3$ and evolution-corrected 
$r$-band absolute magnitude of -22. This result is consistent with the value of $A_{\rm I} = (0.066\pm 
0.008)\rho_{\rm crit}^{-1}$ found by the more recent study by \citet{SMM14}, which we 
discuss below. Note that the yellow contour in \Cref{fig:JMA11_fig14} is for the combination of the MegaZ-LRG and SDSS Main samples (rather than the SDSS LRGs), which yields weak and degenerate constraints because this combination cannot tell redshift from luminosity scaling. 

To study deviations from the redshift dependence of the signal predicted by the 
non-linear alignment model, \cite{JMA+11} used several samples at a range of redshifts from $z \sim 0.1$ to 
$z \sim 0.55$ and found $\eta_{\text{other}} = -0.3 \pm 0.8$.  Luminosity-dependence was analysed through fits to different luminosity subsamples. The L3 and L4 SDSS samples are defined with absolute magnitude cuts of $-20 \leq M_{r} + 5 \log_{10}h < -19$ and $-21 \leq M_{r} + 5 \log_{10}h < -20$ respectively and colour cuts detailed in \citet{JMA+11}. L3 contains 66,312 galaxies, with a mean redshift of $\langle z \rangle = 0.07$, L4 contains 118,618 galaxies, with a mean redshift of $\langle z \rangle = 0.11$. The power-law slope of the dependence 
of the alignment signal with galaxy luminosity was constrained to be  $\beta = 1.1^{+0.3}_{-0.2}$, 
also in agreement with the value of $1.3\pm 0.27$ obtained by  \citet{SMM14}.

\citet{SMM14} have presented the most comprehensive study of large-scale intrinsic alignments in 
early-type galaxies to date. They studied 
the intrinsic alignments of the low-redshift LRG sample in the Baryon 
Oscillation Spectroscopic Survey (BOSS) survey, called LOWZ, using data from DR11 \citep{DSA+13,AAA+15}. 
Unlike the original SDSS-I/II LRG sample studied in early works by \cite{HMI+07} and \cite{JMA+11}, 
this sample goes to lower luminosity, with a comoving number density that is three times as high as the earlier sample within the same redshift range ($0.16<z<0.36$). The sample used by \citet{SMM14} to trace the 
density field contains 173,855 galaxies, of which 159,621 have good shape measurements, and are 
further divided into subsamples based on colour, luminosity, and environment.

\citet{SMM14} measured the $w_{\rm g+}$  signal from the cross-correlation of the shape and density 
samples and modelled the signal using the non-linear alignment model for $r_{\rm p}>6 $Mpc/$h$ and with a fitting formula 
for the halo model from \cite{SB10} on small scales ($r_{\rm p}<1.5 $Mpc/$h$).  This provided a set 
of large-scale and small-scale intrinsic alignment amplitudes that were studied as a function of sample properties. 

\begin{figure}[ht!]
  \begin{flushleft}
    \centering
       \includegraphics[width=4in]{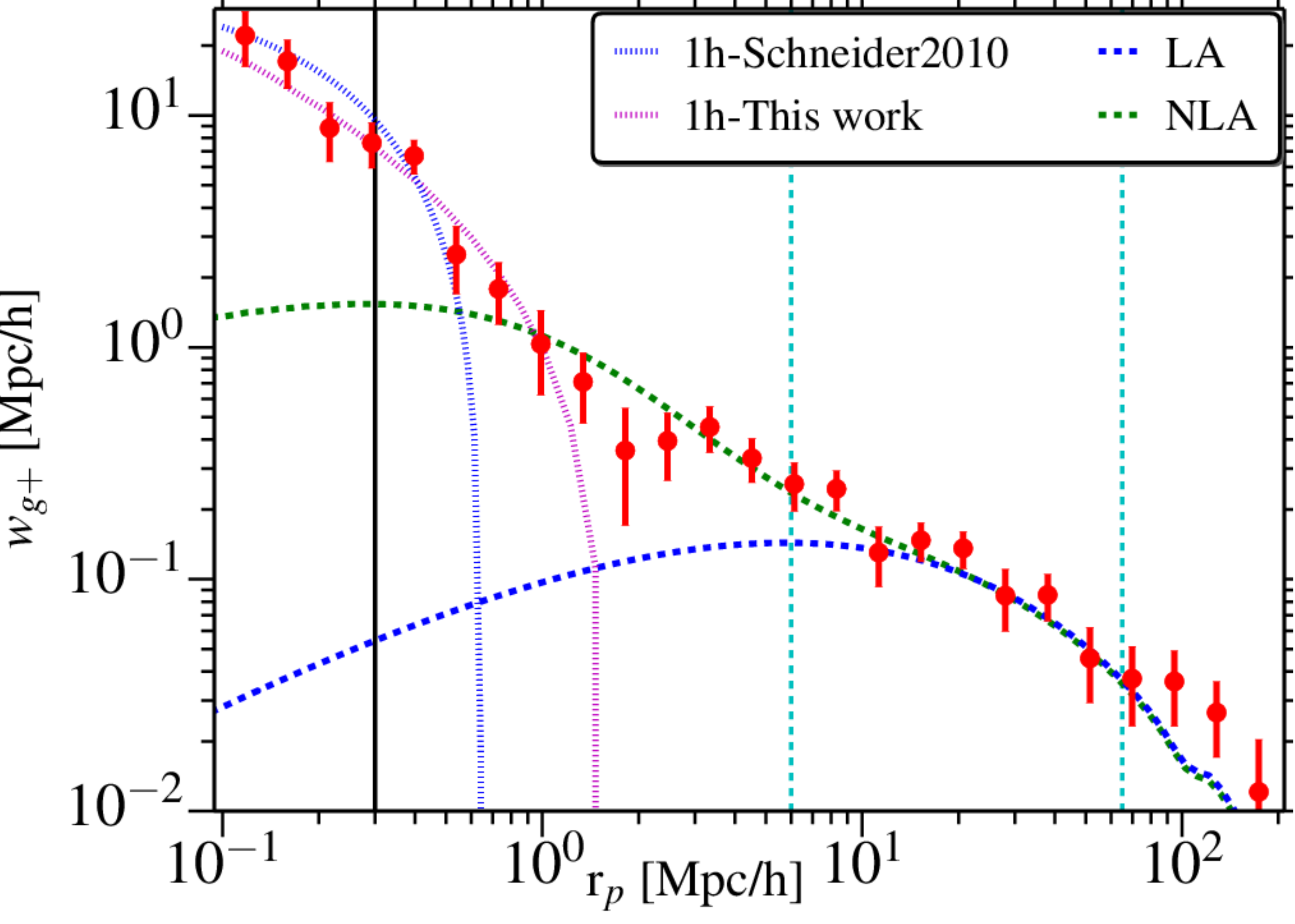}
\vspace{0mm}
\caption{The density-shape correlation function, $w_{\rm g+}$ for the SDSS-III BOSS LOWZ sample used
  in \citet{SMM14}. The data used covers the redshift range $0.16 < z < 0.36$. The red points are
  the measurements from the data, the dashed green lines are the non-linear alignment model, and the dashed blue
  lines are the linear alignment model. The non-linear alignment model is fitted only in the range $6 \textrm{Mpc}/h < r_{\rm p}
  < 65$ Mpc/$h$ (shown by dashed vertical lines), while the linear alignment model is shown with the same
  parameters as the non-linear alignment model. The dotted purple and blue lines show halo model fits to
  $w_{\rm g+}$ at small scales, see text for more details. The black solid line shows the SDSS fibre collision scale at $z =
  0.36$. \permmn{SMM14}
    }
\label{fig:SMM14_fig4}
  \end{flushleft}
\end{figure}

The measurements of $w_{\rm g+}$ for the full sample agree well with the linear alignment model at 
scales larger than $\sim 10 $Mpc/$h$,  as shown in \Cref{fig:SMM14_fig4}. Below this scale, the non-linear alignment 
model provides a good fit to the data above $1$ Mpc/$h$ (though there is a notable dip in the measured signal at $\sim 2 $Mpc/$h$). Note that the non-linear alignment model includes an optional smoothing scale, and the selection of this scale can, in principle, affect the match at small scales. Below $\sim 1 $Mpc/$h$ they apply two versions of the intrinsic alignment halo model. An implementation of the fitting formula and parameter values from \citet{SB10} (dotted blue line) does not fit the data well. However, when some of the fitting function parameters are modified to better suit the SDSS LOWZ sample used in this work (dotted purple line), the halo model is shown to fit the data well on the 
smallest scales, from $0.3 -1$ Mpc/$h$. The vertical black line at $0.3$ Mpc/$h$ shows the SDSS fibre collision scale, below which the difficulty of placing optical fibres in close proximity makes measuring clustering statistics difficult. Fitting the non-linear alignment model to scales larger than $6$ Mpc/$h$ yields 
an intrinsic alignment amplitude $A_{\rm I} = 4.6 \pm 0.5$ (with the galaxy clustering suggesting an 
average linear galaxy bias of  $b_{\rm g} = 1.77 \pm 0.04$). It is interesting to note that the best 
fit to the data appears to be the non-linear alignment model at large and intermediate scales combined with the halo model at 
small scales. There is no physical motivation for this combination (compared to the linear alignment+halo model) but 
it suggests that more modelling and simulations work needs to be done to understand behaviour at intermediate scales, $2 < 
r_{\rm p} < 10 $Mpc/$h$. Recent work in this direction has been done by \citet{BVS15} which presents all relevant non-linear corrections at one-loop order, under the tidal alignment paradigm.

\citet{SMM14} also presented results which explore intrinsic alignments as a function of galaxy 
properties and galaxy environment.  As mentioned earlier, the luminosity-dependence of large-scale 
intrinsic alignments for this sample is consistent with the results from previous work 
\citep{JMA+11}, but with smaller statistical uncertainties. 
No variation of the 
intrinsic alignment amplitudes with redshift was found, though given the narrow range of redshift 
for this sample, the ability to study redshift evolution is quite limited. Importantly, the 
amplitude of the signal did not depend on the colour of the red-sequence galaxies, which supports 
the 
tendency in the literature to predict intrinsic alignments for future surveys using a single 
estimate for intrinsic alignments for all red-sequence galaxies regardless of their exact colour.

In terms of environmental dependence, using dark matter haloes identified with multiple LRGs and dividing the 
galaxies into centrals (i.e. Brightest Group Galaxies; BGGs),
satellites, and field galaxies, \citet{SMM14} found that satellite galaxies exhibit no detectable 
large-scale intrinsic alignments, but the radial alignments of the satellite galaxy semi-major axes towards 
the centres of their host haloes are detected at high significance.  Central galaxies show
both small- and large-scale intrinsic alignments, with a higher strength than for field galaxies, 
consistent with their host halo
masses and luminosities being larger. See \citet{Paper2} and attendant references for more discussion on the use of dark matter haloes to model intrinsic alignments and their identification in simulations. 

\begin{figure}[ht!]
  \begin{flushleft}
    \centering
       \includegraphics[width=6in,height=4in]{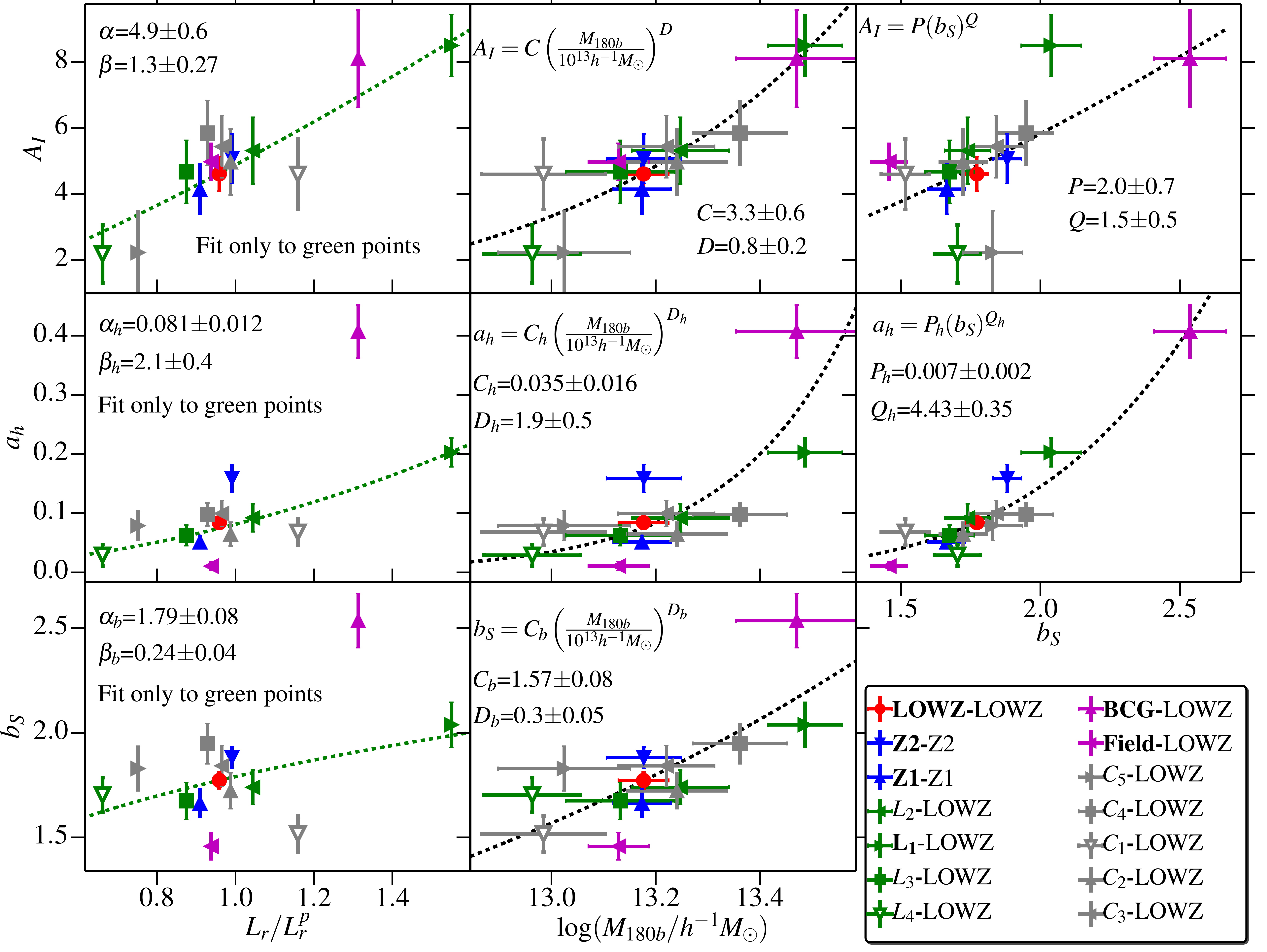}
\vspace{6mm}
\caption{Intrinsic alignment amplitudes and bias from the SDSS-III BOSS LOWZ data of \citet{SMM14}. 
Results are shown for various shape samples, as a function of different galaxy properties of the 
shape sample. Z1 refers to $0.16 < z < 0.26$ and Z2 to $0.26 < z < 0.36$. Note that the full LOWZ 
sample was used as the density sample, except in the cases of Z1 and Z2 where these redshift subsets 
are used instead. L1 contains the brightest 20\% of galaxies, L2 the next brightest 20\%, L3 the 
next brightest 20\% and L4 the faintest 40\%. The bins C1-C5 split the sample by colour. Each C-bin 
contains 20\% of the total sample from bluest (C1) to reddest (C5). The top row shows $A_{\rm I}$ as a 
function of different properties of shape sample. $A_{\rm I}$ shows clear evolution with luminosity 
$(L_{\rm r})$ 
as well as mass, where $M_{180b}$ is the halo mass from weak lensing, and bias, $b_{\rm S}$, with negligible evolution in redshift. The green dotted line in 
$A_{\rm I}$ vs. $L_{\rm r}$ 
shows a power-law fit to the luminosity samples (green points). Similarly in $A_{\rm I}$ vs 
$\log(M_{180b}/h^{−1}M_{\odot}$), the black 
dotted line is the power-law fit, using all the points. The middle row shows the halo model 
amplitude, $a_{\rm h}$ 
as a function of different galaxy properties. For cases where the density sample is fixed to LOWZ, 
the effects of the non-linear bias of the density sample is the same. 
The black dotted line in $a_{\rm h}$ vs $M_{\rm r}$ is 
the power-law fit to the luminosity samples (green points), and the dotted line in $a_{\rm h}$ vs 
$b_{\rm S}$ is the power-law fit using all the points. Remember that the galaxy properties shown on the $x$-axis 
are correlated, for example, more luminous galaxies also have higher bias and live in more massive 
haloes. \permmn{SMM14}}
\label{fig:SMM14_fig8}
  \end{flushleft}
\end{figure}

\citet{SMM14} detail a number of scaling relations for intrinsic alignments with respect to luminosity, dark matter halo mass and galaxy bias, shown in \Cref{fig:SMM14_fig8}. A key finding of \citet{SMM14} is that dark matter halo mass and galaxy luminosity seem to be equally good (low scatter) predictors 
of the large-scale intrinsic alignment amplitude ($A_{\rm I}$, the non-linear alignment model amplitude) for a given shape sample, whereas the linear bias does not do as well, having a large scatter with respect to intrinsic alignment amplitude.  
In contrast, the lowest scatter predictor of the {\em small-scale} intrinsic alignment amplitude ($a_{\rm h}$, the halo model amplitude) for a given shape sample, is the linear bias. However, the small-scale amplitude has a non-trivial 
dependence on the choice of density tracer sample. See \Cref{fig:SMM14_fig8} for a summary of the 
results of \citet{SMM14} in terms of intrinsic alignment amplitudes and galaxy bias as a function of different galaxy 
properties (of the shape samples). The intrinsic alignment amplitude, $A_{\rm I}$, shows a clear 
dependence on luminosity, galaxy mass and bias but little change with redshift. Small-scale 
$w_{\rm g+}$ is fit with a halo model prescription characterised by an amplitude $a_{\rm h}$ and the 
bias of the galaxy sample $b_{\rm s}$.

Comparison with other papers on intrinsic alignments in SDSS reveals a possible dependence of the 
measured intrinsic alignment signal on the way that the galaxy shape is estimated.  \citet{OJL09} 
measured large-scale intrinsic alignments of 83,773 LRGs from the SDSS DR6. Interpretation of this data by 
\cite{BMS11} in terms of the non-linear alignment model reveal a systematically higher amplitude compared to 
measurements for comparable samples in \cite{HMI+07} and \cite{JMA+11}.  One key difference is that the measurements of \citet{OJL09} use an estimate of the shape based on a low surface 
brightness isophote, instead of
using centrally-weighted PSF-corrected shapes used in \cite{HMI+07} and \cite{JMA+11}.  \cite{HKF+11} found signatures of systematics in the 
alignments of galaxies when measured using isophotal shapes which could be the cause of the extra signal for galaxies in clusters.  
However, it is not implausible that the outer isophotes of galaxies truly are more strongly aligned 
with 
large-scale structure, which could mean that the higher alignment amplitude results from a real 
physical effect.  A direct comparison of measurements using the exact same methodology and intrinsic 
alignment estimator, but different ellipticity estimates, would be necessary to fully understand 
this discrepancy.

One interesting aspect of the interpretation of the results in \citet{OJL09} is that instead of directly 
comparing the data to an analytic model such as the linear alignment model, they compared with predictions for 
dark matter halo alignments from $N$-body simulations, and used the lower signal in the data to 
infer a typical stochastic misalignment between halo and galaxy position angles of $\sigma_{\theta} 
= 35.4^{+4.0}_{\ -3.3}$ degrees. The authors thus concluded that central LRGs are preferentially but 
not perfectly aligned with their parent haloes, since perfect alignment would produce an intrinsic 
correlation signal four times higher than what is measured.

\citet{BMS11} used the measurements of \citet{OJL09} to calculate the $w_{\rm g+}$ statistic and the intrinsic alignment auto-correlation functions, $w_{++}$, 
using LRGs from SDSS DR6 and the New York University value added catalogue \citep{BSS+05}. The paper is worth 
noting here because the authors also went beyond the $w_{\rm g+},w_{++}$ approach by calculating the 
correlation functions of curl-free E-modes and  divergence-free B-modes for intrinsic alignments (see \Cref{sec:2_two-point} for a description of these observables). In addition they presented a different statistic, first introduced by \citet{FLW+09}, called the alignment correlation function, $w_{\rm 
gg}(r_{\rm p},\theta_{\rm p})$, which describes the dependence of clustering on both projected separation, $r_{\rm p}$, and the galaxy orientation angle, $\theta_{\rm p}$, that is the angle between the major axis of a galaxy and the axis of separation. 
When comparing their measured statistics $w_{\rm g+}$ and $w_{++}$ with the linear alignment model, they found 
both fit well at large transverse scales (above $10$ Mpc/h). From their fit to $w_{\rm g+}$ they 
found a best-fit amplitude of $C_{1}\rho_{\textrm{crit}} \approx 0.13 \pm 0.02$, consistent with the result of \citet{JMA+11}.
They found E-modes largely in agreement with theory 
and B-modes consistent with zero above $10$ Mpc/$h$. Below this scale other alignment processes are 
expected to generate B-modes. 

At linear order the tidal 
alignment model predicts that the angular dependence of 
$w_{\rm gg}(r_{\rm p},\theta_{\rm p})$ is $w_{\rm gg}(r_{\rm p})\cos(2\theta_{\rm p})$. \citet{BMS11} demonstrated 
that the alignment correlation function, $w_{\rm gg}(r_{\rm p},\theta_{\rm p})$, was thus completely 
described by $w_{\rm gg}(r_{\rm p})$ and $w_{\rm g+}(r_{\rm p})$. Nevertheless it is useful to 
measure $w_{\rm gg}(r_{\rm p},\theta_{\rm p})$ and compare to predictions of the linear alignment model as a check on 
observations of $w_{\rm g+}(r_{\rm p})$.
\citet{BMS11} found that $w_{\rm gg}(r_{\rm p},\theta_{\rm p})$ increases with luminosity, in agreement with \citet{JMA+11} and \citet{SMM14}. 
The authors hazarded an explanation for the trend of increased alignment with luminosity, namely that 
``more luminous objects have formed more recently and have had less time to misalign from the tidal 
axis along which they formed.'' More detailed interpretation of these trends can be found in \citet{Paper2}.

\subsection{Other Large-Scale Measurements}
\label{sec:observations_largescale_indirect}

We have so far considered direct measurements of the galaxy-ellipticity correlation of both late- and 
early-type galaxies at large scales, as well as observations of the correlation of disc galaxy spin 
vectors and the correlation of major axis with separation vector. Other papers have studied large-scale shape correlations using more indirect methods. 
Instead of using a dataset to directly measure the alignment of shapes with the galaxy distribution, or 
with each other, it is possible to learn about intrinsic alignments through the simultaneous modelling 
of lensing and intrinsic alignments. In these cases the intrinsic alignment signal is sometimes seen as a ``contaminant'' of the cosmic shear signal. This contamination must be characterised to avoid bias, which can provide information about the intrinsic alignments themselves.

This was done by \cite{BMS+12} when they presented a formalism for simultaneous modelling of cosmic shear and intrinsic alignments in the context of galaxy-galaxy lensing, where 
background sources are lensed by specific massive foreground structures associated with galaxies. 
When using photometric redshifts, as they did in this paper, it is not possible to perfectly separate the foreground and 
background populations, hence there will be some intrinsic alignment signal sourced by physically 
close galaxies. The authors exploited this property to measure intrinsic alignments by removing the
galaxy-galaxy lensing contaminant.  While this issue was addressed to some extent by
\cite{JMA+11}, who used data with high quality photometric redshifts
for which the lensing contamination was present at the $\sim 10$\%
level, in \citet{BMS+12} the formalism was designed to accommodate the case where
the photometric redshifts are of more typical quality and therefore the
weak lensing contamination is the dominant component of the radial 
alignment signal, even for sets of galaxies chosen so that they are
supposed to be near the lenses in redshift space.  
\cite{BMS+12} 
produced a null
detection, 
interpreted as meaning
that the intrinsic alignment contamination of the galaxy-galaxy
lensing signal is limited to be $<10$\% for $0.1\le r_{\rm p} \le
10$ Mpc/$h$. 
Under the
assumption that the non-linear alignment model is valid at
describing the radial dependence of density-shape alignments of these
source galaxies, the constraints become considerably tighter, with the
contamination expected to be $<1$--$2$\% 
on those scales.  

\cite{CMS+14} also used this new formalism, extended to allow for full photometric redshift probability distributions, $p(z)$, to place constraints on
intrinsic alignments of a deeper sample of source galaxies in the SDSS
stripe 82 region (using coadds from \citealt{HHM+11} and coadd photometry from \citealt{ASS+14}). 
They used galaxy clusters as density tracers and measured
the tangential shears of sources around and behind those clusters to
compute the lensing contamination to the intrinsic alignments signal.
Using the non-linear alignment model to define the scaling of intrinsic alignments with
transverse separation, \cite{CMS+14} constrained the contamination
fraction for a galaxy-galaxy lensing measurement to lie between $-18\%$ and $+23\%$ ($95$\% 
confidence level), using this cluster
sample and source sample below $1$ Mpc/$h$, more discussion of this paper is given when environment-dependent observations are covered in \Cref{sec:observations_smallscale}.

A more truly ``indirect'' measurement was made by \citet{HGH+13}. Here the goal was to measure the 
weak gravitational lensing cosmic shear signal. To this end the expected intrinsic alignment 
contribution was modelled out before any cosmological inferences were made. In this approach the 
amplitude of the alignment signal was a free parameter, included as part of the model that was compared to 
the data. The authors thus marginalised over intrinsic alignment contamination as part of a 
tomographic cosmic shear analysis using six redshift bins.

\citet{HGH+13} used the non-linear alignment model with a single free parameter, $A_{\rm I}$, parameterising both the II and 
GI amplitudes, citing the null result of \citet{MBB+11} for late-type galaxies as justification for using this relatively 
inflexible (one free parameter) parameterisation. Note that the same intrinsic alignment model is applied to both early- and late-type galaxies in this analysis. The observable under consideration is the real-space ellipticity two-point 
correlation function, i.e.  the sum of the cosmic shear signal due to weak gravitational lensing 
{\it and} both the II and GI signals. 

\begin{figure}[ht!]
  \begin{flushleft}
    \centering
       \includegraphics[width=3.5in,height=2.5in]{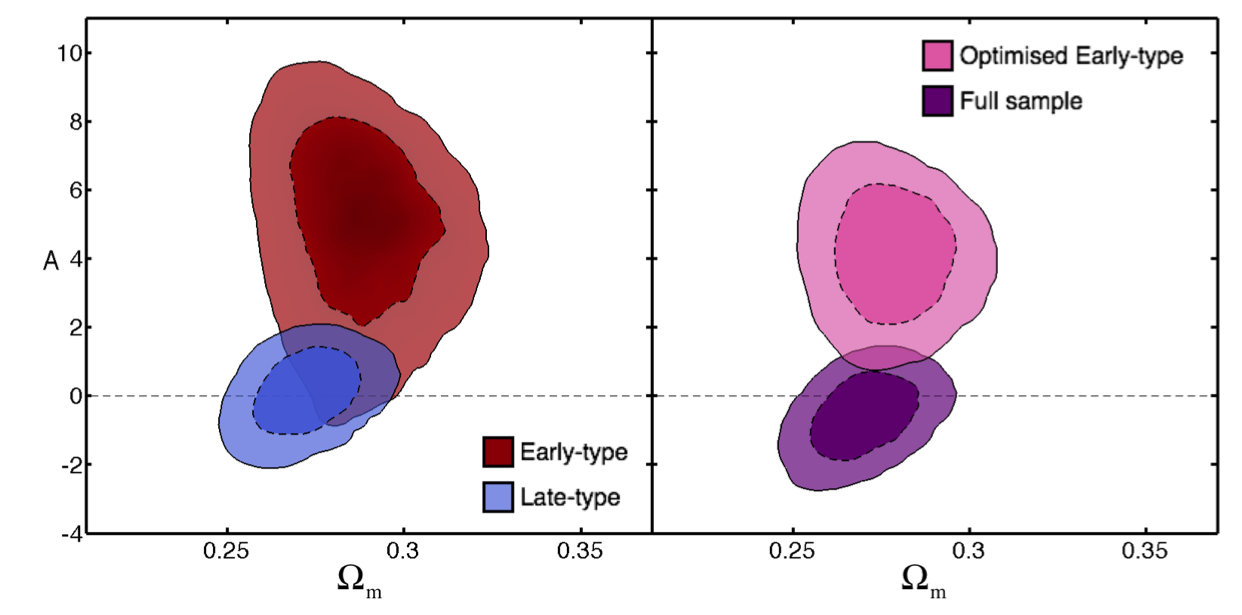}
\vspace{0mm}
\caption{Joint parameter constraints on the amplitude of the intrinsic alignment model ($A$, which is the same as our $A_{\rm I}$)
and the matter density parameter, $\Omega_{\rm m}$, from CFHTLenS combined with WMAP7, BOSS and the \citet{RMC+11} supernova sample as 
presented in \citet{HGH+13}. \textbf{Left panel:} Constraints for two galaxy samples split by spectral energy distribution 
type (early-type in red and late-type in blue). \textbf{Right panel:} Constraints from an optimised 
analysis to enhance the measurement of the intrinsic alignment amplitude of early-type galaxies 
(pink). The full sample, combining early and late-type galaxies, produces an intrinsic alignment 
signal that is consistent with zero (shown purple). A flat $\Lambda$CDM cosmology is assumed. 
\permmn{HGH+13}}
\label{fig:cfhtlens_constraints}
  \end{flushleft}
\end{figure}

\citet{HGH+13} split galaxies by the best-fit spectral energy distribution type, as determined by the photometric redshift 
algorithm 
into late-type spiral galaxies, which constitute $\sim\!80\%$, and early-type galaxies, which make up 
the remaining $\sim\!20\%$. Various combinations
were considered: full sample, early-type, late-type and optimised early-type. Optimised early-type uses the same sample as the early-type analysis but includes the full sample (early- and late-type) in the background redshift bins used for tomographic cross-correlations. This optimisation is designed 
to overcome 
the noise due to the small sample size of early-type galaxies without changing the intrinsic alignment contribution, which is sourced by the foreground bin population. The intrinsic alignment amplitude, $A_{\rm I}$, is 
constrained simultaneously with cosmological parameters. 
\Cref{fig:cfhtlens_constraints} shows constraints on $A_{\rm I}$ and $\Omega_m$ marginalised over 
the  standard cosmological parameters. Note that these results also include auxiliary cosmological 
 information from outside the CFHTLenS survey.

The results showed a strong dependence on galaxy type. The late-type sample has an alignment signal 
consistent with zero, as does the full sample, with amplitudes 
of $A_{\rm I,late} = 0.18^{+0.83}_{-0.82}$ and $A_{\rm I,all} = -0.48^{+0.75}_{-0.87}$, 
respectively. A tentative signal is detected for the
early-type sample, with $A_{\rm I,early} = 5.15^{+1.74}_{-2.32}$, but the null result is inside the 
$2\sigma$ confidence region. The optimised early-type sample is less ambiguous, with a best-fit 
amplitude of $A_{\rm I} = 4.26^{+1.23}_{-1.39}$. For tomographic surveys like CFHTLenS the 
linear alignment model has the (negative) GI contribution dominating over the (positive) II signal. A 
negative value of $A_I$, as marginally preferred by the full sample, would suggest that the data prefer a more positive intrinsic alignment 
contribution. It should be noted however that \citet{HGH+13} applied the linear alignment model to all their galaxy 
samples, even though intrinsic alignment of late-type galaxies is not expected to be sourced by the 
same mechanism (see the introduction to thsi section and \citet{Paper2} for more detail). 
Fitting this simple model to a joint population of late- and early-types, as they do in the full 
sample analysis, may not always be appropriate, nor will ignoring luminosity evolution in analyses with larger samples.

\citet{DIP+15} re-analysed the CFHTLenS data in the context of joint constraints on intrinsic 
alignments and deviations from general relativity. When they assumed general relativity, their results agreed with 
those of \citet{HGH+13}. 
Scale independent 
modifications to general relativity have no major effect on the intrinsic alignment constraints. However, when the authors allow 
scale dependent modifications to general relativity, the constraints on $A_{\rm I}$ weaken and the 95\% confidence 
contours of the $A_{\rm I}$ constraint from the optimised early-type sample are consistent with the 
null result. Effectively the data are not sufficient to simultaneously constrain intrinsic alignments and deviations from general relativity. It should also be noted that \citet{DIP+15} assumed that modifications to GR did not modify the form of the IA signal, which may not be entirely justified.

\section{Environmentally dependent alignments}
\label{sec:observations_smallscale}

In this section, we review measurements of the alignments of galaxies and groups/clusters of 
galaxies which have considered the specific environment in which such systems reside. At megaparsec scales, 
the cosmic web can be roughly divided into four different types of environments: voids, 
sheets, filaments, and knots. In numerical simulations, these environments are defined according to 
the number of gravitationally collapsed dimensions: 0, 1, 2, and 3, respectively. Voids correspond 
to the emptiest parts of the sky (in terms of galaxy density): underdense regions with scales of 
$\gtrsim10$ Mpc; while knots correspond to massive galaxy groups and clusters 
\citep[e.g.,][]{HCP+07}.

Alignment measurements that take into account the different environments started in the late 1960s 
with photographic plates and only in the 21st century did measurements start to be performed with 
modern CCD cameras and the resulting high accuracy of shape measurements. In the following sections we discuss the most recent results; a full historical 
review is presented in \cite{Paper1}, which also contains extensive referencing of the field.

In this section (particularly in \Cref{sec:environment_field,sec:environment_clustercluster}) we take 
the distribution of satellite galaxies as an observational proxy for the shape of the host dark 
matter halo, which is not accessible to observations. This is a reasonable, although not exact,
approximation for dark matter haloes with elliptical central galaxies, i.e. mostly groups and 
clusters \citep[see][]{Paper2}. It is probably not applicable to spiral (late-type) centrals; for 
such galaxies a coherent satellite distribution may hint to a common tidal origin of these 
satellites \citep[e.g.][]{PK13}.

For these observations it is of course important that galaxy type and the morphology of the local 
large-scale structure (groups, filaments, voids, sheets) are determined to high 
accuracy when attempting to make statements about galaxy shape correlation for a given population.
\cite{BPN+08} assessed different selection criteria for isolated and group galaxies using SDSS and the 
Millenium simulation \citep{SWJ+05} and showed that most studies correctly identified only 
$\sim30-40\%$ of isolated galaxies (and their satellites) in their samples, with the rest 
typically being incorrectly identified as members of galaxy groups. 
Improvements in identification are very important for the future of these studies. More discussion of the characterisation of morphology in $N$-body and hydrodynamical simulations can be found in \citet{Paper2}.

Some groups have made interesting attempts to use quasar polarisation as a tracer of alignment \citep{Hutsemekers98,HL01,HCL+05,PH15}. The unified picture of Active Galactic Nuclei (AGN), which include quasars, sees them as sourced by the accretion of matter onto a central supermassive black hole. The polarisation is believed to be either parallel or perpendicular to the accretion disc, depending on inclination with respect to the line of sight, based on studies of low redshift AGN \citep{SRA+04}. \cite{HBP+14} found that quasar polarization in galaxy groups at $z\sim1.3$ is either 
parallel or perpendicular to the principal axis defined by group galaxies, as well as being correlated with the polarisation vectors of their neighbouring quasars.

\subsection{Galaxy position alignments in the field and the Local Group}
\label{sec:environment_field}

\cite{Holmberg69} originally found that the distribution of galaxies that are satellites around  
field spirals tend to be located along the minor axis of the central galaxy. He looked at edge-on 
spiral host galaxies, whose minor axes are easier to identify, and restricted himself to satellites 
at radii smaller than 50 kpc. 
Subsequent studies using larger samples of galaxies were not able to confirm this result 
\citep[e.g.][]{HP75,SLW79}. \cite{ZSF+97} did find evidence for this ``Holmberg effect'', 
but only for satellites at distances between 300 and 500 kpc from their host galaxy. Later 
studies using SDSS have found that satellites of spiral galaxies are distributed isotropically 
\citep[e.g.][]{APP+07,BPN+08}. In contrast, satellites of early-type centrals are located 
preferentially along their host's \textit{major} axes, i.e. an anti-Holmberg effect\footnote{See \citet{Paper1} for details of the history of these observations and the messy nomenclature that resulted.} \citep{Brainerd05,APP+07,SL09,NAT+11}. 

Contrary to the results for field galaxies, there is evidence of a strong Holmberg effect for Milky Way satellites 
\citep[e.g.][]{LyndenBell76,KD76,KTB05,PK13}.
M31 is probably the only galaxy other than the Milky Way for which the three-dimensional distribution 
of satellites can be mapped with  
accuracy because distances can be measured precisely. 
\cite{KG06} found that early-type dwarf satellites of M31 are located in a polar plane, only 16 kpc 
thick, that is only $\sim6^\circ$ from the pole of M31. Similar findings have been presented by 
\cite{CLI+13} and \citet{ILC+13}. \cite{PKJ13} have suggested that two similar structures can be found for 
non-satellite galaxies in the Local Group as a whole, at roughly equal distances to the Milky Way 
and M31.

As discussed by \cite{BPN+08} these results need not be contradictory, for two reasons: 
firstly, the large spirals in the Local Group are not isolated as defined in the above SDSS 
studies, since M31 and the Milky Way are too close to one another to define either of them as 
``isolated''; secondly, the satellites of these galaxies are much fainter than the 
typical satellites of SDSS field galaxies. Indeed, \cite{BPN+08} found a minor-axis alignment of 
sufficiently faint satellites around their hosts in hydrodynamical simulations.

\subsection{Galaxy alignments within galaxy groups and clusters}

\begin{figure}[ht!]
 \centerline{\includegraphics[width=3.5in]{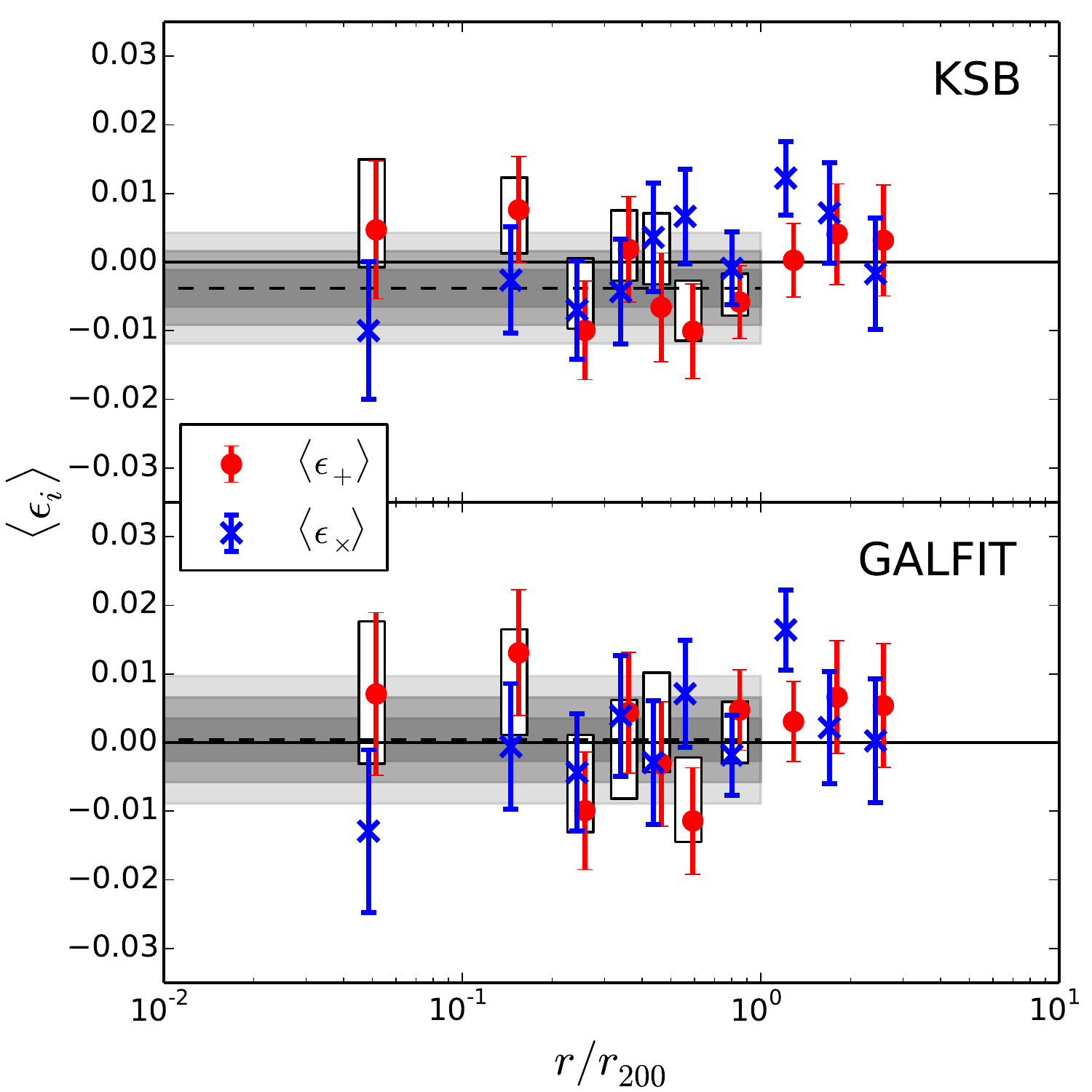}}
\caption{The average ellipticities, $\langle \epsilon_i\rangle$ (with $i$ denoting the $+$ or $\times$ ellipticity component), of galaxies in 
galaxy clusters from 14,250 spectroscopically-confirmed cluster members, using two different shape 
measurement methods on data from the 
CFHT, as a function of projected distance from the BCG, normalised to the cluster radius, $r_{200}$ (note that $r$ as used here is the same as $r_{\rm p}$ used throughout this paper). 
KSB measure quadrupole moments of the brightness distribution \citep{KSB95}, Galfit is a model-fitting method \citep{PHI+02}.
Positive (negative) red circles represent radial (tangential) alignments, while the blue crosses 
show the cross component of the ellipticity. Dashed lines and shaded regions show the weighted mean 
and its 1, 2, and 3$\sigma$ uncertainty regions for all galaxies within $r_{200}$, and the white 
boxes show the results when including photometrically-selected red sequence members. \permaa{SHC+15}
}
\label{fig:clusters_sifon}
\end{figure}

Galaxy clusters 
can host hundreds of 
galaxies and have therefore attracted considerable attention since the pioneering works of 
Fritz Zwicky \citep[e.g.,][]{Zwicky33,Zwicky37}. SDSS data have been used to generate 
catalogues 
that contain hundreds of thousands of groups and clusters 
\citep[e.g.,][]{KMA+07,HMK+10,WHL12,RRB+14}.

There is consistent evidence that the major axes of centrals in galaxy clusters, typically referred to as 
brightest cluster galaxies, BCGs, but different from these in some non-negligible fraction of cases \citep{SCM+15,HLL+15}, seem to produce an inverted Holmberg effect; that is, the major 
axes of the central galaxy and the satellite distribution (which serves as an observational 
proxy for the shape of the cluster's dark matter halo) coincide to a good degree. This tendency has been 
evident since the very first measurements 
\citep{Sastry68,RS72,AP74,Dressler78,CM80,Binggeli82,Struble87}. The most recent measurements have 
come 
from analyses of large galaxy group/cluster catalogues from SDSS spectroscopic 
\citep{YBM+06,APP+07,FLM+07,WYM+08,WPH+10,SRG09} or photometric 
\citep{NSD+10,HKF+11} data, all of which have confirmed this alignment, which is typically found to 
be stronger for early-type, or red, central galaxies. 
The brightest satellite galaxies' shapes are also aligned with the parent halo \citep{LWY+13,SMM14}, and 
there is evidence that BCGs are also aligned with the shape of the X-ray emission from the 
intracluster medium \citep{HHB08}. Using a sample of 405 Abell clusters, \cite{Struble88} showed 
that pairs of brightest cluster galaxies tend to be aligned parallel to each other and 
perpendicular to the separation vector. This is at odds with more recents results which show alignment between clusters, discussed in \Cref{sec:environment_clustercluster} below.

On the other hand, as reviewed in \citet{Paper1}, historically there has not been 
a strong consensus about the alignments of satellite galaxy shapes within their host haloes (or the 
lack thereof). However, the latest measurements seem to agree that, on scales from small galaxy 
groups ($M\sim10^{13}M_\odot$) to massive galaxy clusters ($M\gtrsim10^{15}M_\odot$), the 
orientations of satellite galaxies are consistent with an isotropic distribution \citep{SHC+15,CMS+14} (probably with the exception of the brightest 
satellites, as described in the results above).

The first modern (i.e., using CCD observations), statistical measurements of galaxy alignments in 
galaxy groups were presented by \cite{PK05}, who used SDSS photometric and spectroscopic 
observations and found anisotropic galaxy orientations at the $4\sigma$ level. 
Some other authors have also found that satellite galaxies are aligned either radially from the position of the central galaxy \citep{AB06a,FLM+07} or aligned with the major axis of the central galaxy \citep{YBM+06}. However, \citet{HKF+11} showed that these detections correspond to systematic effects in the isophotal measurements from SDSS. Since \citet{HKF+11}, all measurements have been consistent with no alignments \citep{HE12,SCF+13,CMS+14,SHC+15}. \cite{CMS+14} 
presented a modelling of intrinsic alignments in photometrically selected groups and clusters, 
improving upon the method of \cite{BMS+12} by accounting for photometric redshift errors using the 
full redshift probability distribution of each galaxy. In a complementary work, \cite{SHC+15} used 
a large sample of spectroscopically-confirmed galaxy cluster members to directly measure the 
average alignment of satellite galaxies. 

The work of \cite{CMS+14} has the advantage of being 
applicable to photometric data, with great potential for large photometric surveys such as the Large 
Synoptic Survey Telescope \citep[LSST,][]{LSST09} and Euclid \citep{LAA+11}, while that of 
\cite{SHC+15} is a cleaner and more direct measurement but depends on large spectroscopic datasets 
which are less readily available. Both works found no evidence for galaxy alignments in clusters, 
constraining the average intrinsic ellipticity signal in galaxy groups and clusters to be 0.5\% or lower.
The measurements of \cite{SHC+15} are reproduced in \Cref{fig:clusters_sifon}; they used two 
different shape measurement methods with different radial weighting schemes (GALFIT, by
\citealt{PHI+02}, and KSB, by \citealt{KSB95}, see \Cref{sec:obs_measuring_shapes} above for more discussion of the general principles of shape measurement), both of which gave consistent results, suggesting 
that the non-detection of alignments in clusters is robust to differences in shape measurement. 
Additionally, thanks to their large, clean sample of cluster members, \cite{SHC+15} directly 
measured the alignment \textit{between} cluster satellites, which they also found to be consistent 
with zero. The fact that the measurement of the cross component, $\langle \epsilon_{\times} \rangle$, was consistent with zero acts as a null test and demonstrates that the analysis of \citet{SHC+15} is robust to, at least some types of, systematics. Also using spectroscopically-confirmed cluster members but with reduced constraining 
power due to lower image quality of single-pass SDSS observations (as opposed to deep SDSS Stripe 82 
data and deep, better-seeing CFHT/MegaCam images used by \citealp{CMS+14} and \citealp{SHC+15}, 
respectively), \cite{SCF+13} constrained the average shear signal of satellite galaxies in Galaxy and Mass Assembly (GAMA) survey\footnote{\texttt{http://www.gama-survey.org/}} \citep{DBB+09} groups to 
$\lesssim2\%$. \cite{SCF+13} note the disagreement between their null detection and results from $N$-body simulations, suggesting that this is evidence for misalignment between baryonic and dark matter shapes.

As discussed by \cite{SMM14}, the observed radial alignments of the most luminous satellite 
galaxies by \cite{LWY+13} and \cite{SMM14} would lead to null detections of fainter satellites as 
observed by, for example, \cite{CMS+14} and \cite{SHC+15}. This assumes that the scaling of satellite alignments with 
luminosity estimated by \cite{SMM14} for bright red galaxies (specifically LRGs) is extrapolated 
below the luminosity limit explored by the latter authors, whose sample also incuded blue galaxies which would further dilute any signal. Therefore, to current precision, the 
latest observations show a consistent picture, in which very luminous (red) satellites align towards the 
central galaxy and progressively fainter (red and blue) satellites align less and less until the alignment signal 
of faint satellites is below the current detection limit.

\subsection{Galaxy alignments with voids}

Voids are an attractive reference against which alignments of galaxies can be 
measured. They have a higher degree of symmetry than filaments and, in contrast to clusters, tend to 
become more spherical as they evolve gravitationally -- the effect of the strongest inward forces 
acting along the shortest axis in an aspherical overdensity is reversed for underdense voids 
\citep{SW04}. On the downside, large voids are rare objects and are characterised by the absence of 
luminous structures, so detecting them requires a galaxy survey covering a large, contiguous volume with 
densely sampled spectroscopic redshifts. 

Such a dataset only became available with the advent of SDSS, which additionally supplied imaging 
of sufficient quality to measure galaxy morphology with high accuracy. Thus it is not 
surprising that the observational study of void alignments closely follows the development of SDSS, 
with three publications based on DR3 \citep{TCP06}, DR6 \citep{SW09}, and DR7 
\citep{VBT+11}. These works shared a lot of their methodology, in particular the algorithm to find 
and define voids (the HB algorithm described in \citealp{PBP+06} who also give an overview of 
other methods), but obtained strikingly different results. It is possible that details of the implementation, for example assumed limiting magnitudes, which differed between implementations of the same algorithm are responsible, underlining the sensitivity of void-finding to the method employed.

\citet{TCP06} searched for alignments among galaxies on the surface of voids using data from SDSS DR3 and the Two-Degree Field Galaxy Redshift Survey (2dFGRS). They defined voids as spheres of radius larger than $10$ Mpc$/h$ (clearly an idealistic assumption, there will generally be some confusion over void boundaries) within the SDSS survey 
boundaries that contain no galaxies brighter than $M_{b_J}=-19.32 + 5 \log h$. Here, $b_J$ denotes 
a blue photometric band used for the target photometry of the 2dF Galaxy Redshift Survey 
\citep{CDM+01}. Disc-dominated galaxies were selected in a shell of $4$ 
Mpc$/h$ thickness on the surface of the voids, using only objects that are nearly face-on or 
edge-on. The latter step avoids an ambiguity in the inclination of the disc as it is impossible to 
decide which are the near and far \lq edges\rq\ of the galaxy if only the axis ratio of the image 
is known.

In total, \citet{TCP06} used 178 voids and 201 galaxies with estimates of their spin axes. 
From these they derived the probability distribution of the 
angle $\theta$ between the spin vector of a galaxy and the vector connecting the void centre with 
the position of the galaxy. Their measurement is inconsistent with random galaxy orientations at the 
$99.7\,\%$ level, preferring an orientation of the spin vector perpendicular to the void radius 
vector. The signal is well described by a simple model based on tidal torque theory, assuming
\eq{
\label{eqn:lee_spin_align}
\ba{ \hat{J}^i \hat{J}^j } = \frac{1+a_{\rm T}}{3}\, \delta_{ij} - a_{\rm T}\, \sum_k \hat{T}^{ik} \hat{T}^{jk} \;,
}
where $\hat{J}^i$ is the normalised spin vector, $\hat{T}^{ij}$ the normalised traceless shear tensor, $i,j$ denote pairs of galaxies, and 
$a_{\rm T} \in \bb{0;1.0}$ a correlation parameter measuring the alignment of the shear and inertia tensors \citep{LP00}. $\delta_{ij}$ is the Kronecker delta. Based on this model for spin correlations, 
and assuming a Gaussian distribution of the spin vector elements, \citet{Lee04} derived a general 
result for the probability distribution of angles of galaxy spin vectors relative to eigenvectors 
of the tidal shear field which was able to qualitatively recover the observed inclinations of spiral galaxies in the local supercluster.  
\citet{TCP06} measured $a_{\rm T}=0.7^{+0.1}_{-0.2} (1\sigma)$.

This detection is in marked contrast to that of \citet{SW09} who reported a null detection with 
a constraint $a_{\rm T}<0.13$ at $3\sigma$. Their analysis 
only differed in that it was updated to the larger and more homogeneous DR6 and used an absolute magnitude limit\footnote{Note the two works use different conventions for the Hubble constant.} of $M_r=-20.23+5 
\log h$ 
 as the threshold for 
void detection. The larger area and higher filling factor for galaxies inside the survey volume resulted in a significantly larger sample of voids. Additionally, their definition of the 
galaxy sample on which spin measurements were made was slightly different, but even when 
approximately recovering the selection of \citet{TCP06}, no signal was detected.

\begin{figure}[ht!]
\centering
\includegraphics[width=4.5in,height=3.5in]{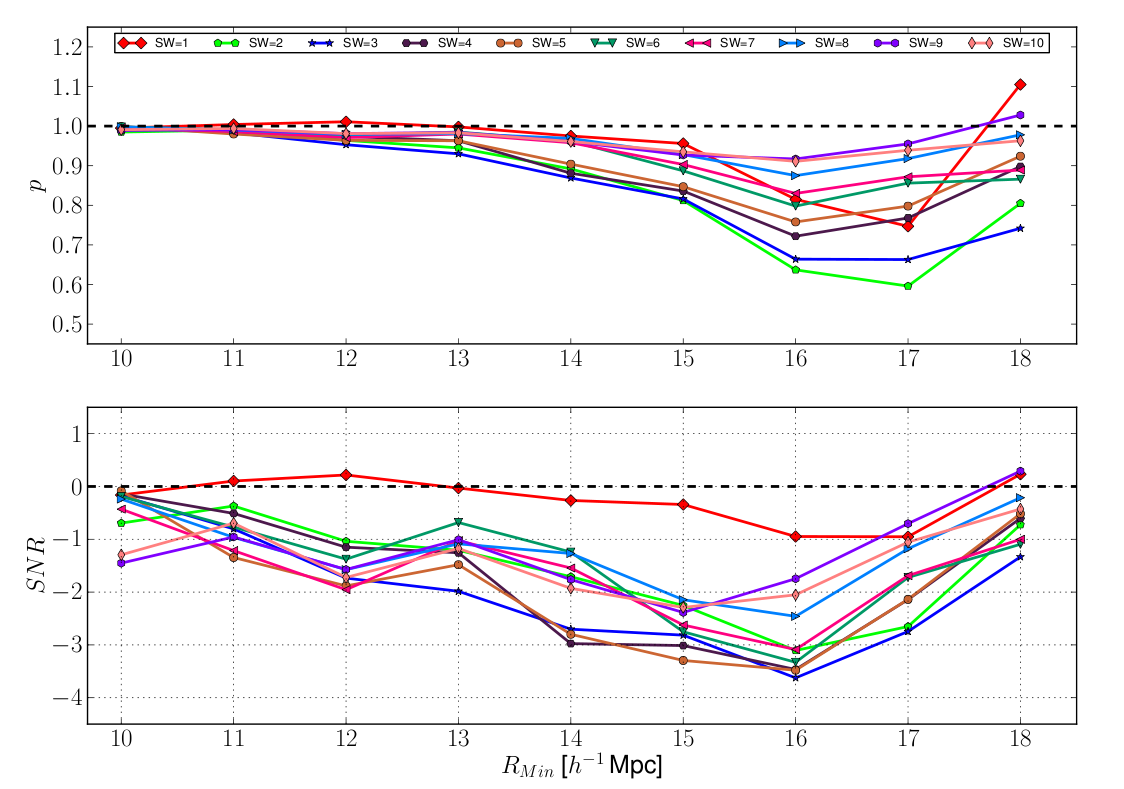}
\caption{\textbf{Top:} Alignment parameter, $p=\langle | \cos \theta | \rangle^{-1}-1$, as a function 
of minimum void radius $R_{\rm min}$. See \Cref{eqn:alignment} for more details of these quantities. The different lines correspond to the width (SW) in Mpc/$h$ 
of 
the shell in which galaxy orientations are measured, in the range 1 to 10 Mpc/$h$ as indicated 
in the legend. \textbf{Bottom:} As above but showing signal-to-noise ratio (SNR) of the 
alignment. Negative SNR and $p<1$ indicates that the galaxy spin preferentially lies parallel to 
the line connecting the void centre and the galaxy. \permapj{VBT+11}}
\label{fig:void_alignment_varela}
\end{figure}

In the most recent analysis \citet{VBT+11} used galaxies from the SDSS DR7 and included disc galaxies with all inclinations via a fit to a 
thick-disc model using morphological classifications from Galaxy Zoo. The ambiguity in inclination 
angle was accounted for statistically. Employing a luminosity threshold of $M_r=-20.17 + 5 \log h$, 
voids with a range of minimum radii between $10\,{\rm Mpc}/h$ and $18\,{\rm Mpc}/h$ were 
identified, 
with shells for spin measurement between $1\,{\rm Mpc}/h$ and $10\,{\rm Mpc}/h$ in thickness. The 
signals were fitted to the same model as used in the preceding works, but with a different free 
parameter $p$ which is related to the mean angle and the parameter $a_{\rm T}$ of 
\Cref{eqn:lee_spin_align} as follows:
\eqa{
\ba{|\cos \theta|} = \frac{1}{1+p}\;;\\
a_{\rm T} = \frac{2 (p^2-1)}{2 (p^2-1)+3}\;. \label{eqn:alignment}
} 
Hence $p=1$ implies random orientations of galaxies, while $p>1$ if galaxy spin is preferentially 
perpendicular to the vector from void centre to galaxy. Best fits for $p$ and the significance of 
the 
alignment signal are shown in \Cref{fig:void_alignment_varela}. For a minimum void radius of 
$10\,{\rm Mpc}/h$ and a shell width of $4\,{\rm Mpc}/h$, \citet{VBT+11} find no alignment, in agreement 
with \citet{SW09}, and corresponding to a constraint $a_{\rm T} = -0.003^{+0.020}_{-0.021}\; (1\sigma)$. 
Using only voids with radii in excess of $\sim 15\,{\rm Mpc}/h$, alignments with $a_{\rm T} \approx -0.5$ were 
found around the $3\sigma$ level. Note that $a_{\rm T}<0$ indicates a breakdown of the assumptions underlying the model in \Cref{eqn:lee_spin_align}, which states that the spin vector preferentially aligns with the intermediate axis of the tidal shear tensor. Indeed, \citet{VBT+11} found a tendency for parallel alignment with the radius vector of the void, i.e. the major axis of the tidal shear tensor at the surface of the void. This is exactly the opposite finding from \citet{TCP06}.

Hence the observational evidence for galaxy alignments with the surface of voids remains unclear. 
\citet{VBT+11} stated that the results presented in \citet{TCP06} were selected \emph{a posteriori} as the 
configuration with the strongest detection, and re-computed the significance to be less than 
$2\sigma$. Direct comparison with simulations is hindered by observational selection effects and the 
difficulty of relating galaxy morphology to halo shape and angular momenta. For instance, 
\citet{CPK+08} found, in $N$-body simulations, for a similar setup as in the observational studies, $a_{\rm T} \approx 0.04$ for 
spin alignment and $a_{\rm T} \approx -0.17$ for minor axis alignment of haloes. \citet{Paper2} provide a detailed discussion of this and other simulation results.

\subsection{Galaxy alignments with filaments and sheets}

The notion that galaxies are not randomly distributed on the sky but tend to be concentrated in 
large, elongated structures predates the demonstration of the extragalactic nature of 
galaxies \citep{Reynolds20}. Studies of galaxy position and alignment with respect to the local part of the cosmic web were made from the early 20th century onwards, but results were 
generally inconclusive and often contradictory.
The review by \citet{HWS+06} 
provides an overview of alignment studies, both early and recent, in the local supercluster.

We first discuss two papers \citep{LP02,LE07} which, while they do not discuss filaments or sheets explicitly, do look at the alignment of galaxy spin with the shear field eigenvectors, which in turn are very often used to define filaments, as discussed below. \citet{LP02} claimed the first observational evidence for the alignment of galaxy spin with the tidal shear field at 
the position of the galaxy. 
They estimated the matter density field from the Wiener-filtered 
positions of galaxies from the infrared astronomical satellite (IRAS) all-sky survey\footnote{\texttt{http://irsa.ipac.caltech.edu/Missions/iras.html}} \citep{SCS+06}. The tidal shear field was then derived via explicit 
integration and differentiation of the gravitational potential. Galaxy position angle and axis 
ratio measurements for about $10^4$ spiral galaxies were taken from the photographic plate-based Uppsala General catalogue and its southern counterpart \citep{N73}. With these data, \citet{LP02} used the thin-disc approximation to calculate galaxy spin. They rejected a random 
orientation of galaxy spins at $99.98\,\%$ confidence and found $a_{\rm T}=0.17 \pm 0.04\; (1\sigma)$, 
again 
using the ansatz in \Cref{eqn:lee_spin_align}.

\citet{LE07} kept the morphology data but included a correction for disc thickness in the 
calculation of the spin vector and used the 2 Micron All Sky Survey (2MASS) Redshift Survey\footnote{\texttt{http://www.ipac.caltech.edu/2mass/}} to determine the tidal field. The 
2MASS galaxy positions were expanded on a three-dimensional grid in terms of Fourier-Bessel 
functions, the tidal field obtained via Fast Fourier Transformation, and then interpolated to the 
galaxy position. These authors also found a strong detection of spin alignments, obtaining $a_{\rm T}=0.08 
\pm 0.01$ on average, increasing with increasing overdensity. In the context of tidal torque 
theory, a correlation with $a_{\rm T}>0$ means that galaxy spin is aligned with the intermediate axis of the 
tidal shear tensor (see the argument in \citealp{LP01}), i.e. their spin vectors tend to be perpendicular to filaments and lie in the plane of sheets.

\citet{JWA10} selected edge-on galaxies with axis ratio $< 0.2$ in SDSS DR5 and constructed the 
matter density field via Delaunay tessellation, using the eigenstructure of the tidal shear tensor 
to identify filament candidates. These were then inspected visually to select a clean sample of 67 
filaments, containing only 69 galaxies with spin measurements. Nonetheless, the authors claimed that the 14 objects among those 67 filaments which were oriented with spins perpendicular to the filament direction ($\cos\theta<0.2$, where $\theta$ is the angle between the filament axis and the spin axis) constitute a statistically significant detection of this type of alignment.

Based on SDSS DR8, \citet{TSS13} fitted three-dimensional bulge+disc models to the light 
distribution of galaxies at low redshift ($z<0.2$) to infer the spin vector direction. They approximated the filamentary 
structure by a network of cylinders, defined by elongated overdensities of galaxies. For their full 
sample of spiral galaxies, no alignment was found, while for a subsample of spirals with $M_r < -20.7$ 
the spin axis tended to be parallel to the filament direction. In a sample of flattened early-type 
galaxies, mostly composed of lenticulars, \citet{TSS13} observed a significant alignment of spin 
perpendicular to the filament axis, which again was stronger for the brightest galaxies. The 
luminosity dependence could have been due to a physical trend or due to selection effects as, for example, 
fits to the photometry were more reliable for bright objects. 
The authors argued that, under the 
assumption that spiral (S0/elliptical) galaxies mostly live in low-(high-)mass haloes, their 
findings agree with the established simulation result that massive haloes have their spin aligned 
orthogonally to filaments, while low-mass haloes show preferentially parallel alignment 
\citep[e.g.][]{TLB13}.

This work was extended by \citet{TL13}, who modified the filament-finding algorithm and also 
included large-scale structure sheets, identified as \lq flattened filaments\rq, i.e. a filament 
whose detection probability extends into a plane. The authors found no correlation between the spin 
axis of early-type galaxies (assumed to be the same as the short axis of the galaxy ellipsoid) and the sheet, whereas the spin of spiral galaxies weakly aligns with 
filaments and tends to avoid pointing away from the filament into the plane of the sheet. These 
latter signals vanish inside a 200 kpc$/h$ 
radius around the filament central axis. It is possible that these results are linked to gas infall along the sheets onto the filaments, where angular momentum 
is generated with a rotation axis along the filament direction.

\citet{Zhang13} quantified the alignment of the major axes of galaxy images in SDSS DR7 with sheets 
and filaments. The analysis was based on a catalogue of close to $5 \times 10^5$ \lq groups\rq\ 
with a minimum estimated halo mass of $10^{12} M_\odot/h$, found using an adaptive halo-based 
group finder \citep{YMB+05,YMB+07}. These groups could consist of a single 
galaxy. Following the method of \citet{WMY+12}, the group haloes were used to reconstruct the 
matter density field via a halo bias model extracted from simulations. Based on this 
reconstruction, \citet{Zhang13} identified those groups that resided in a filament or sheet 
environment via the eigenvalues of the tidal shear tensor at that point (one negative and two positive eigenvalues corresponding to a filament; two negative and one positive eigenvalue corresponding to a sheet), and calculated the angle between the projected major axis of the galaxy and the 
projected axes of the filament and the sheet normal vector, respectively. Note however, that this 
study employed the isophotal shapes provided by the SDSS pipeline, which have been flagged as potentially 
unreliable\footnote{\texttt{https://www.sdss3.org/dr8/algorithms/classify.php\#photo\_iso}}; see 
\Cref{sec:introduction} for more discussion on isophotes and shape measurement.

\begin{figure}[ht!]
\centering
\includegraphics[width=3.25in,height=4in]{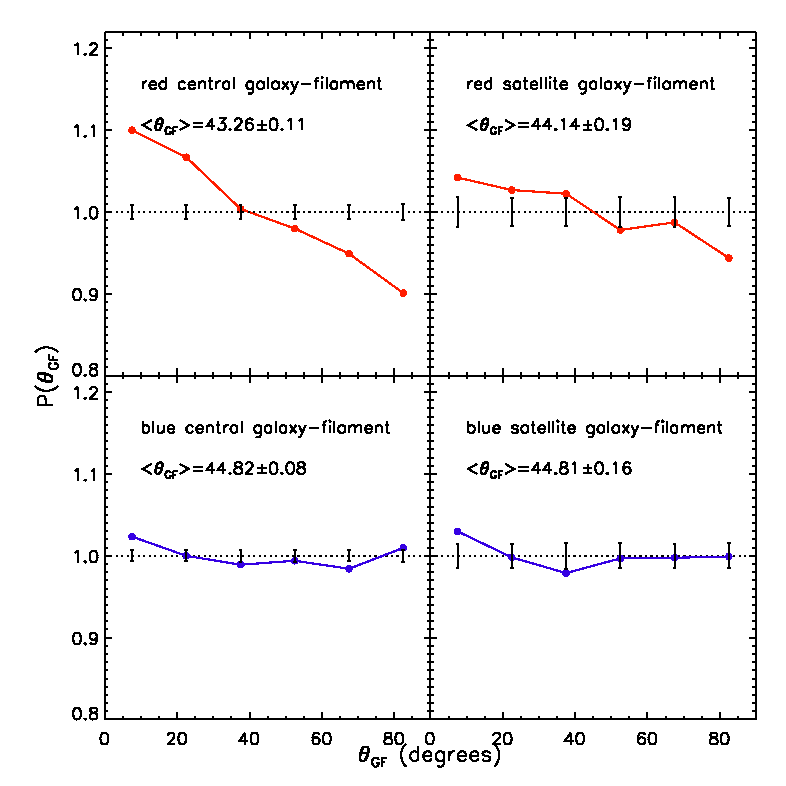}%
\includegraphics[width=3.25in,height=4in]{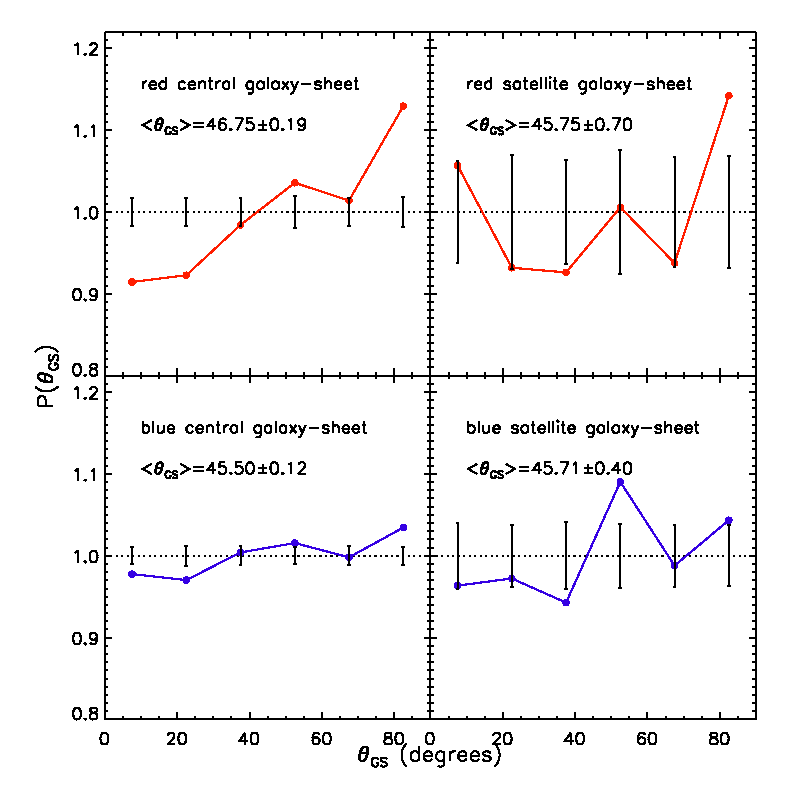}
\caption{\textbf{Left:} Renormalised probability distribution of the angle $\theta_{\rm GF}$ 
between the projected galaxy major axis direction and the filament axis. For random galaxy 
orientation a constant value of unity is expected, as indicated by the black dotted line. Error 
bars are obtained from 100 Monte-Carlo realisations of randomised galaxy orientations. Data are 
shown for red/blue and central/satellite subsamples, as shown in the legends. \textbf{Right:} Same 
as on the left, but for the angle $\theta_{\rm GS}$ between the projected galaxy major axis 
direction and the sheet normal. \permapj{Zhang13}}
\label{fig:lss_alignment_zhang}
\end{figure}

\Cref{fig:lss_alignment_zhang} shows the probability distribution of alignment angles between 
galaxy and filament ($\theta_{\rm GF}$) and galaxy and sheet normal ($\theta_{\rm GS}$), for 
subsamples split into blue/red 
and 
central/satellite (centrals are defined as the brightest group galaxies). Generally, galaxies 
preferentially align with the direction of the filament (small $\theta_{\rm GF}$) and lie within 
the sheet ($\theta_{\rm GS}$ close to a $90^\circ$ angle). The signals are weak for blue galaxies and highly 
significant for central (and bright) red galaxies. These dependencies prevail for the alignments of 
galaxy major axes with the eigenvectors of the tidal tensor. Here, \citet{Zhang13} find 
preferentially orthogonal alignment with the largest eigenvector, parallel alignment with the 
smallest eigenvector and none with the intermediate eigenvector. These results are in good agreement with 
$N$-body simulations for red galaxies. The null detection for blue galaxies is somewhat in tension with $N$-body studies: for example \citet{WMJ+11} found significant correlation between the spin axes of dark matter haloes and the eigenvectors of the tidal field. This tension lessens if there is some mismatch between the spin of the overall dark matter halo and that of the galaxy at the centre, which the subsequent study of \citet{ZYW+14} indicates may be the case.

\citet{ZYW+14} conducted the equivalent study for galaxy spins, reconstructing the three-dimensional direction of the spin vector for galaxies via a simple thick-disc model for galaxies that were classified as spiral by Galaxy Zoo, and as central in their group catalogue. They saw only weak evidence for an alignment with the intermediate axis of the tidal shear tensor (as predicted by tidal torque theory), and therefore for galaxy spins to be preferentially perpendicular to filaments and parallel to the normal of sheets. Comparing their results to $N$-body simulations, they found better agreement with the observations when calculating the spin only in the inner part of dark matter haloes, i.e. closer to the scales of the bright parts of a galaxy. Since the surfaces of voids and sheets classified via the tidal shear tensor should define similar environments, the results of \citet{ZYW+14} are qualitatively consistent with those of \citet{VBT+11}. Note both studies are restricted to low redshifts ($z<0.2$ and $z<0.12$ respectively), additional studies will be required to increase this baseline.

\subsection{Alignments between galaxy groups and clusters}
\label{sec:environment_clustercluster}

As with galaxy alignments, the alignments of galaxy groups and galaxy clusters with the large-scale 
structure have received considerable attention. \cite{Binggeli82} originally discovered that 
neighbouring galaxy clusters tend to point towards each other. He found that all pairs of clusters 
within 30 Mpc of each other (11 out of 30 clusters studied) pointed towards one another, with a 
misalignment of at most $45^\circ$. \cite{West89} showed that this alignment can also be seen for 
less massive galaxy groups out to similar scales, and \cite{Plionis94} found that it decreases with 
distance and is stronger for clusters residing in the same supercluster.

More recently, \cite{WPY+09} used a large sample of galaxy groups to show that the orientations of 
groups and their galaxies are strongly correlated. They confirmed the result of \cite{West89} that 
groups tend to point to their nearest neighbours, in addition to showing that group central galaxies 
point to the nearest group, and that both have preferentially parallel major axes. These effects 
are strongest for early-type centrals in more massive groups and decline slowly with distance 
between the groups. Moreover, \cite{PSM+11} showed that this alignment extends to the surrounding 
large-scale structure, in the sense that groups with masses $M\gtrsim6\times10^{13}M_\odot$ point 
towards galaxy overdensities in general, and that more massive groups do so more strongly.

\begin{figure}[ht!]
\centering
  \includegraphics[width=13cm]{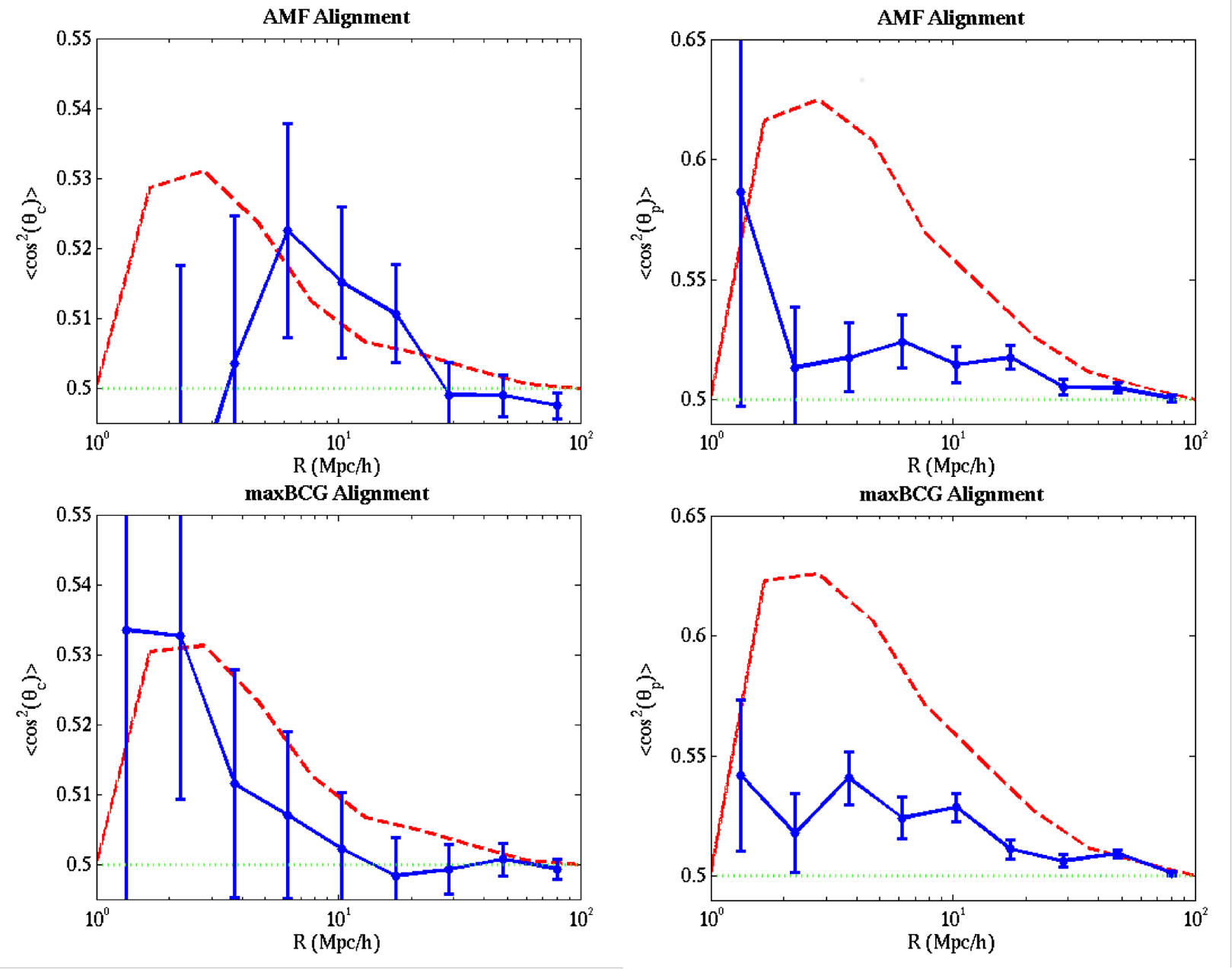}
\caption{The cluster correlation angle alignment $\langle \cos^{2} \theta_{C}\rangle$ (\textbf{left panels}) and 
the cluster pointing angle alignment $\langle \cos^{2} \theta_{p}\rangle$ (\textbf{right panels}), as a 
function of comoving pair projected separation $\mathrm{R}$ in 2D. Note $\mathrm{R}$ in this plot is therefore equivalent to the 2D comoving separation that we elsewhere denote as $r_{\rm p}$. These angles are described in the text. The blue 
points with error bars are the observational results; the red dashed lines are theoretical 
predictions from N-body simulations assuming a $\Lambda$CDM cosmology \citep{Hopkins05}, corrected for photometric redshift uncertainties; and the green dotted 
horizontal lines indicate purely random cluster orientations. The errors between the different bins 
in $R$ are independent. \textbf{Top panels:} Uses the Adaptive Matched Filament (AMF) catalogue \citep{DPG+08}. \textbf{Bottom panels:} The maxBCG  catalogue \citep{KMA+07}. \permmn{SMB+12} 
}
\label{fig:SMB12}
\end{figure}

In the most recent attempt, \cite{SMB+12} measured alignments of galaxy cluster pairs out to 
$100$ Mpc/$h$ and $z<0.44$ using two galaxy cluster catalogues extracted from SDSS data. The right-hand panels 
of \Cref{fig:SMB12} show their measurements of radial alignment between cluster pairs. They compute these using the ``pointing angle'', $\theta_{\rm p}$, which is the angle
on the sky between the projected cluster major axis and the line connecting one cluster to the other cluster in the pair. A positive correlation is 
detected out to $\sim 100$ Mpc/$h$. The left-hand panels show a measurement of their 
common orientations, using the ``correlation
angle'', $\theta_{\rm c}$, that is the angle between the projected major axes
of the two clusters. This alignment is marginally detected for pairs 
separated by less than $20$ Mpc/$h$. The top/bottom panels use different galaxy cluster catalogues, the Adaptive Matched Filament (AMF) \citep{DPG+08} and maxBCG \citep{KMA+07} catalogues respectively.

The observed signals, while significant, are weaker than the predictions from simulations. This discrepancy may be due to idealised assumptions that go into the simulations. Other possibile sources of difference are systematic effects that weaken the observed signal, particularly errors in line-of-sight redshift selection due to the fact that photometric redshifts were used. The photometric redshift errors correspond to typical separations of $\sim 50 \mathrm{Mpc}/h$ along the line-of-sight, leading to contamination from clusters which are really unassociated, thus diluting the signal. \cite{SMB+12} note that this effect alone cannot explain the full discrepancy; they discuss additional sources of error including centroiding and noise from both the small number of cluster members and clusters that are nearly round.

\section{Impact on cosmology \& Mitigation}
\label{sec:impact}

Intrinsic alignments lead to additional contributions to the observed ellipticity correlation 
functions, thus compromising
a simple cosmological interpretation of the results. The same is true for instrumental effects, such as contamination of the galaxy shapes by residual (uncorrected) PSF anisotropy.
The impact of any systematic effect on a measurement of parameters of interest is to change the 
likelihood distribution  for those parameters away from that which would have been observed if no 
systematic were 
present. If the systematic effect is sufficiently large, this can lead to parameter 
inferences that differ significantly, in a statistical sense, from the ``true'' values of those parameters (i.e. those that would 
have been found in a 
perfect experiment). Furthermore, the shape of the likelihood may change completely, for example 
from a surface with no curvature to  something with significant curvature or even multi-modal 
features. This scenario may possibly occur in the case of  
intrinsic alignments, where different galaxy populations may have different intrinsic alignment 
signal. A joint analysis of all 
galaxy types could result in a multi-peaked likelihood surface in the direction of the amplitude of 
the intrinsic alignment effect.

Systematics can both change the measured confidence levels for a particular parameter constraint (either increasing or decreasing them) and ``bias'' the measurement of a parameter, that is shift the maximum likelihood away from where it would be found in the absence of the systematic. The shift in the maximum likelihood, the biasing, is a general feature of any 
likelihood analysis in which the incorrect model is used -- in the case of intrinsic 
alignments either because the effect was not included or mitigated at all, or because the assumed 
model is not correct. 
The change in confidence levels, or the errors on the parameters, is more complicated and can lead 
to an increased sensitivity (smaller 
error bars), or decreased sensitivity (larger error bars) on parameters depending on the nature of 
the assumptions made. For 
example, including a model for a systematic that depends also on the parameters of interest may 
increase sensitivity compared 
to a model that does not. 

The observational evidence presented in \Cref{sec:observations_largescale,sec:observations_smallscale} 
 suggests that the amplitude of the intrinsic alignment signal 
is such that it will lead to significant biases in ongoing and future cosmic shear surveys. In \Cref{sec:quantifying_impact} we quantify this impact for a representative cosmic shear survey. The simplest way to deal with intrinsic alignment contamination would be to measure the cosmic shear signal and then subtract the 
intrinsic alignment
contribution (both II and GI), leaving a ``pure'' signal. This would require perfect knowledge of 
the true intrinsic alignment signal as well as total confidence in the classification of galaxies and
measurement of 
redshifts. It is therefore not considered feasible now or in the foreseeable future. This means that it is 
necessary to mitigate bias from intrinsic alignments in more imaginative ways. Most of these utilise the different redshift dependences of the GG, II and GI signals, as discussed in \Cref{sec:using_z_dependence}. The use of ``nuisance parameters'' to absorb the intrinsic alignment signal is discussed in \Cref{sec:nuisance_parameters} and the use of auxiliary data for ``self-calibration'' is summarised in \Cref{sec:self_calib}. We discuss cosmic shear three-point statistics in \Cref{sec:three_point} and novel probes of the unlensed galaxy shape in \Cref{sec:probes_of_unlensed_shape}.

\subsection{Quantifying Impact}
\label{sec:quantifying_impact}

The importance of intrinsic alignments for weak lensing studies was recognised early on and various 
studies have examined
the expected impact. As early as \citet{KAM97}, novel mitigation schemes were being proposed at the same time as measurement of intrinsic 
alignments was being discussed. 

Especially after the first detection of the cosmic shear signal \citep{BRE00,KWL00,WME00,WTK00}, much effort was spent on 
quantifying the impact of intrinsic alignments.
\citet{HRH2000} used $N$-body simulations to show that the impact of intrinsic alignments on cosmic  
shear correlation functions, 
as measured in their simulations, could be mitigated. They suggested that deep weak lensing surveys 
could be 
used to calibrate the level of intrinsic alignments because the broader source redshift 
distribution of sources in deeper surveys reduced the relative importance of intrinsic alignments. However, this suggestion neglected the importance of GI correlations and thus cannot be relied on.
\citet{CM00} also studied intrinsic alignments in $N$-body simulations, and whilst coming to similar 
conclusions as \citet{HRH2000}, the magnitude of the effect appeared to be more problematic. They 
suggested that the signal could be measured (calibrated) using relatively small surveys of only a 
few thousand galaxies at low redshift, where intrinsic alignments dominate. This signal could then be applied to wider, deeper surveys, where the shear signal could be 
measured. The type of observations discussed in \Cref{sec:observations_largescale} might be suitable for such an approach but they should not be regarded as a sufficient substitute for measurements of intrinsic alignments over the redshift range representative of the full cosmic shear survey of interest (proper coverage of galaxy types is also very important). These, and other relevant results, are discussed in more detail in 
\citet{Paper2}. 

\cite{CKB01} considered several ways to discriminate between weak gravitational lensing and the intrinsic alignment signal. The first method 
that they proposed was simple, based on their particular model for intrinsic alignments, which had a strong 
ellipticity dependence: the impact of  intrinsic alignments could be reduced by using sources with 
smaller intrinsic 
ellipticities, but this would pose serious problems for ellipticity measurement. 
More practically, they also
suggested the use of density correlations, such as the $w_{\rm g+}$ measurements 
presented in this review, around galaxy clusters and the use of morphological information to remove intrinsically 
aligned galaxies.

An attempt at total removal of intrinsic alignments is complicated, however, because, as we have already discussed, the observed ellipticity correlation is the sum of a gravitational lensing term, GG, an intrinsic alignment term, II, and a cross-term, GI,
\begin{equation}
C^{(ij)}_{\rm \epsilon\epsilon}(\ell) = C^{(ij)}_{\rm GG}(\ell) + C^{(ij)}_{\rm II}(\ell) + C^{(ij)}_{\rm GI}(\ell),
\label{eqn:cls_GG_II_GI}
\end{equation}
where we have expressed each as a projected angular power spectrum $C^{(ij)}(\ell)$ in Fourier space for a pair of tomographic bins in spherical harmonic 
space and $\ell$ denotes the angular frequency, the Fourier variable on the sky. 
The superscripts $i,j$ denote a redshift bin pair for the tomographic analysis. Note that, if galaxies are well separated in redshift, any IG term is expected to be zero. The importance of the GI correlation was not fully appreciated in the earliest literature discussing mitigation \citep{CM00,HRH2000,CKB01}. This is a serious drawback as the GI term is not only more difficult to remove, it can also dominate over the II contribution for many tomographic bin pairs in realistic cosmic shear surveys. The details of the modelling of 
intrinsic alignments are given in our companion paper \citep{Paper2} but, for a linear 
alignment model, each of the 2D projected angular power spectra in \Cref{eqn:cls_GG_II_GI} can be constructed from the 
integration of the 3D power spectrum multiplied by the appropriate redshift distribution or lensing 
weight functions,
\begin{align}
C^{(ij)}_{\rm GG}(\ell) &= \int^{\chi_{\textrm{H}}}_{0} d\chi \frac{q^{(i)}(\chi)q^{(j)}(\chi)}{f_{\rm K}^{2}(\chi)} 
P_{\delta\delta}\left(\frac{\ell}{f_{\rm K}(\chi)},\chi\right), \label{eqn:GG_limber}\\
C^{(ij)}_{\rm II}(\ell) &= \int^{\chi_{\textrm{H}}}_{0} d\chi \frac{p^{(i)}(\chi)p^{(j)}(\chi)}{f_{\rm K}^{2}(\chi)} 
P_{\rm II}\left(\frac{\ell}{f_{\rm K}(\chi)},\chi\right), \label{eqn:II_limber} \\
C^{(ij)}_{\rm GI}(\ell) &= \int^{\chi_{\textrm{H}}}_{0} d\chi \frac{q^{(i)}(\chi)p^{(j)}(\chi)}{f_{\rm K}^{2}(\chi)} 
P_{\rm \delta I}\left(\frac{\ell}{f_{\rm K}(\chi)},\chi\right).
\label{eqn:GI_limber}
\end{align}
Here $f_{\rm K}(\chi)$ is the comoving angular diameter distance, given by 
\eq{
\label{eqn:fk2}
f_{\rm K}(\chi)=\left\{ \begin{array}{ll}
  1/\sqrt{K} \sin\left(\sqrt{K}\chi\right)    &K>0 ~(\mbox{open})\\
  \chi                             &K=0 ~(\mbox{flat})\\
  1/\sqrt{-K} \sinh\left(\sqrt{-K}\chi\right) &K<0 ~(\mbox{closed})\;,
\end{array} \right.
}
where $1/\sqrt{|K|}$ is interpreted as the curvature radius of the spatial part of spacetime.

$p^{(i)}(\chi) = p^{(i)}(z)\mathrm{d}z/\mathrm{d}\chi$, where $p^{(i)}(z)$ is the galaxy redshift distribution of bin $i$, $q^{(i)}(\chi)$ is the lensing 
weight function of bin $i$ \citep{JB10},
\eq{
\label{eqn:weightlensing}
q^{(i)}(\chi) = \frac{3 H_0^2\, \Omega_{\rm m}}{2\, c^2} \frac{f_{\rm K}(\chi)}{a(\chi)}
\int_{\chi}^{\chi_{\rm H}} \dd \chi'\; p^{(i)}(\chi')\; \frac{f_{\rm K}(\chi' -
  \chi)}{f_{\rm K}(\chi')},\;
}
 and $\chi_{\rm H}$ is the 
comoving distance to the horizon. Ideally the tomographic bins do not overlap, which is possible in the case of 
spectroscopic redshifts. This is, however, not feasible in the case of cosmic shear surveys, which 
rely on photometric redshifts. Due to limitations in the precision with which photometric redshifts 
can be determined, as well as catastrophic outliers due to misidentification of features in the 
spectral energy distribution, the bins partially overlap in practice.

Examples of projected angular power spectra of the different GG, II, and GI terms for a projected spherical harmonic tomographic 
analysis of a fiducial wide-field survey are shown in \Cref{fig:cornerplot:dk2012}, alongside other 
terms related to number counts (see below). The figure shows forecasts originally made in \citet{KRH+12}. The II terms are positive (by 
definition), while the GI terms are negative in amplitude to match observations. \Cref{fig:cornerplot:dk2012} illustrates 
how intrinsic alignment terms add (through the II term) and subtract (through the GI term) to the weak lensing GG power 
spectrum. Being a local effect, the II correlation is strongest for redshift bin auto-correlations, where the number of physically close 
pairs is largest. As this plot is for a photometric cosmic shear survey the redshift cross-terms do 
not have zero II contribution as there is usually some overlap in redshift between bins. In contrast 
the GI term is strongest for bins separated in redshift where the redshift distribution of the ``I'' 
bin overlaps with the lensing kernel of the ``G'' bin. In general the relevant weight functions 
overlap differently for different combinations of tomographic bins, affecting both the amplitude and 
effective scale dependence of each contribution to the measured shear or galaxy position 
correlation.

\begin{figure}[ht!]
   \centering
   \includegraphics[width=18cm]{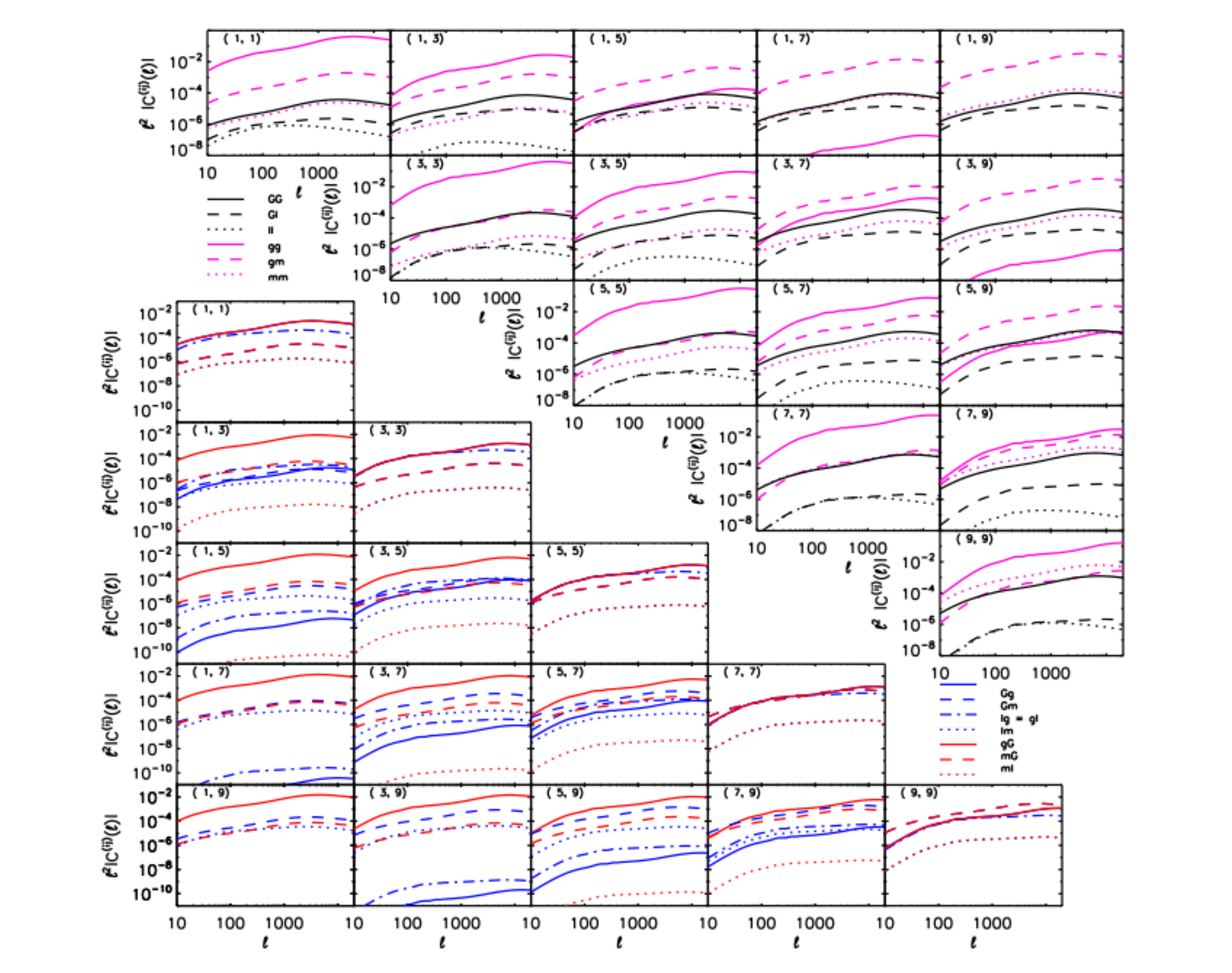} 
   \vspace{-0.5cm}
   \caption{Forecast projected angular power spectra, $C^{(ij)}(\ell)$, for a tomographic analysis of a wide-field survey based on the Euclid mission design \citep{RAK+10}. \textbf{Upper Right Panels:} Gravitational lensing and intrinsic alignment terms related to the observed ellipticity 
auto-correlation (GG, GI, II) and galaxy clustering and cosmic magnification terms related to 
the number count auto-correlation (gg, gm, mm).  \textbf{Lower Left:} Terms related 
to the cross-correlation of ellipticity and number counts, including contributions from intrinsic alignment and magnification (gG, mG, gI, mI). The absolute value of these power spectra are shown but it should be remembered that the GI, gI and mI contributions are negative in amplitude. See \Cref{sec:self_calib} for more details on these power spectra. The numbers in the top right corner of each panel denote the tomographic bin pair being considered. There are 10 bins in total, split so each has roughly the same number density of source galaxies; bin 1 is the lowest redshift bin, while bin 10 is the highest redshift bin. See \Cref{sec:quantifying_impact,sec:self_calib} for detailed descriptions of each term. \permmn{KRH+12}
}
   \label{fig:cornerplot:dk2012}
\end{figure}

\begin{figure}[ht!]
  \begin{flushleft}
    \centering
       \includegraphics[width=6.5in,height=3.0in]{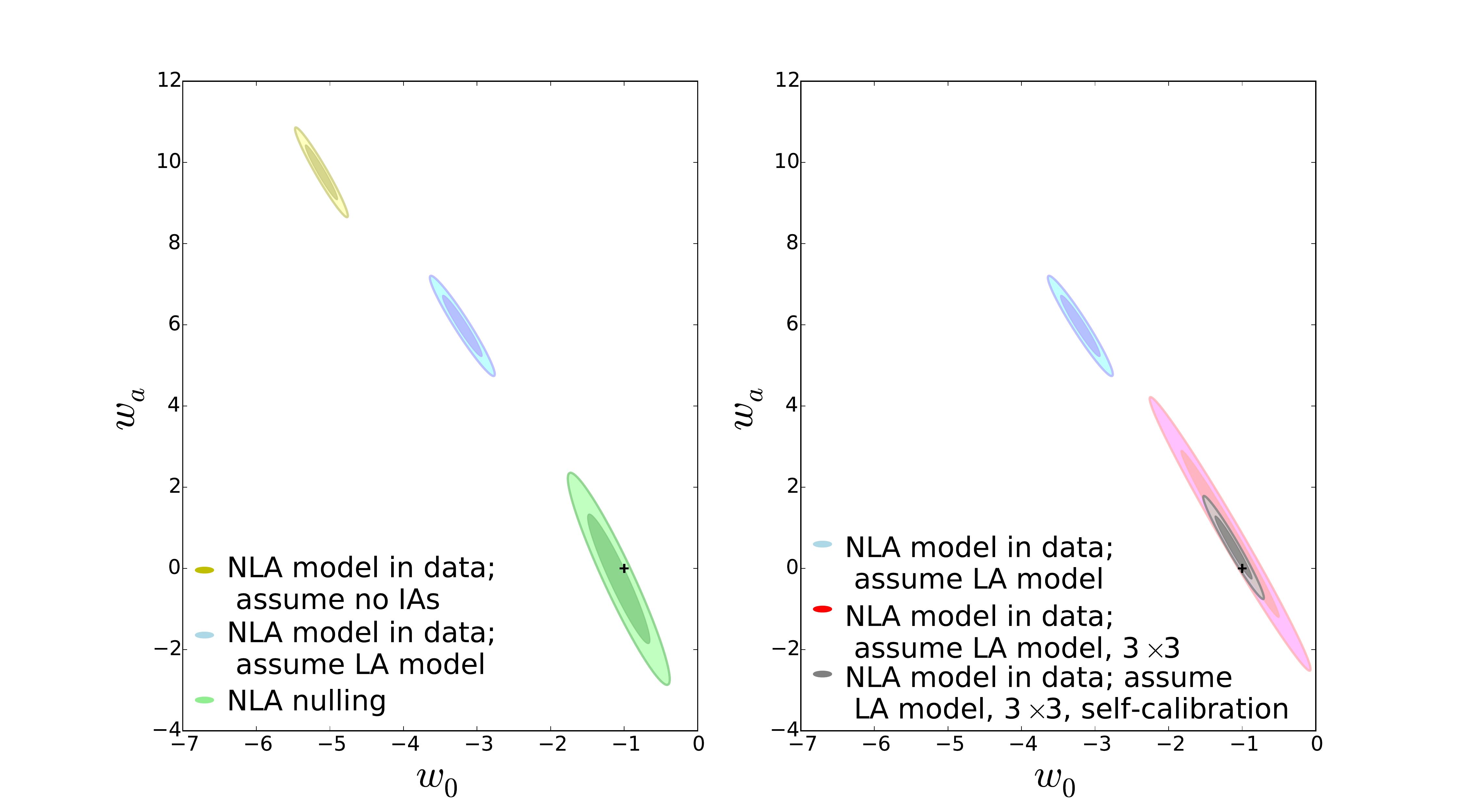}
\vspace{-3mm}
\caption{Forecast cosmological constraints for a generic Euclid-like survey, making different assumptions about intrinsic alignments. 95\% confidence ellipses are shown for 
the dark energy equation of state parameters, $w_0$ and $w_{\rm a}$. 
Constraints shown have been marginalised over $\Omega_m$, $h$, $\sigma_8$, $\Omega_{\rm b}$, $n_s$ 
and nuisance parameters where appropriate, see \Cref{sec:quantifying_impact} for more details. \textbf{Left Panel:} Impact of incorrect model choice. 
True model assumed is the non-linear alignment model \citep{HS04,BK07}. The yellow contour shows constraints and bias on 
$w_{0},w_{\rm a}$ when intrinsic alignments are ignored. The blue contour assumes the (incorrect) linear alignment model. The green contour 
shows the constraints from nulling, see \Cref{sec:nulling} for more details. \textbf{Right Panel:} 
Impact of marginalising over a robust grid of nuisance parameters in redshift and angular scale and self-calibration with galaxy 
clustering information. Each contour uses the non-linear alignment model as the ``truth''. The blue contour is the same 
as in the left-hand panel i.e. it assumes the (incorrect) linear alignment model. The red contour also assumes the linear alignment 
model, marginalised over a $3\times3$ grid of nuisance parameters in redshift and angular scale. The grey contour shows the same 
scenario (assume linear alignment, $3\times3$ nuisance grid) with the inclusion of galaxy clustering information 
i.e. self-calibration, see \Cref{sec:self_calib} for more details. The black crosses show the fiducial values of $w_{0},w_{\rm a}$.
}
\label{fig:FMs_bias}
  \end{flushleft}
\end{figure}

Some features of the impact of intrinsic alignments on two-point statistics, as well as simple 
mitigation techniques, are brought together in \Cref{fig:FMs_bias}. Here we forecast constraints on 
cosmology from a generic weak gravitational lensing survey, modelled on the European Space Agency 
Euclid mission\footnote{\texttt{http://www.euclid-ec.org/}} \citep{RAK+10,LAA+11}. This generic survey covers 15,000 deg$^2$ with a galaxy density of 30 arcmin$^{-2}$, 
split into 10 tomographic redshift bins over the range $0 < z < 2.0$. A Gaussian
total shape noise contribution of $\sigma_{\epsilon}=0.35$ is assumed. Our results are shown as 95\% 
confidence contours in the dark energy equation of state parameters $w_0$ and $w_a$. These describe the amplitude and time-evolution of the dark energy equation of state, $w_{\rm de}(z) = w_{0} + w_{a}(1-a)$. All constraints are shown marginalised 
over the cosmological parameters: $\Omega_m$, the dimensionless matter density, $\Omega_b$, the dimensionless baryon density, $\sigma_8$, the amplitude of the density perturbations, $h$, the Hubble parameter, and $n_s$, the spectral index of the density perturbations. Euclid 
is an example of a Stage-IV survey as defined by the Dark Energy Task Force \citep{detf}.

The constraints are calculated by the Fisher matrix technique \citep{F35}, assuming a Gaussian likelihood function and covariance matrix, independent of the fiducial cosmological parameter values. The Fisher matrix approach can be extended to make an estimate of the 
bias on cosmological parameters, $\Delta p_{\alpha}$, when an incorrect cosmological model is 
assumed (\citealp{HTB+06}; see also 
\citealp{amara08} and Appendix A of \citealp{JMA+11}):
\begin{equation}
\Delta p_{\alpha} = \sum_{\beta} F_{\alpha\beta}^{-1} \sum_{\ell} \sum_{i\leq j; m \leq n} \Delta C_{\epsilon\epsilon}^{(ij)} (\ell)
\left( {\rm Cov } \left[ 
C_{\epsilon\epsilon}^{(ij)}(\ell),C_{\epsilon\epsilon}^{(mn)}(\ell) \right] \right)^{-1} \frac{\partial C_{\epsilon\epsilon}^{(mn)}(\ell)}{\partial p_{\beta}},
\label{eqn:biascl}\end{equation} 
where $ \Delta C_{\epsilon\epsilon}^{(ij)}(\ell)$ is the difference between the power spectra for the true and assumed models; 
$F$ is the Fisher matrix,
\begin{equation}
F_{\alpha\beta} = \sum_{\ell} \sum_{i\leq j; m \leq n} \frac{\partial C_{\epsilon\epsilon}^{(ij)}(\ell)}{\partial p_{\alpha}} \left( {\rm Cov } \left[ 
C_{\epsilon\epsilon}^{(ij)}(\ell),C_{\epsilon\epsilon}^{mn}(\ell) \right] \right)^{-1} \frac{\partial C_{\epsilon\epsilon}^{mn}(\ell)}{\partial p_{\beta}},
\label{eqn:deltacl}\end{equation} 
Cov is the covariance matrix \citep{TJ04}; $i,j,m,n$ count over tomographic redshift 
bins and $\alpha,\beta$ count over some set of cosmological (and nuisance) parameters. $p_{\alpha}$ 
refers to a particular cosmological or nuisance parameter, hence $\Delta p_{\alpha}$ is the 
resulting bias on that parameter. The equations refer to the ellipticity-ellipticity auto-correlation, $C_{\epsilon\epsilon}^{(ij)} (\ell)$, and the bin pairs are restricted to $i\leq j$ because the symmetry of the observable means that these pairs exhaust the available information. The formalism is easily extendable to the galaxy position-position correlation, $C_{\rm nn}^{(ij)} (\ell)$, and the position-ellipticity cross-term, $C_{\rm n\epsilon}^{(ij)} (\ell)$, see \Cref{sec:self_calib} below for more details.

We can use this to show the importance of a well-modelled intrinsic alignment contribution to the measured cosmic 
shear signal. We do not know the true intrinsic alignment model but the left-hand panel of \Cref{fig:FMs_bias} 
shows the bias on cosmological parameters when we take the non-linear alignment model (with an amplitude of $C_{1}=5 \times 10^{-14}(h^{2}\mathrm{M}_{\odot} \mathrm{Mpc}^{-3})^{-1}$ \citep{BK07} and, for simplicity, no dependence on galaxy type or luminosity) as the (true) observed signal but use 
either no intrinsic alignments (yellow contour) or the linear alignment model (blue contour) in our analysis. The assumed intrinsic alignment amplitude is based on the SuperCOSMOS normalisation \citep{BTH+02,BK07} and is consistent with the lower end of current observational constraints for early-type galaxies \citep{JMA+11,HGH+13}, making the bias predictions realistic for current and future cosmic shear surveys. It is clear that the 
results are catastrophically biased. The true fiducial cosmology is indicated by a black cross and 
the forecast contours are off by several standard deviations. The contour that assumes the linear alignment model (blue)
is less biased than that which ignores intrinsic alignments completely because the linear alignment model replicates the non-linear alignment phenomenology at linear scales.

Cosmic shear in tomographic redshift slices is an approximation of a more general formalism called 3D cosmic shear, the most notable being the Limber approximation, a binning in redshift. 3D cosmic shear uses the one-point shear transform coefficients that are calculated using a spherical-Bessel transform of the data \citep{H03}. 
The impact and mitigation of intrinsic alignments in 3D cosmic shear analysis has been studied in \citet{KTH08,MS13,KAH+14,KHD14}. Results and strategies presented below are framed for tomographic analyses but can in principle also be applied to the 3D cosmic shear methodology.

\subsection{Exploiting redshift dependence}
\label{sec:using_z_dependence}

Taking a conservative approach, one can assume a complete lack of knowledge about the physics  
underlying intrinsic alignments and thus the form of the II and GI spectra. In that case the only 
reliable information left to separate the weak lensing signal from intrinsic alignments is the 
redshift dependence of the signals, which is governed by the redshift distribution of the galaxy 
samples and their lensing weight functions (see \Cref{eqn:GG_limber,eqn:II_limber,eqn:II_limber}). Here we describe some appraoches which exploit this information:

\begin{itemize}
\item{\textbf{Downweighting:}} \citet{KS02} proposed an algorithm to suppress the II term in 
non-tomographic weak lensing surveys 
with photometric redshift information. They demonstrated that incorporating a Gaussian 
kernel that downweights galaxy pairs close in redshift is effective at reducing the intrinsic 
alignment contamination while moderately reducing the effective number of galaxies in the analysis. 
In a similar analysis \citet{HH03} derived statistically optimal weights for suppressing the II 
term 
if either photometric or spectroscopic redshift information is available. They concluded that 
high-quality photometric redshifts would be a necessity for the analysis of future weak lensing 
surveys.

Using redshifts to divide the galaxy sample into tomographic bins, \citet{KS03} fitted a linear 
combination of generic template functions to II and GG correlation functions. If the redshift overlap of neighbouring tomographic bins is small, it may well be sufficient to 
simply discard all redshift auto-correlations from analysis, which causes a $10\,\%$ increase in 
errors on cosmological parameters when at least five tomographic bins are used \citep{TW04}. This approach was 
extended to include the GI term by \citet{King05}, assuming independence of the II and GI terms, 
and 
demonstrated on toy models.

\item \textbf{Nulling:} \label{sec:nulling} \citet{JS08,JS09} introduced a nulling technique for the GI term. For a 
given set of tomographic 
two-point statistics in $N_z$ tomographic bins, one can construct new measures, the nulled power spectrum $\zeta$, via linear combinations of the standard statistics, e.g. in the case of projected angular power spectra,
\eq{
\zeta^{(i)}_{[q]}(\ell) = \sum_{j=i+1}^{N_z} T_{[q], j}\; C^{(ij)}(\ell)\;,
}
with weights $T_{[q],j}$, for every foreground redshift bin $i$. The weights are 
orthogonal to each other, and orthogonal to $1 - \chi(z_i)/\chi(z_j)$, which is an approximation to 
the kernel of the GI term in the limit of redshift bins with narrow redshift distributions; compare to \Cref{eqn:GI_limber}. Depending on implementation, the redshifts $z_i$ and $z_j$ can correspond to the mean or median redshifts in each tomographic bin. In this 
way one can construct $N_z-i-1$ independent statistics $\zeta^{(i)}_{[q]}$ for every foreground bin, numbered by the index $q$ in square brackets, in which the intrinsic alignment signal should be strongly suppressed.

\citet{JS09} showed that nulling, combined with a suppression of redshift auto-correlations, 
reduces 
the bias due to the combination of II and GI by at least an order of magnitude on well-constrained 
cosmological parameters. This was achieved robustly over a wide range of photometric redshift 
parameters, for random scatters up to $0.1(1+z)$ and catastrophic outlier (galaxies whose redshifts are comprehensively misidentified) fractions up to $10\,\%$. 
However, due to the similar redshift scaling of the GI and GG signals, 
the robust removal of GI contamination comes at the price of 
substantially reducing the statistical power. Marginal errors on cosmological parameters increase 
by 
typically a factor of two, and by a factor of three in case of the dark energy equation of state 
parameters $w_0$ and $w_a$, 
whose constraints rely strongly on the redshift evolution of the 
lensing signal. While an order of magnitude loss in the dark energy figure of merit is intolerable 
for cosmological surveys, nulling techniques and their kin can still serve as a robust validation 
test for intrinsic alignment mitigation strategies that rely on more assumptions about the nature of these 
signals.

We show schematically the power of nulling in the left panel of \Cref{fig:FMs_bias}. The green contour shows the 
results of a nulling analysis of the same experimental scenario as for the other contours. Nulling succesfully reduces the 
bias to within the $1\sigma$ contour but at the cost of reducing usable information, hence the 
ellipse is broader i.e. less constraining.

\item \textbf{Boosting:} As was already noted in the early works \citep[e.g.][]{KS02,KS03}, any 
procedure to suppress the intrinsic alignment contribution can be reversed to boost these signals, 
which enables their study at scales and redshifts where the lensing signal would otherwise dominate. 
\citet{JS10} devised a GI boosting technique, again via linear combinations of tomographic two-point 
statistics and showed explicitly how it links to nulling. The method can turn a GI signal that is  
$10-30\,\%$ of the GG term into being about one order of magnitude stronger than GG for good quality 
photometric redshifts and two orders of magnitude stronger for spectroscopic redshifts. Recently, 
\citet{S14} refined this concept to boost the galaxy-magnification cross-correlation over the galaxy 
clustering signal, which de-biased and tightened cosmological constraints.

\end{itemize}

\subsection{Parameterisation and marginalisation} 
\label{sec:nuisance_parameters}

The most common approach to dealing with intrinsic alignments in the 
literature is to introduce a number of free parameters that describe the amplitude, 
redshift/scale/colour dependence etc. of intrinsic alignments and allow these parameters to vary 
within some prior 
range as part of a cosmological likelihood analysis. 
These intrinsic alignment ``nuisance'' parameters can be marginalised over when quoting constraints on cosmological parameters. 
This marginalisation will make cosmological parameter constraints weaker but, if the nuisance 
parameterisation is sufficiently flexible that it captures the full range of the intrinsic alignment signal, 
the 
resulting constraints will be unbiased. Note that the model being parameterised can be based on an assumed physical mechanism for intrinsic alignments and is therefore not distinct from the linear alignment or non-linear alignment models discussed above. For example \citet{HGH+13} uses the non-linear alignment model with a single free amplitude parameter, which is marginalised over whenever constraints on cosmological parameters are discussed. See \Cref{sec:observations_largescale_indirect} for a more detailed discussion.

The benefit of the parameterisation/marginalisation approach is that it can be implemented 
simultaneously with the cosmological parameter estimation. This means that the same procedure produces constraints on cosmological parameters and the parameters of the intrinsic alignment model.   
The downside is that a higher dimensional parameter space must be explored, sometimes significantly higher, which is computationally expensive. 
There is also some ambiguity in the statement that cosmological constraints are unbiased 
``provided the parameterisation is sufficiently flexible.'' In the absence of very strong 
constraints on intrinsic alignments there is no definitive statement about either how many nuisance 
parameters are 
required or what their prior ranges should be. 
With this approach, it is easy to update the analysis as more precise 
measurements of intrinsic alignments become available, either through a joint likelihood analysis or 
by altering the 
priors of the initial analysis.  

In the right-hand panel of \Cref{fig:FMs_bias} we show an attempt to reduce bias through 
nuisance parameters and marginalisation for our toy survey. We also show the use of self-calibration 
to recover information through full exploitation of joint gravitational shear and galaxy position 
information, described in \Cref{sec:self_calib} below. Each contour takes the non-linear alignment model to be the 
true description of intrinsic alignments but models them (incorrectly) with the linear alignment model. The blue 
contour is the same as in the left hand panel, showing the bias this produces in the simple case.

The red contour is the result of the same analysis with additional nuisance parameters included. A 
grid of nuisance parameters which can vary in scale and redshift is employed: $3\times3$ parameters 
in $z \times k$, where $k$ is the Fourier space wavenumber, 
are used for both the amplitude of the II and GI terms, as well as free amplitudes for each. This 
leads to a total of $2(3\times3)+2 = 20$ nuisance parameters. Marginalising over these new 
parameters reduces the precision, as shown by the increased size of the contour, but it also 
reduces the bias in the inferred cosmological parameters to within the $1\sigma$ 
area. For a more detailed example of marginalisation using this parameterisation see \citet{JB10}.

A goal of the parameterisation and marginalisation approach is to include information from intrinsic alignment measurements as physically motivated priors on the intrinsic alignment nuisance parameters. This inclusion of prior information from observations is not yet mature in the intrinsic alignment literature but, for example, \citet{SHC+15} used measurements of intrinsic alignments in clusters to inform parameters for their implementation of the halo model. They aimed to test how strong a deviation from the non-linear alignment model at small scales was allowed by observations. They found an allowed deviation significantly lower than the fiducial model assumed in \citet{SB10}.


\subsection{Self-calibration}
\label{sec:self_calib}

All cosmic shear surveys contain information beyond the correlation of galaxy ellipticities. Even a survey with photometric-quality redshifts can be used to study galaxy clustering (i.e. position-position correlations), and additional information is contained in the cross-correlation between position and ellipticity. Exploitation of this additional information can regain some of the constraining power lost when marginalising over intrinsic alignment nuisance parameters. \citet{Bernstein09} outlined the formalism to treat a range of two-point correlations available with cosmic shear survey data. We summarise relevant aspects below and refer readers to that paper for more details. 

Exploitation of this information is an example of ``self-calibration'' because the uncertainties due to intrinsic alignments are being calibrated by information contained in the cosmic shear survey itself.
The effect of magnification is important in both galaxy position-galaxy position clustering and the galaxy position-ellipticity cross-correlation, so omitting it can bias results \citep{DJH+14}. We can now write down a set of three observables, each made up of multiple contributions, analogous to \Cref{eqn:cls_GG_II_GI},
\begin{eqnarray}
C^{(ij)}_{\epsilon\epsilon}(\ell) &= C^{(ij)}_{\rm GG}(\ell) + C^{(ij)}_{\rm II}(\ell) + C^{(ij)}_{\rm GI}(\ell) + 
C^{(ij)}_{\rm IG}(\ell), \\
C^{(ij)}_{\rm n\epsilon}(\ell) &= C^{(ij)}_{\rm gG}(\ell) + C^{(ij)}_{\rm gI}(\ell) + C^{(ij)}_{\rm mG}(\ell) + 
C^{(ij)}_{\rm mI}(\ell), \\
C^{(ij)}_{\rm nn}(\ell) &= C^{(ij)}_{\rm gg}(\ell) + C^{(ij)}_{\rm gm}(\ell) + C^{(ij)}_{\rm mg}(\ell) + C^{(ij)}_{\rm mm}(\ell). 
\label{eqn:cl}
\end{eqnarray}

\noindent These are the set of power spectra shown in \Cref{fig:cornerplot:dk2012}. As before ``G'' denotes gravitational lensing and ``I'' intrinsic alignment, we use ``g'' and ``m'' to refer to galaxy clustering and the change in number density due to magnification respectively (see below for more details). $\epsilon\epsilon$ is then the observed ellipticity correlation signal from both weak lensing and intrinsic alignment contributions while nn is the observed galaxy position correlation signal including the magnification contributions. $\mathrm{n}\epsilon$ is the observed cross-correlation.

The gG term is the cross-correlation of galaxy clustering and gravitational lensing. As galaxies are biased 
tracers of the underlying matter distribution, we would expect a foreground galaxy population to be 
correlated with the lensing of background source galaxies. This is often referred to as 
galaxy-galaxy lensing, especially on scales where the lensing is dominated by the haloes of the 
foreground galaxies 
\citep[e.g.][]{MSK+06,vUHV+11,VvUH+14}. It is clear that a galaxy-intrinsic alignment term, gI, 
appears in the galaxy-shear cross-correlation in an analogous way to the gravitational lensing-intrinsic 
alignment GI term. Consequently, intrinsic alignments are an important contamination in 
galaxy-galaxy lensing if the source and lens populations cannot be perfectly separated \citep{BMS+12,CMS+14}. We can write the projected angular power spectra of these terms as integrals over the three dimensional matter power spectrum and the appropriate window functions under the Limber approximation:
\begin{align}
C^{(ij)}_{\rm gg}(\ell) &= \int^{\chi_{\textrm{H}}}_{0} d\chi \frac{p^{(i)}(\chi)p^{(j)}(\chi)}{f_{\rm K}^{2}(\chi)} 
b_{\rm g}^{2}\left(\frac{\ell}{f_{\rm K}(\chi)},\chi\right)P_{\delta\delta}\left(\frac{\ell}{f_{\rm K}(\chi)},\chi\right), \\
C^{(ij)}_{\rm gG}(\ell) &= \int^{\chi_{\textrm{H}}}_{0} d\chi \frac{p^{(i)}(\chi)q^{(j)}(\chi)}{f_{\rm K}^{2}(\chi)} 
b_{\rm g}\left(\frac{\ell}{f_{\rm K}(\chi)},\chi\right)P_{\rm \delta G}\left(\frac{\ell}{f_{\rm K}(\chi)},\chi\right), \\
C^{(ij)}_{\rm gI}(\ell) &= \int^{\chi_{\textrm{H}}}_{0} d\chi \frac{p^{(i)}(\chi)p^{(j)}(\chi)}{f_{\rm K}^{2}(\chi)} 
b_{\rm g}\left(\frac{\ell}{f_{\rm K}(\chi)},\chi\right)P_{\rm \delta I}\left(\frac{\ell}{f_{\rm K}(\chi)},\chi\right).
\end{align}

Weak gravitational lensing, as well as distorting 
the shape of galaxy images, changes their apparent sizes while the surface brightness is unchanged. 
This means that the flux of galaxies is changed and galaxy images can be either magnified or 
demagnified. This is the cosmic magnification contribution referred to by the subscript m. For a flux-limited survey this can mean galaxies are promoted or demoted across the 
detection threshold, changing the observed number density and the clustering statistics. Magnification also causes a change in effective area, which also modulates the observed number density. This is 
apparent in the mm term that contributes to the observed galaxy correlation, but there are also 
cross-correlations between magnification and galaxy counts, gm, mg, and cross-correlations between 
magnification and gravitational lensing, mG, and indeed, magnification-intrinsic alignment 
correlations, mI.

Magnification due to lensing arises due to the same gravitational potential responsible for the shear, therefore 
the magnification signal can be related to the correlation functions that involve the shear. The 
amplitude of the effect depends on $\alpha^{(i)}$, the power-law slope
of the observed number counts for the galaxies in the $i$th tomographic bin. The form of the 
magnification terms is
\begin{align}
C^{(ij)}_{\rm mm}(\ell) & = 4(\alpha^{(i)} - 1)(\alpha^{(j)} - 1) C^{(ij)}_{\rm GG}(\ell), \\
C^{(ij)}_{\rm gm}(\ell) & = 2(\alpha^{(i)} - 1) C^{(ij)}_{\rm gG}(\ell), \\
C^{(ij)}_{\rm mG}(\ell) & = 2(\alpha^{(i)} - 1) C^{(ij)}_{\rm GG}(\ell), \\
C^{(ij)}_{\rm mI}(\ell) & = 2(\alpha^{(i)} - 1) C^{(ij)}_{\rm GI}(\ell). 
\end{align}
These magnification terms are important because results from an analysis that ignores them can be significantly biased \citep{DJH+14}.

Exploitation of some of this set of observables has been proposed in different ways \citep{Zhang2010,TI12a,TI12b} and has been shown to be an effective self-calibration strategy in forecasts \citep{Bernstein09,JB10,LBK+12,KLB+13}. Even if  intrinsic alignment model ignorance is aggressively marginalised over, self-calibration can recover the bulk of the 
information present in a naive cosmic shear analysis that ignored intrinsic alignments, but without 
the bias on cosmology. This approach has yet to be attempted as part of a real data analysis because the statistical power of existing datasets does not warrant such a thorough treatment.

An example of the power of this self-calibration approach is shown in the forecasts of \Cref{fig:FMs_bias}. We have already shown how marginalisation over 
nuisance parameters can remove the cosmological bias due to intrinsic alignments, at the cost of constraining power. 
The grey contour in the right-hand panel uses the same approach but includes galaxy clustering information and galaxy + weak 
gravitational lensing cross-correlation, all from the same (photometric) survey as the cosmic shear 
signal. In the case where multiple probes are considered we replace the simple $C^{(ij)}(\ell)$ with a data vector, for example $\mathcal{D}^{(k)} = \{C_{\rm \epsilon\epsilon}^{(ij)} (\ell),C_{\rm n\epsilon}^{(ij)} (\ell),C_{\rm nn}^{(ij)} (\ell)\}$, where $k$ counts over bin pairs. Note that the inclusion of the $C_{\rm n\epsilon}^{(ij)} (\ell)$ cross-correlations breaks the bin pair symmetry and each bin pair $i,j$ must be considered separately. The same grid parameterisation is extended to galaxy bias and the galaxy + weak 
gravitational lensing cross-correlation amplitude, giving an extra 20 nuisance parameters for a 
total of 40. Even with these new nuisance parameters, the self-calibration result is much tighter 
than that of weak gravitational lensing alone while remaining unbiased at $1 \sigma$. Indeed the 
self-calibration with 40 nuisance parameters is on a par with the naive weak gravitational lensing 
analysis without nuisance parameters in terms of constraining power.

Similarly, it has been shown that information contained in changes to the size of lensed galaxy 
images due to magnification can be exploited in parallel with that from image shape distortion 
\citep{HAJ13,AKH+14}. Using this magnification information mitigates the impact of intrinsic 
alignments by $\sim 20\%$, even in a pessimistic scenario. This information can also be included as 
part of a general self-calibration scheme, along with galaxy clustering information.

Self-calibration exploits information beyond shape measurements present in any cosmic shear survey. It is of course possible to calibrate uncertainty, from intrinsic alignments or other sources, by utilising additional data from beyond the optical weak lensing survey in question. This could mean assuming priors on cosmological parameters from cosmic microwave background (CMB) experiments, the use of spectroscopic redshift surveys to calibrate photometric redshift estimates or cross-correlation with a different dataset. The potential for measuring galaxy shapes in radio surveys has been noted \citep{BBC+15}. This could be used to make a weak lensing measurement in the radio or to calibrate shape measurement systematics or the intrinsic alignment signal in an optical weak lensing survey \citep{KBB+15,KBA+15,PHM+15}. A note of caution should be drawn from \citet{PBB+10} however, who found no evidence of correlation between shapes measured in optical and radio surveys. Further work is required to determine the true relationship between galaxy shapes in optical and radio wavelengths. If they were truly uncorrelated, they could not be used for mutual calibration but the effective number density of the combined surveys would be increased by a factor of two. We discuss novel approaches to intrinsic alignment mitigation using radio surveys in \Cref{sec:probes_of_unlensed_shape} below.

Another interesting cross-correlation is between weak lensing of galaxy images and weak lensing of the CMB \citep{HLD+15,TI14,HT14}. Light from the CMB is lensed on its way to the observer, just like light from 
galaxies. In the case of the CMB there is a single source plane, the surface of last scattering, 
and the light pattern contains the imprinted signal of all the structure formation from last 
scattering until today.  
There is no equivalent to intrinsic alignments in the source of the CMB. 
However, when one 
cross-correlates the weak gravitational lensing signal from galaxies with that from the CMB, which we indicate by $\epsilon\mathrm{G^{CMB}}$, there 
will also be a cross-term from 
galaxy intrinsic alignments and the CMB lensing:
\begin{equation}
C^{(i)}_{\epsilon\mathrm{G^{CMB}}}(\ell) = C^{(i)}_{\rm G\mathrm{G^{CMB}}}(\ell) + C^{(i)}_{\rm I\mathrm{G^{CMB}}}(\ell),
\end{equation}

where $i$ labels a weak gravitational lensing tomographic bin (the CMB lensing has only one source plane). The contribution of this term to the observed cross-correlation signal has been estimated to be 
$\sim 15\%$ \citep{HT14} in a survey combination such as the ACTPol/CFHT (Atacama Cosmology 
Telescope-polarisation/Canada France Hawaii Telescope) Stripe 82 cross-correlation of 
\citet{HLD+15}. \citet{TI14} suggested a self-calibration technique for the CMB-intrinsic alignment correlation using 
a scaling relation between the intrinsic alignment information in the weak gravitational lensing and the combined 
weak gravitational lensing and CMB observables. They claim that this self-calibration allows one to reduce the impact of intrinsic alignments in the cross-correlation by a factor of 20 or more in all redshift bins.

\subsection{Higher-order cosmic shear statistics}
\label{sec:three_point}

The three-point statistics of the shear field can in principle be used to mitigate the intrinsic alignment 
systematics in 
the two-point analysis, because the dependencies of the three-point  
intrinsic alignment terms on cosmological parameters is different from those in the two-point measurement \citep{VLW+10,TI12a,TI12b}.
In the three-point case, the correlation terms that arise have the form GGG, IGG, GII and III (and 
combinatorial 
variations thereof). The origin is analogous to that of two-point intrinsic alignments; here GGG is the pure gravitational lensing three-point, IGG contains one intrinsic alignment term, GII contains two and III is the three-point intrinsic alignment auto-correlation.

The most recent observations of three-point cosmic shear were presented in \cite{fu14}. They inferred constraints on three-point intrinsic alignments by including it in their model and found a slight improvement in cosmological parameter measurements when intrinsic alignments were 
included. The 
amplitude of the three-point intrinsic alignment signal was tested using simulations in 
\cite{semboloni11}, who 
found that third-order weak lensing statistics are typically more strongly contaminated by intrinsic alignments than 
second-order shear measurements, which leads to the possibility 
of using three-point 
statistics to measure the intrinsic alignment amplitude and constrain intrinsic alignment models. This knowledge of intrinsic alignments could then be used to improve the accuracy of two-point cosmic shear 
measurements. \cite{SJS10} applied the nulling method to three-point statistics and showed that 
a factor of ten suppression can be 
achieved in the GGI/GGG ratio.  
\cite{Valageas2014} found that source-lens clustering can affect both two- and three-point statistics, and 
that 
the intrinsic alignment bias is typically about 10\% of the signal for both two-point and 
three-point 
statistics. 

Higher than second-order statistics are essential to studies of non-Gaussianity in the weak lensing signal. One way to access the full information encoded in weak lensing is to produce reconstructed maps of convergence or mass \citep{KS93} from cosmic shear surveys. These maps can then be analysed statistically, for example the counting of peaks in the convergence distribution is a source of cosmological information \citep[e.g.][]{BT03,vWBE+13}. \citet{PLS12} showed that these peak counts capture rich non-Gaussian information beyond the skewness and kurtosis of the distribution, while \citet{DH10} showed significant statistical improvement when peak counts were combined with two-point statistics. Alternatively, topological features can be exploited by decomposing the map into Minkowksi functionals \citep{KLW+12,PHH+13,SY14}. The impact of intrinsic alignments on mass reconstruction, peak counts and Minkowski functionals is a subject of current study. \citet{STH09} included intrinsic alignments in their mass reconstruction estimator, based on 3D cosmic shear, while a new peak counts model has been proposed by \citet{LK15} which can be adapted to include intrinsic alignments.

\subsection{Probes of the unlensed galaxy shape}
\label{sec:probes_of_unlensed_shape}

One potentially very useful avenue for the control of intrinsic alignments is the use of observables 
that can access
information about the intrinsic (or ``unlensed'') galaxy shapes and/or
orientations. Here we discuss two possible observables of this type: the polarised
emission from background galaxies observed in the radio (\Cref{sec:radio}) and estimates of the 
rotation
velocity axis of disc galaxies (\Cref{sec:rotational_velocity}).

\subsubsection{Radio polarisation as a tracer of intrinsic orientation}
\label{sec:radio}

Several authors have demonstrated from a theoretical perspective that, 
for observations using radio telescopy, the local 
plane of
polarisation is not altered by gravitational light deflection
effects~\citep{KDB+91, DS92, Faraoni93,SH99}. This property was
first exploited by \cite{KDB+91}, who developed and applied a
technique for measuring gravitational lensing using polarisation
observations of the resolved radio jets of distant quasars.

\cite{AS99} demonstrated how the integrated polarisation
emission from ``normal'' star-forming galaxies could in principle be
included in weak lensing estimators of the parameters describing intervening
lenses. The fundamental assumption underlying the technique is that
the orientation of the integrated polarised emission from a background
galaxy is a noisy tracer of the intrinsic
structural position angle of the galaxy. \cite{KDB+91} and \cite{AS99} focused on the
reduction in uncertainty on estimates of the lensing shear signal that
is afforded by including the polarisation information. 

It was further shown in \cite{BB11b} that the use of polarisation 
may potentially be a powerful tool to help separate 
intrinsic alignments from weak lensing distortions of galaxy shapes. The idea is that optical surveys provide a measure of the intrinsic ellipticity plus the shear from weak lensing, while the radio polarisation provides information about the unlensed galaxy orientation. In combination, 
these estimators effectively provide a measure of the difference
between the intrinsic and observed orientations of individual
galaxies. They are by construction insensitive to contamination by
intrinsic alignments in the limit of a perfect relationship between
the orientation of the polarised emission and the structural position
angle. 
The weak lensing analysis could be conducted using radio images alone \citep{BBC+15} but this type of measurement has been limited to date by the low source counts available \citep{CRH04}.

\begin{figure*}[ht!]
  \begin{flushleft}
    \centering
       \includegraphics[width=4.5in,height=3in]{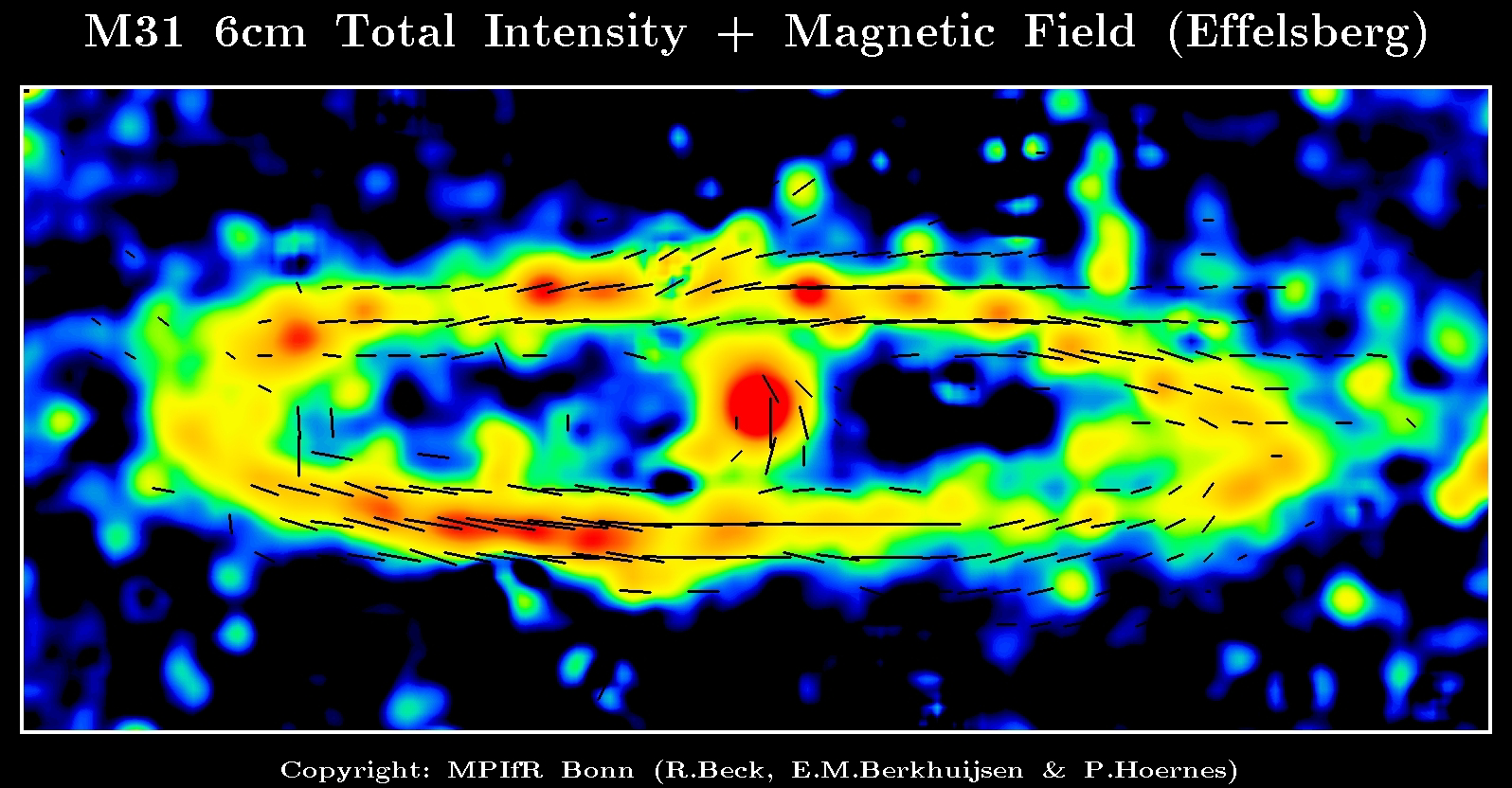}
\vspace{-1mm}
\caption{Total emission (colour scale) and polarised emission (black vectors) from M31 at $\lambda=$6.2 cm, 
smoothed to an angular resolution of 5 arcmin. Contour
levels are 5, 10, 15, 20, 30, 40 and 60 mJy/beam area. 
\permaa{BBH03} 
}
\label{fig:bb11b_fig1}
  \end{flushleft}
\end{figure*}

\Cref{fig:bb11b_fig1} shows an apparent alignment between 
polarisation pseudo-vectors and total intensity from radio wavelengths \citep{BBH03} for the galaxy M31.  \citet{SKB+09} further suggested alignment between the polarisation orientation and the emission at optical wavelengths. If this relationship was found to generally hold, one could 
potentially use the radio polarisation orientation as a proxy for the
optical intrinsic position angle in order to mitigate intrinsic
alignments in overlapping optical and radio surveys. However in order
to fully exploit this potential, in addition to understanding the
relationship between polarisation and structural position angles, one
would also require a much deeper understanding of the correlation
between the shapes of galaxies as measured in optical data and their
shapes as measured from radio observations. At the current time, the
situation with respect to this issue is not clear and more work is required in advance of results from the SKA, Euclid and LSST \citep{PBB+10,BBC+15,PHM+15}. One benefit of shape measurements from both radio and optical surveys is that shape measurement systematics are expected to be very different for each, so cross-correlation of measured shapes should reduce the impact of these measurement systematics \citep{BBA+15}. See \citet{WBB15} for a recent overview of the potential uses of radio observations for intrinsic aligment purposes.
  
\subsubsection{Rotational velocities as a tracer of intrinsic orientation}
\label{sec:rotational_velocity}
\begin{figure}[ht!]
\vspace{-10mm}
\centering
\includegraphics[width=13cm, clip]{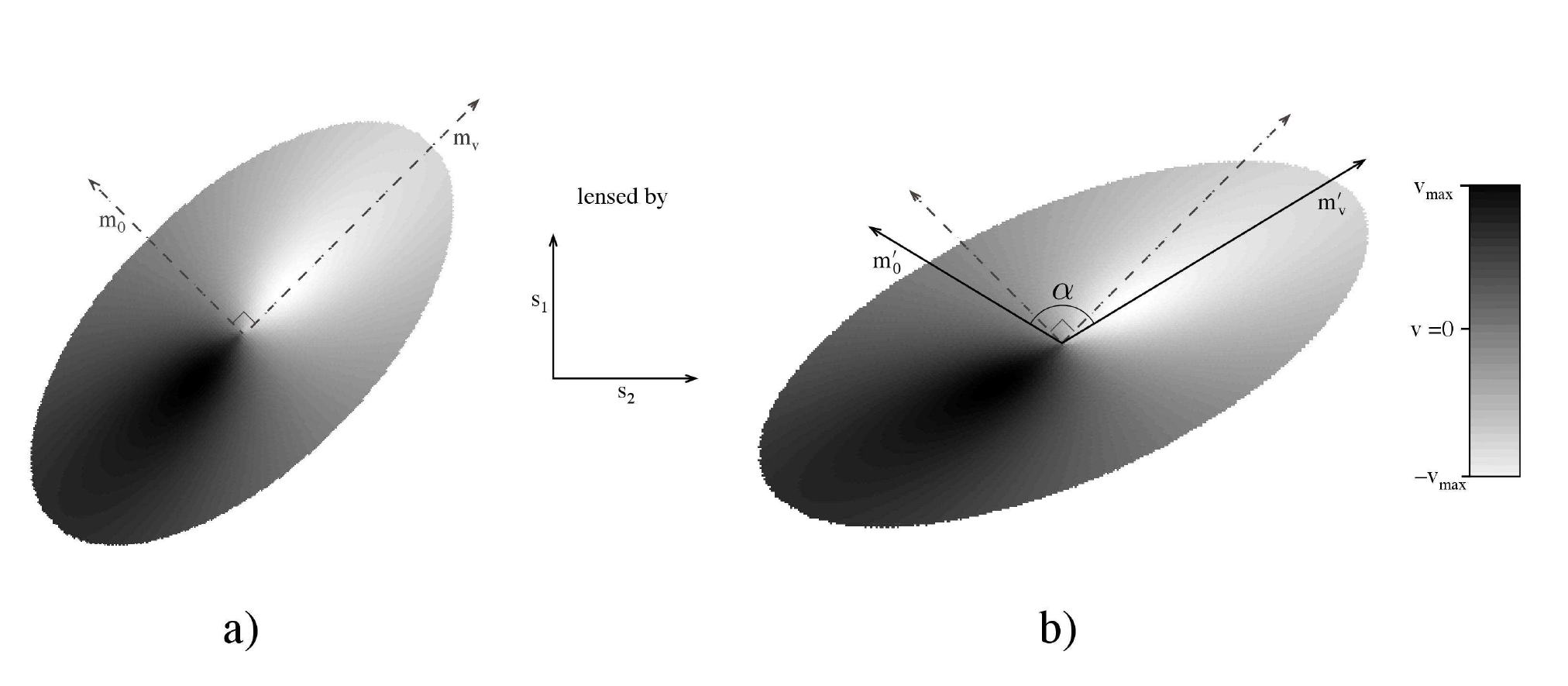}
\caption{An illustration of the rotational velocity lensing
  technique. The grey scale indicates the observed radial velocity. A
  model galaxy in the absence of lensing is shown in the left-hand
  panel, where the zero velocity axis ($m_0$) and the axis defined by
  the maximum radial velocity ($m_v$) are perpendicular. The
  application of a shear (here applied along the $s_1$ and $s_2$ axes) breaks this perpendicularity. The observed angle $\alpha$
directly measures the shear component. \permapj{Morales06}}
\label{fig:Morales06_fig1}
\end{figure}

A second novel approach is to
use rotational velocity measurements to provide information about the
intrinsic shapes of galaxies. The idea, first suggested by
\cite{Blain02} and \cite{Morales06}, is to measure the axis of
rotation of a disc galaxy and to compare this to the orientation of
the major axis of the galaxy disc image. In the absence of lensing,
these two orientations should be perpendicular, so measuring the
departure from perpendicularity directly estimates the shear field at
the galaxy's position. The basic technique is illustrated in
\Cref{fig:Morales06_fig1}. 

We note that this technique can in principle be applied in both the
radio and optical bands, making use of spectral line observations in
the former and spectroscopic observations in the latter. It is worth
pointing out that many future radio surveys plan to conduct 
HI line observations alongside continuum-mode observations
(which can be done at no extra cost in terms of telescope time). 

The rotation velocity technique shares many of the characteristics of
the polarisation approach described above -- in the limit of perfectly
well-behaved, infinitely thin disc galaxies, it is also free of shape noise and it can
also be used to remove the contaminating effect of intrinsic galaxy
alignments. In practice, the degree to which the rotational velocity
technique improves on standard methods will be dependent on
observational parameters. 
First, one would need to account for the fact that
the HI line emission of galaxies is much fainter than the broad-band
continuum emission and so the number of galaxies will be reduced. 
Secondly, for a population of real disc galaxies, there will again be
some scatter in the relationship between the rotation axis and the
major axis of the galaxy disc. 
Recently \cite{HKE+13}
proposed to extend this technique using the Tully-Fisher relation to
calibrate the rotational velocity shear measurements and thus reduce
the residual shape noise even further. This approach would require overlapping photometric and spectroscopic survey data.

\section{Summary \& Outlook}
\label{sec:summary}

Historically, studies of galaxy shape alignments have focused on the role of the environment on the formation and evolution of galaxies. However, the diversity of results in the literature also highlight the difficulty of measuring the observational signatures. Closer examination shows that the findings are expected to depend on the methods used to quantify galaxy shapes. The importance of intrinsic alignments for the interpretation of weak lensing measurements has renewed interest in this field of research, with larger data sets expected to become available soon.

Shape estimation itself continues to be an area of active development because improvements on current shear measurement pipelines \citep[e.g.][]{MRA+14} are necessary to cope with the statistical accuracy of next-generation surveys. The determination of the 
intrinsic alignment signal depends critically on the adopted algorithm for shape measurement. Consequently, results from early measurements of galaxy shape alignment cannot be readily applied to  current studies. On the other hand, it is also not clear that the algorithms used in lensing studies are optimal for the study of environment-dependent galaxy shape alignments. Regardless, a careful accounting for a range of observational biases is essential, and progress in either field will benefit from advances in the other.  The last decade has also seen the development of a sophisticated set of statistics and estimators that allow for the measurement of ellipticity-density correlation on large scales and the development of a theory which matches the data at linear scales. Results have so far largely been limited to two-point statistics, but with the advent of larger data sets, which can probe smaller physical scales, we expect that the study of higher-order statistics will gain further interest.

The existence of intrinsic galaxy alignments is now well established, to a large extent thanks to modern wide area surveys, such as  
the SDSS, where we need to distinguish between galaxy type and the physical scales involved. For early-type galaxies significant
detections have been reported: they show a clear intrinsic alignment signal, in good agreement with the linear alignment model at scales $>10$ Mpc/$h$. At intermediate scales the non-linear alignment model, which uses the non-linear power spectrum to boost the amplitude of the predicted signal, provides a good fit to data. We stress, however, that there is currently no theoretical justification for this phenomenology and more work on these quasi-linear scales is required. At  smaller scales, $<2-6$ Mpc/$h$, these models fail to explain the strength of the observed correlation and attempts to model local alignments through, for example, the halo model are required. The halo model tends to match the observations well, though the number of free parameters provides a good deal of freedom in the fit. There is no significant evidence 
for redshift-dependence of the signal beyond that which is already included in the linear alignment model, but there exists strong evidence for a dependence of intrinsic alignment on luminosity, with the brightest galaxies exhibiting stronger alignments. 

In contrast, for late-type galaxies the situation is less clear. Observational constraints are more limited because of the limited spectroscopic coverages for large samples, and no statistically significant evidence for shape alignment has been detected from such surveys. This null signal is consistent with the quadratic alignment model at linear scales, though a higher number density in observations would be desirable to reduce the statistical errors of current measurements. Some papers that studied the alignment of disc galaxy spin vectors, as an alternative to direct measurement of the ellipticity correlation of spirals, did see evidence, albeit at low significance, of correlation at small scales $\lesssim 
1$ Mpc/$h$. 

Hence, the picture at large scales, where correlations between large, diverse galaxy samples of galaxies are considered, is starting to become clear as powerful datasets become available. The results at scales where the morphology of the local large-scale structure becomes important are more ambiguous. This is not surprising as the intricacies of astrophysics, galaxy formation physics and galaxy evolution history are complicated and far from perfectly understood. 
The existing literature tends to classify the alignment of galaxy shapes on small scales by reference to the local morphology of the large-scale structure: galaxies located in voids, sheets, filaments or knots (groups and clusters) are expected to exhibit different shape alignment properties. In addition, the influence of galaxy type and history remains relevant at small scales.

These complications hamper a clear interpretation and comparison of the results. For instance the study of alignments for galaxies on the surfaces of voids has produced measurements that are consistent with no alignment, a significant alignment parallel to the void surface and the opposite, alignment perpendicular to the void surface. This highlights a need for more observations aimed at determining environment-dependent alignment. Even for well-defined structures such as groups and clusters of galaxies contradictory results have been reported,
while we note that all the most recent studies in galaxy clusters find no evidence for shape alignments, neither from studies using spectroscopic redshifts \citep{SHC+15} nor those using photometric  redshifts \citep{CMS+14}. In such high-density environments the presence of nearby galaxies can affect shape measurements, especially those based on the shapes of isophotes, which may have biased earlier measurements.

A main driver of current research into galaxy alignments is the promise of cosmic shear as a powerful probe of cosmology. For this applications, the intrinsic alignments compromise a straightforward interpretation of the measurements and thus represent a dominant astrophysical source of bias. If ignored, the biases in the resulting cosmological parameter estimates are much larger than the statistical uncertainties afforded by future wide-field cosmic shear surveys using photometric redshifts such as Euclid, LSST and WFIRST \citep{LAA+11,LSST09,SGB+15}. Consequently, the desire to constrain the nature of dark energy will drive much development in this field in terms of observations and the development of shape measurement and analysis pipelines. 

These data will not be optimal for the study of environment-dependent alignments because of the relatively crude redshift precision.
To clear up the uncertainties about the relation of alignment to morphology, and hence learn about galaxy formation and evolution, we need surveys with (near-)spectroscopic redshifts that have high galaxy density down to relatively faint magnitudes. Such data will allow a good determination
of the local morphology required for  unambiguous measurements of shape alignments. 
Many of the future spectroscopic surveys plan to survey the brightest, easiest target galaxies, which may not provide the type of data we need for environment-dependent intrinsic alignment studies but there is some reason to expect progress in the right direction. For instance, the Dark Energy Spectroscopic Instrument (DESI) bright survey \citep{LBB+13} or Subaru Prime Focus Spectrograph (PFS) \citep{TEC+14} may extend the sample to fainter magnitudes and higher redshifts. Wide-area observations using a large number of narrow-band filters, such as PAUCam \citep{MMC+14}, provide another avenue. These yield photometric redshifts that are much more precise than those obtained using broad-band observations, and do so for a wide range of galaxy types.

Despite the expected progress in measuring alignments, the much smaller uncertainties from future surveys with larger area will require significantly improved strategies for mitigation if we are to produce unbiased measurements of cosmology. To this end, much effort is spent on exploring general approaches that seek to exploit the different redshift dependencies of the GG, II and GI contributions. The most promising alternative involves the use of a flexible model of the intrinsic alignment contribution that includes a number of variable ``nuisance parameters'' which can be explored in tandem with the cosmological parameter space under consideration. The nuisance parameters are then marginalised over. If the model is sufficiently flexible, then the resulting cosmological constraints will be unbiased, albeit with a loss in overall constraining power. 
It has been demonstrated that, even after marginalising aggressively over uncertainties in intrinsic alignments, useful cosmological information can be gained from a photometric cosmic shear survey \citep{JB10}.
In this context, the goal of observational studies of intrinsic alignments, as they relate to cosmic shear systematics, can be thought of as applying more rigorous priors to the intrinsic alignment nuisance parameters. The large- and small-scale observations quoted in this review are an excellent start to this process, though it is worth noting that no paper seeking to marginalise over intrinsic alignments in the pursuit of cosmic shear has, thus far, explicitly employed priors derived from dedicated observations of intrinsic alignments. 

For this reason it is also worthwhile to examine the value of complementary data to reduce or calibrate the intrinsic alignment signal. An interesting application in the near term is the cross-correlation between galaxy cosmic shear surveys and weak lensing of the CMB. Further ahead, the large density of sources at radio wavelengths that will arrive with the forthcoming SKA survey offers a number of exciting possibilities for the study of intrinsic alignments. In this case information from radio polarisation or rotational velocity measurements could allow the intrinsic alignment and weak lensing information to be separated cleanly. These constitute extremely powerful datasets, complementary to those from optical weak lensing surveys, and, if the many practical difficulties of radio weak lensing can be surmounted, they will provide important advances in our understanding of intrinsic alignments across all galaxy types, particularly much improved statistical uncertainties for late-type galaxies.

In the meantime a number of important open questions remain, including: are late-type galaxies really free of intrinsic alignment? How do intrinsic alignments evolve with redshift? Can we predict the alignments for a mix of morphological type? Addressing these questions requires additional measurements with larger number densities, covering higher redshifts. They should be a priority, as answering these questions also sheds light on the relative importance of the physical mechanisms that give rise to the alignments. The greater uncertainty at small scales is somewhat offset by the presence of competing systematics like non-linear clustering and the influence of baryon physics on the matter power spectrum. On the one hand, this means that the pressure to fully understand intrinsic alignments at these scales is reduced, as the other sources of uncertainty may make these scales less useful regardless of our intrinsic alignment knowledge. Nevertheless, future weak gravitational lensing surveys such as Euclid or LSST aim to exploit cosmic shear down to scales of $1.5$ Mpc/$h$ \citep{KHD14}, so a dedicated programme of intrinsic alignment measurements at small scales would be beneficial and may require auxiliary datasets in addition to those planned for standard cosmic shear analysis. For example, good spectroscopic redshifts for reasonably well-sampled  galaxies representative of cosmic shear survey galaxies would be invaluable in making accurate measurements of the relevant intrinsic alignment contamination signal. 

In this review we have provided an overview of the current status of observations of intrinsic alignments, perhaps with a bias towards the impact on cosmic shear. It is clear, however, that the data that are due to become available over the next decade offer exciting opportunities to test methods for intrinsic alignment measurement and mitigation much more rigorously. The larger number density will allow us to measure intrinsic alignments in large shear catalogues with more precision, particularly for late-type galaxies, while deeper surveys will push our baseline for intrinsic alignment measurements to higher redshift. Despite all the progress, the most unclear part of the current intrinsic alignment observational landscape is certainly still the dependence on environment of the alignments on quasi- and non-linear scales. Although planned surveys may help in this regard, dedicated efforts to resolve this situation will be needed.

\section*{Acknowledgements}

We acknowledge the support of the International Space Science Institute Bern for two workshops at which this work was conceived.
We thank S. Bridle and J. Blazek for stimulating discussions.

MLB is supported by the European Research Council (EC FP7 grant number
280127)  and  by a  STFC  Advanced/Halliday  fellowship (grant  number
ST/I005129/1).

HH, MC and CS acknowledge support from the European Research Council under FP7 grant number 279396.

BJ acknowledges support by an STFC Ernest Rutherford Fellowship, grant reference ST/J004421/1. 

TDK is supported by a Royal Society URF.

RM acknowledges the support of NASA ROSES 12-EUCLID12-0004.

MC was supported by the Netherlands organisation for Scientific Research (NWO) Vidi grant 639.042.814.

AC acknowledges support from the European Research Council under the EC FP7 grant number 240185.

AK was supported in part by JPL, run under a contract by Caltech for NASA. AK was also supported in part by NASA ROSES 13-ATP13-0019 and NASA ROSES 12-EUCLID12-0004.

AL acknowledges the support of the European Union Seventh Framework Programme (FP7/2007-2013) under grant agreement number 624151.

\bibliographystyle{apj_title}

{\small
\bibliography{bibliography} 
}

\end{document}